\documentclass[10pt,twocolumn,letterpaper]{article}

\usepackage{cvpr}              

\newcommand{\camready}[1]{#1}

\usepackage[utf8]{inputenc} 
\usepackage[T1]{fontenc}    
\usepackage{url}            
\usepackage{booktabs}       
\usepackage{amsfonts}       
\usepackage{amsmath}
\usepackage{nicefrac}       
\usepackage{microtype}      
\usepackage{xcolor}         
\usepackage{algorithm}
\usepackage{algpseudocode}
\usepackage{appendix}

\usepackage{wrapfig}
\usepackage[font=small]{caption}
\usepackage{soul}

\usepackage{multirow}
\usepackage{multicol}

\makeatletter
\let\ftype@table\ftype@figure

\DeclareRobustCommand\onedot{\futurelet\@let@token\@onedot}
\def\@onedot{\ifx\@let@token.\else.\null\fi\xspace}
\DeclareMathAlphabet{\mathcal}{OMS}{cmsy}{m}{n}
\newcommand{\xe}[1]{\ensuremath{\mathbf{#1}}}

\newcommand*{\affmark}[1][*]{\textsuperscript{#1}}
\newcommand*{\affaddr}[1]{#1} 
\newcommand*{\email}[1]{#1} %

\definecolor{cvprblue}{rgb}{0.21,0.49,0.74}
\usepackage[pagebackref,breaklinks,colorlinks,allcolors=cvprblue]{hyperref}

\usepackage{amsmath,amsfonts,bm}

\def\eqref#1{equation~\ref{#1}}

\def\1{\bm{1}}

\def\vc{{\bm{c}}}

\def\vf{{\bm{f}}}

\def\vx{{\bm{x}}}

\DeclareMathAlphabet{\mathsfit}{\encodingdefault}{\sfdefault}{m}{sl}
\SetMathAlphabet{\mathsfit}{bold}{\encodingdefault}{\sfdefault}{bx}{n}

\newcommand{\mbf}[1]{\bm{\mathbf{#1}}}

\def\tP{{\mbf{P}}}

\def\paperID{6663} 
\def\confName{CVPR}
\def\confYear{2025}

\title{VideoSPatS: Video SPatiotemporal Splines for Disentangled Occlusion, Appearance and Motion Modeling and Editing}

\author{%
Juan Luis Gonzalez\affmark[1], Xu Yao\affmark[1], Alex Whelan\affmark[1], Kyle Olszewski\affmark[1], Hyeongwoo Kim\affmark[2], Pablo Garrido\affmark[1] \\
\affaddr{\affmark[1]Flawless AI} \quad
\affaddr{\affmark[2]Imperial College London}\\
\small{\email{\{juanluis.gonzalez, xu.yao, kyle.olszewski, pablo.garrido\}@flawlessai.com}} \quad \small{\email{hyeongwoo.kim@imperial.ac.uk}}
}

\begin{document}

\twocolumn[{
\renewcommand\twocolumn[1][]{#1}
\maketitle
\begin{center}\centering
  \vspace*{-4mm}
  \includegraphics[width=0.99\textwidth]{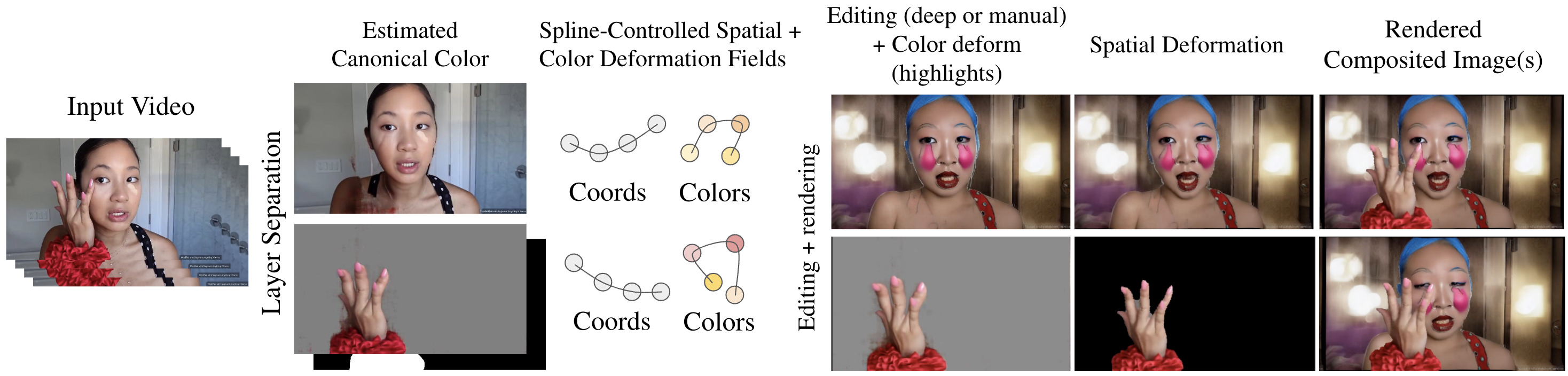}
  \vspace*{-1mm}
  \captionof{figure}{Time varying appearance (e.g. highlights, shadows, etc.) create optical flow ambiguities for video implicit representations. By adopting spatial and color deformation spline fields, our proposed method can disentangle occlusions, appearance, and motion in videos, allowing for consistent video editing, even in the presence of temporally varying texture appearance and occlusions.
  }
  \label{fig:opening_img}
\end{center}
}]

\begin{abstract}
We present an implicit video representation for occlusions, appearance, and motion disentanglement from monocular videos, which we call Video SPatiotemporal Splines (\textit{VideoSPatS}).
Unlike previous methods that map time and coordinates to deformation and canonical colors, our \textit{VideoSPatS} maps input coordinates into Spatial and Color Spline deformation fields $\mathcal{D}_s$ and $\mathcal{D}_c$, which disentangle motion and appearance in videos. With spline-based parametrization, our method naturally generates temporally consistent flow and guarantees long-term temporal consistency, which is crucial for convincing video editing.
Using multiple prediction branches, our \textit{VideoSPatS} model also performs layer separation between the latent video and the selected occluder. 
By disentangling occlusions, appearance, and motion, our method enables better spatiotemporal modeling and editing of diverse videos, including in-the-wild talking head videos with challenging occlusions, shadows, and specularities while maintaining an appropriate canonical space for editing.
We also present general video modeling results on the DAVIS and CoDeF datasets, as well as our own talking head video dataset collected from open-source web videos. Extensive ablations show the combination of $\mathcal{D}_s$ and $\mathcal{D}_c$ under neural splines can overcome motion and appearance ambiguities, paving the way for more advanced video editing models. Visit our project site\footnote{\url{https://juanluisg-flwls.github.io/videospats-website/}}.

\end{abstract}
    
\section{Introduction}
\label{sec:intro}


Implicit neural representations have shown promising results for modeling images~\cite{sitzmann2020implicit,tancik2020fourier,mildenhall2021nerf,muller2022instant} and videos~\cite{kasten2021layered,ye2022deformable,ouyang2024codef}. 
The implicit representations of videos are typically modeled as continuous functions that map spatial and temporal coordinates into color values.
This becomes a challenging modeling task when various types of motion, lighting variations, and occlusions are present.
By increasing the number of parameters in the implicit functions,
perfect video reconstructions are achievable.
However, without the implicit function disentangled, applying consistent editing to the whole sequence remains an open challenge.

Recent advances in diffusion models for text-to-video generation, such as Sora~\cite{videoworldsimulators2024}, Mochi~\cite{genmo2024mochi} and CogVideoX~\cite{yang2024cogvideox}, have made video modeling an increasingly relevant task.
Although these models have succeeded in generating impressive quality video content with high-fidelity motions, the question of how to perform semantics-aware and disentangled editing still prevails. Several existing approaches~\cite{lu2020layered,lu2021omnimatte,kasten2021layered,ye2022deformable,ouyang2024codef} propose to learn a canonical representation for a video, such that edits can be applied in this canonical space, and then propagated through the entire sequence. However, these approaches present several limitations in the wild.
First, the canonical representation is typically modeled as a static image, which struggles to capture objects with temporally varying appearance. Moreover, when large motions or occlusions are present, existing methods~\cite{kasten2021layered,ouyang2024codef} often produce distorted canonical images, making it challenging to perform semantics-aware editing. 

To address the aforementioned issues, we propose a novel approach to learning an implicit video representation that disentangles occlusion, appearance, and motion. Our method is inspired by existing work that models spatiotemporal deformation with neural spline fields~\cite{ye2022deformable,chugunov2024neuralsplines}.
Unlike raw neural representations that map coordinates directly into colors, neural spline field networks are trained to map coordinates into spline control points, which are then interpolated at sample timestamps to form color values. While previous works~\cite{lu2020layered,lu2021omnimatte,kasten2021layered,ye2022deformable,ouyang2024codef} only model a deformation field and a static canonical image, our method learns both a spatial spline deformation field and a color spline deformation field for the temporal-aware canonical space, allowing us to model time-dependent appearance in videos. Moreover, our approach handles occlusions more effectively, as neural splines generate temporally consistent flow without the need for any explicit regularization. 
Additionally, our novel approach opens up new applications, such as motion editing, since the splines can be easily edited by modifying the control points, which is a non-trivial task with previous methods based on raw neural representations. Our main contributions can be summarized as follows:

\begin{itemize}
\setlength\itemsep{0em}

\item We propose a novel implicit video representation that disentangles occlusions, appearance, and motion from monocular videos.

\item We introduce, to our knowledge, the first temporal canonical space for modeling time-dependent appearance.

\item We achieve improved editability through a more consistent, state-of-the-art canonical space representation.
\end{itemize}

\section{Related Work} 
\label{sec:relatedwork}

\noindent\textbf{Layered Video Decomposition.} 
Factorizing appearance and motion in a video to decompose the video into layers is a longstanding problem in computer vision~\cite{wang1994representing,jojic2001learning,zitnick2004high,pawan2008learning}. 
The seminal work by Rav-Acha~\etal~\cite{rav2008unwrap} models video frames as a 2D-to-2D mapping from an object’s texture map to the image, reconstructing an "unwrap mosaic" representation. The editing can be applied to the mosaics and then composited back to the original sequences using the learned mapping. With recent advances in deep learning, several works propose using neural networks to decompose videos into layers. Lu~\etal~\cite{lu2020layered,lu2021omnimatte} propose learning a layered video representation in which each frame is decomposed into separate RGBA layers that represent the appearances of different people in the video. Kasten~\etal~\cite{kasten2021layered} propose decomposing a video of a dynamic scene into a set of layered neural atlases, with a single atlas per object, using a coarse mask identifying the object of interest as input.
Ye~\etal~\cite{ye2022deformable} address this problem in an unsupervised manner by decomposing the video into layers of persistent motion groups without requiring object masks. Ouyang~\etal~\cite{ouyang2024codef} model a video as a canonical content field and a temporal deformation field using a hash-based architecture for warping and reconstruction, significantly reducing training time.  Omnimatte RF~\cite{OmnimatteRF} synthesizes fully-visible layers of individual objects with their associated effects from a video. 
Although these approaches achieve good reconstruction quality, they still have several limitations. First, the canonical representation is usually modeled as a static image, which sometimes fails to capture objects with appearance changes over time. Moreover, existing methods often produce distorted canonical images for videos with complex motions, making semantic-aware editing difficult and artifact-prone. On the contrary, our method learns a temporally aware canonical space and generates a more regularized canonical image, making it better suited for semantics-aware editing.

\noindent\textbf{Occlusion-Aware Editing.} Many videos contain occlusions in real-life scenarios, and performing occlusion-aware editing on such videos is still an unsolved problem in the existing literature. This holds true since edits must retain the original occluder-background relationship and preserve temporal consistency.
A straightforward approach to address this is to use occlusion detection methods~\cite{liao2020occlusion,park2022handoccnet} to segment out the occluded regions, turning it into a video inpainting problem~\cite{zhou2023propainter,yang2023deep}. Nevertheless, occlusions can vary widely in type and extent, making it challenging to train a single model to detect them all.
Several studies~\cite{yuan2023make,xu2024n} attempt to address editing on occluded faces using 3D-aware GAN inversion to perform editing in the latent space. However, GAN inversion techniques often fail to achieve perfect reconstruction and are prone to artifacts. \camready{Diffusion-based editing approaches, such as frame-guided video editing~\cite{I2VEdit} or motion editing~\cite{revideo}, are also constrained in generation and reconstruction by the underlying model.}
Our work is inspired by recent research on burst image fusion by Chugunov~\etal~\cite{chugunov2024neuralsplines}, where motion is modeled using neural spline fields. Although our work shares similarities with that of Chugunov~\etal in its use of neural spline fields to model motion, our methodology stands out in several ways. 
First, we use neural spline fields to model continuous videos rather than sparse burst images. Furthermore, we introduce both a spatial deformation spline field and a color spline deformation field, allowing us to handle temporally varying appearances and much larger motions.

\noindent\textbf{Neural Scene Representations.} The flexibility of neural representations have proven effective for representing content in a variety of domains, such as 2- or 3-dimensional scenes, without the limitations of traditional discrete representations such as pixels and voxels, or explicit surface representations such as meshes.
This enables the mapping of continuous coordinates to a variety of learned signals such as 2D image content~\cite{tancik2020fourier}, 3D surfaces~\cite{sitzmann2020implicit,martel2021acorn}, 3D volumes~\cite{mildenhall2021nerf,Gao-ICCV-DynNeRF, park2021nerfies, cao2023hexplane}, and combined representations of 3D structure and appearance~\cite{sitzmann2019srns}. Our work adopts implicit representations to model 2D motion separation and dynamic appearance changes over time. To the best of our knowledge, we are the first work addressing this problem using an implicit video representation.  

\section{Method}
\label{sec:method}

Our aim is to disentangle occlusions, motion, and appearances in videos, focusing mainly on talking face videos (\eg, Fig.~\ref{fig:opening_img}). Given a video and a user-provided mask indicating the object of interest, we model each object (background talking face and foreground occluder) with a canonical representation $C_{rep}$ and a deformation field $\mathcal{D}$, akin to \cite{kasten2021layered, ouyang2024codef}.
However, unlike the previous methods that adopt a spatiotemporal model to infer a deformation field, ours is estimated by a cubic spline interpolation, whose \textit{control points} are estimated by our deformation model.
Moreover, our motion canonical representation is further parameterized with \textit{color deformation spline fields} to predict color control points, enabling efficient modeling of lighting changes.

Proper disentangling of video content changes due to varying motion and appearance is a crucial aspect of video processing and editing methods, and it has proven challenging for previous work in this area. 
Common but straightforward motion representations, \eg optical flow, fail due to simplifying assumptions like the constancy of the brightness and spatial gradient of content as it moves throughout the image~\cite{horn1981determining,brox2004high}.
Such assumptions do not hold when varying surface illumination causes complex appearance changes, such as specular highlights or shadows.
This can cause failures to identify when content is in motion rather than when its appearance is changing, or for these transient effects to be ignored in favor of more common features of the extracted content's appearance.
If these time-dependent effects are not properly captured and applied to the moving surfaces in the image, noticeable disparities between the original and edited content will occur.
Our method addresses this with the extraction of a base color representation for this content, akin to the surface albedo used when rendering lighting effects in computer graphics.
This enables us to apply time-dependent changes to this base color before the content's transformation by our deformation fields.

\subsection{Preliminaries}
\label{subsec:preliminaries}

\textbf{Neural Spline Fields.}
Neural Spline Fields (NSF)~\cite{chugunov2024neuralsplines} are used to learn to integrate content from multiple images captured at different times and locations.
By learning NSF models capturing the appropriate transformation between the points in each real image and separate, reconstructed canonical images, \eg for the intended capture target and occluding objects, and an alpha matte defining the transmission between these components, the appropriate content can be assigned to the reconstructed images.
Doing this, however, requires learning the appropriate, continuous transformations mapping the real images content to their corresponding points in these reconstructions.

Splines~\cite{bartels1995introduction,farin2001curves} allow for the use of sparse control points $\tP$ to define complex curves as a piecewise polynomial function.
Given these points and the spline formula $S(\tP,t)$, adjusting the interpolation parameter $t \in [a,b]$ (typically $[0,1]$) enables smooth, continuous traversal of this curve, which enables defining the trajectory of a point traveling through a given space, \eg as a function of time.

An NSF model thus consists of a learned mapping $f_{\theta}(u,v): \mathbb{R}^2 \to \mathbb{R}^{N \times D}$ from image coordinates $\vx = (u,v) \in [0,1]$ to a set of $D$-dimensional spline control points $\tP$ defining the deformation $\vx + \Delta_\vx$ at time $t$:
\begin{equation}\label{eq:spline_field}
   \Delta_\vx = (\Delta_u, \Delta_v) = S(\tP = f_{\theta}(u,v),t)
\end{equation}
Given a set of control points $\tP$ and time parameter $t$, we define our cubic Hermite splines $S(\tP,t)$ using the standard formula, as in~\cite{chugunov2024neuralsplines}:
\begin{align}\label{eq:cubic_spline}
    S(\tP, t) &= (2t_r^3 - 3t_r^2 + 1) \tP_{\lfloor t_s \rfloor} + (-2t_r^3 + 3t_r^2) \tP_{\lfloor t_s \rfloor + 1} \nonumber\\
    &+ (t_r^3 - 2t_r^2 + t_r)(\tP_{\lfloor t_s \rfloor} - \tP_{\lfloor t_s \rfloor - 1})/2 \nonumber \\ 
      &+ (t_r^3 - t_r^2)(\tP_{\lfloor t_s \rfloor + 1} - \tP_{\lfloor t_s \rfloor})/2\nonumber \\
    t_r &= t_s - \lfloor t_s \rfloor, \quad t_s = t \cdot |\tP|.
\end{align}

This spline formulation allows for continuity along the resulting path, while enabling efficient evaluation~\cite{bartels1995introduction}.
Note that control points $\tP$ can refer to vectors with two spatial components $(x,y)$ or three color components $(R,G,B)$ to represent the spatial ($\mathcal{D}_s$) and color ($\mathcal{D}_c$) deformation fields, respectively.
For a set of $N$ control points, we use $\tP_s \in \mathbb{R}^{N \times 2}$ for the former and $\tP_c \in \mathbb{R}^{N \times 3}$ for the latter.

\begin{figure*}[t]
  \centering 
  \vspace*{-8mm}
  \includegraphics[trim={0 3cm 0 3cm}, clip,width=1.0\textwidth]{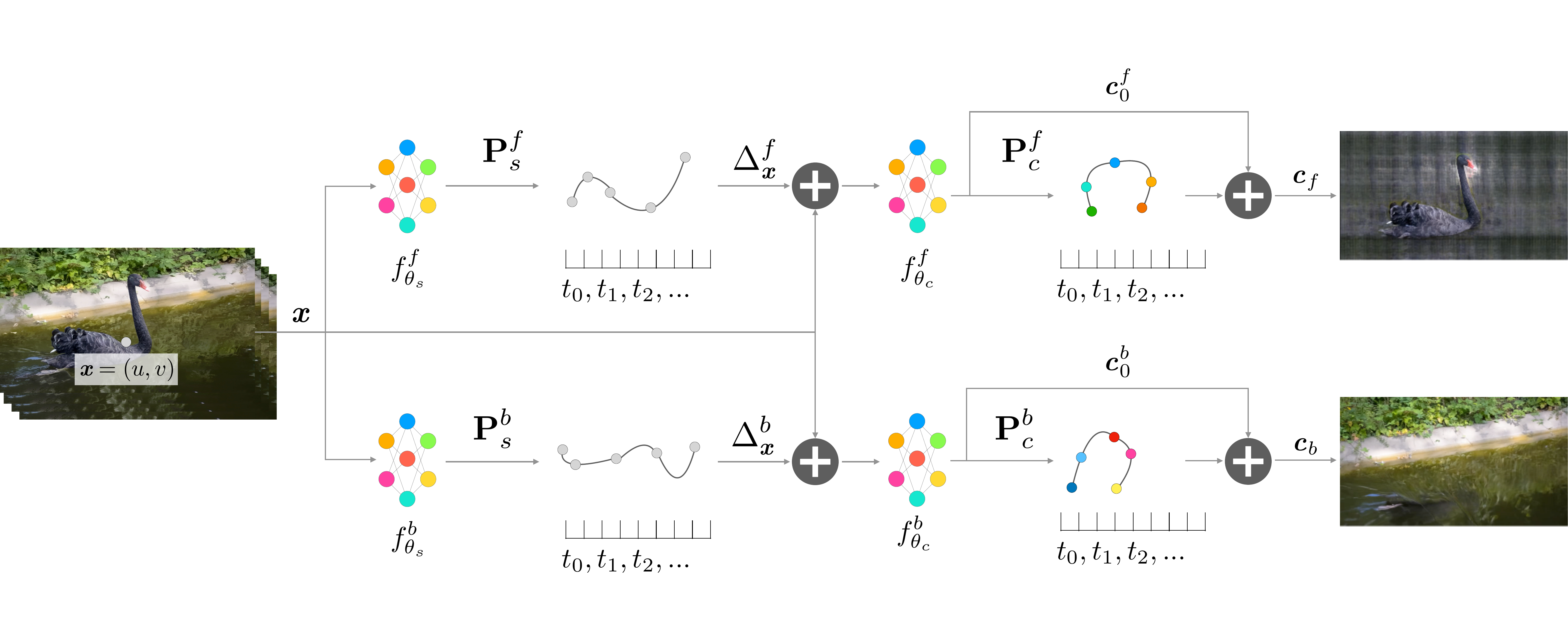}
  \vspace*{-5mm}
  \caption{\textbf{Overview.} Our Video SPatiotemporal Spline model, referred to as \textit{VideoSPatS}, disentangles occlusions, motion, and appearance into editable layers. Using independent branches for the regions of interest, it learns Neural Spline Fields (Sec.~\ref{subsec:preliminaries}) separating the foreground $f$ (\textit{top}) and background $b$ (\textit{bottom}) into editable, canonical representations.
  Given a sequence of video frames (\textit{left}), image coordinates $\vx$ are used as input to train the spatial MLPs $f_{\theta_s}$ for each region to infer spline control points $\tP_s$, which define the trajectories of the corresponding image content through the video frames (\textit{middle left}, Sec.~\ref{subsec:arch}).
  Interpolated points on this path used as input to the color MLPs $f_{\theta_c}$, which infer a base color $\vc_0$ and to infer spline control points $\tP_c$ used to smoothly interpolate the color values along the path (\textit{middle right}).
  An additional MLP, $f_{\theta_\alpha}(\vx + \Delta^f_{\vx})$ (not shown here), is used to predict the alpha matte for foreground/background compositing (Eq. (\ref{eq:c})).
  By reconstructing the input videos, these branches are trained to infer canonical representations of the content and appearance of the foreground and background regions (\textit{right}, Sec.~\ref{subsec:optim}).}
  \label{fig:arch}
  \vspace{-3mm}
\end{figure*}

\vspace*{2mm}
\noindent
\textbf{Alpha Compositing.}
To reconstruct each input image $\vc$, we learn separate NSF models $f_{\theta}^f$ and $f_{\theta}^b$, which map the image content to the appropriate regions in the reconstructed foreground (or occluder) image layer $\vc_f$, and background (or face) image layer $\vc_b$, respectively, as well as the alpha mask $\alpha$ used to composite them:
\begin{equation} \label{eq:c}
\vc = \alpha \vc_f + (1 - \alpha) \vc_b.
\end{equation}

\subsection{Model Architecture}
\label{subsec:arch}
Figure \ref{fig:arch} illustrates our overall model, which takes as input a coordinate $\vx=(u,v) \in [-1,1]$ at time $t \in [0,1]$, and predicts disentangled representation of occlusions, motion, and appearance to render the final scene color $\vc$ (Eq. (\ref{eq:c})).

Our deformation model can be a simple yet effective multi-layer perceptron (MLP) network with either periodical positional embeddings \cite{attentionallyouneed} or 2D-hash encodings \cite{muller2022instant}. For each foreground and background branch, the spatial deformation model $f_{\theta_s}$ maps each image coordinate $\vx$ to a set of spatial deformation spline control points $\tP_s = f_{\theta_s}(\vx)$. Then, to obtain the spatial deformation $\Delta_\vx$, we apply the Hermite spline interpolation (Eq. (\ref{eq:cubic_spline})) at time $t$, described as follows:
\begin{equation} \label{eq:delta_x}
\Delta_\vx = S(\tP_s, t).
\end{equation}

The deformed coordinates $\vx + \Delta_\vx$ are then fed to the color deformation model $f_{\theta_c}$, with a similar structure to $f_{\theta_s}$, which outputs a set of color deformation control points $\tP_c$ and a base color $\vc_0$, as follows:
\begin{equation} \label{eq:c_0}
\vc_0, \tP_c = f_{\theta_c}(\vx + \Delta_\vx).
\end{equation}
Next, we apply the Hermite Spline interpolation on the color space to obtain the final color deformation at time step $t$ and then add it to the base color $\vc_0$. The final color (for the background or foreground branch) is given as follows:
\begin{equation} \label{eq:c_final}
\vc = \sigma(\vc_0 + S(\tP_c, t)),
\end{equation}
where $\sigma(\cdot)$ $\in$ $[0, 1]$ is the sigmoid activation function to ensure outputs with valid color image values.

Finally, an additional implicit model $f_{\theta_\alpha}$ maps the foreground deformed coordinates $\vx + \Delta_\vx$ into an opacity value $\alpha$ as given by
\begin{equation} \label{eq:alpha}
a = \sigma(f_{\theta_\alpha}(\vx + \Delta_\vx, t)),
\end{equation}
where $\alpha$ is utilized to composite the rendered foreground and background colors, $\vc_f$ and $\vc_b$ respectively, into the final predicted color $\vc$ by Eq. (\ref{eq:c}).

\subsection{Objective Functions}
\label{subsec:optim}
We train our neural spline fields with a combination of reconstruction and regularization losses detailed below.

\vspace*{2mm}
\noindent\textbf{Reconstruction Loss $l_{rec}$.} This loss ensures the final composited color $\vc$ matches the corresponding target GT color $\vc^*$ at pixel location $\vx$ and time step $t$. We compute $l_{rec}$ as the component-wise average absolute error, as follows:
\begin{equation} \label{eq:rec_loss}
l_{rec} = ||\vc - \vc^*||_1,
\end{equation}
where $l_{rec}$ is averaged for all input coordinates in a batch.

\vspace*{2mm}
\noindent\textbf{Optical flow guidance loss $l_{fl}$.} 
While this loss is not an absolute constraint, it helps in resolving motion ambiguities caused by large pixel displacements and has extensively been used in prior work~\cite{kasten2021layered, ouyang2024codef}. We can deem optical flow as a dense correspondence map between two consecutive frames $t_0$ and $t_1$, i.e., $\vx_0$ at time $t_0$ corresponds to $\vx_0 + \vf_{0 \rightarrow 1}$ at time $t_{1}$. We leverage this idea and model $l_{fl}$ such that it encourages that corresponding input coordinates are mapped to the same canonical coordinate, as follows:
\begin{equation} \label{eq:flow_loss}
l_{fl} = ||S(\tP_s(\vx_0), t_0) - S(\tP_s(\vx_0 + \vf_{0 \rightarrow 1}), t_{1})||_1,
\end{equation}
where optical flows are obtained from RAFT \cite{teed2020raft} and filtered by cycle consistency following \cite{ouyang2024codef}.

\vspace*{2mm}
\noindent\textbf{Spatial splines deformation regularization loss $l_{\mathcal{D}_s}$.} This regularization loss is composed of two terms: A (i) motion control point smoothness loss $l_{sm}$ and (ii) control point velocity direction loss $l_{pv}$.

The first term $l_{sm}$ encourages neighboring coordinates to be mapped to similar sets of control points, as follows:
\begin{equation} \label{eq:sm_loss}
l_{sm} = ||S(\tP_s(\vx), t)) - S(\tP_s(\vx + (u_0, v_0)), t))||_1,
\end{equation}
where $(u_0, v_0)$ is a $1$-pixel shift in both the horizontal and vertical coordinate axes.

Even when the spline representation already models a smooth curve, we further provide a regularization term that can only be applied to a spline deformation field. The second term $l_{pv}$ encourages that the tangent velocity described by the control points change its direction slowly, as follows:
\begin{equation} \label{eq:pv_loss}
l_{pv} = \partial\left(\tfrac{\partial \tP_s}{\partial t}\odot\left(||\tfrac{\partial \tP_s}{\partial t}||_{u,v}\right)^{-1}\right) \Big / \partial t.
\end{equation}
We remark that regularizing the change in direction (and not the magnitudes) allows more freedom to the learned control points. The final spatial splines deformation regularization loss is then given as the sum of the two terms above, i.e., $l_{\mathcal{D}_s} = l_{sm}+l_{pv}$.

\vspace*{2mm}
\noindent\textbf{Color deformation regularization loss $l_{\mathcal{D}_c}$.} This loss encourages a disentangled representation of motion and appearance. Even though the optical flow loss guidance provides structural consistency, $l_{\mathcal{D}_c}$ prevents large appearance displacements by restricting color deformation control points. $l_{\mathcal{D}_c}$ is given as:
\begin{equation} \label{eq:color_loss}
l_{\mathcal{D}_c} = ||\tP_c||^2_2.
\end{equation}
Note that we use the $\ell^{2}$ norm to penalize large color deformation while allowing small color changes.

\vspace*{2mm}
\noindent\textbf{Layer separation loss $l_{sep}$.} This loss comprises four terms: A guidance loss, a regularization loss, a boundary loss, and an error maximization loss. The first term, $l_{guide}$, encourages the estimated mask to be similar to that of the user provided coarse mask $\alpha^* \in [0, 1]$, where the guidance is defined by $1$-valued elements. $l_{guide}$ is then given by:
\begin{equation} \label{eq:guide_loss}
l_{guide} = \tfrac{N}{\sum_i \alpha_i^*}\alpha^* |\alpha - \alpha^*|,
\end{equation}
where $\tfrac{N}{\sum_i \alpha_i^*}$ normalizes the loss over the valid coordinates $i$, such that only guided values contribute to the loss. $N$ is the number of coordinates in the batch.

The second term, $l_{reg} = m_k \alpha$, minimizes the foreground opacity, where the regularization mask $m_k$ is valued at 1 if $\alpha$ lies at least $k$ pixels away from the guidance mask. The third term ensures a soft transition between $m_k$ and $\alpha^*$ by further regularizing $\alpha$ with $l_{bound} = ((1 - m_k) * (1 - \alpha^*)) \alpha$. 

The last term in $l_{sep}$ aids in reconstructing a detailed $\alpha$ mask from a coarse $\alpha^*$ by maximizing the error between the masked region in the rendered background color and the ground truth color, as given by
\begin{equation} \label{eq:mxe_loss}
l_{mxe} = - \tfrac{N}{\sum_i \alpha_i} \alpha |\vc^* - \vc_b|,
\end{equation}
where $\alpha$ can maximize the error $|\vc^* - \vc_b|$ if it is similar to the corresponding unavailable alpha mask. This is possible because each layer is initially biased to render the semantic class that they are more exposed to, thus, at early iterations alpha converges to a reasonable occlusion mask, given a sufficiently good guidance mask (see supplement for details).

The final layer separation loss is then given as the sum of the four terms above, i.e.,
\begin{equation} \label{eq:sep_loss}
l_{sep} = l_{guide} + \lambda_{reg} l_{reg} + \lambda_{bound} l_{bound} + \lambda_{mxe} l_{mxe},
\end{equation}
where $\lambda_{bound} = 0.01$, $\lambda_{reg} = 0.5$, and $\lambda_{mxe} = 0.1$ are empirically set to be approximately in the same magnitude order as $l_{rec}$.

\vspace*{2mm}
\noindent\textbf{Our final total loss $l_{total}$} can then be summarized as:
\begin{equation} \label{eq:total_loss}
l_{total} = l_{rec} + \lambda_{fl} l_{fl} + \lambda_{\mathcal{D}_s} l_{\mathcal{D}_s} + \lambda_{\mathcal{D}_c} l_{\mathcal{D}_c} + l_{sep},
\end{equation}
where $\lambda_{fl}$, $\lambda_{\mathcal{D}_s}$, and $\lambda_{\mathcal{D}_c}$ are empirically set to have balanced impact with respect to $l_{rec}$.

\begin{figure}[t]
  \centering 
  \includegraphics[width=0.48\textwidth]{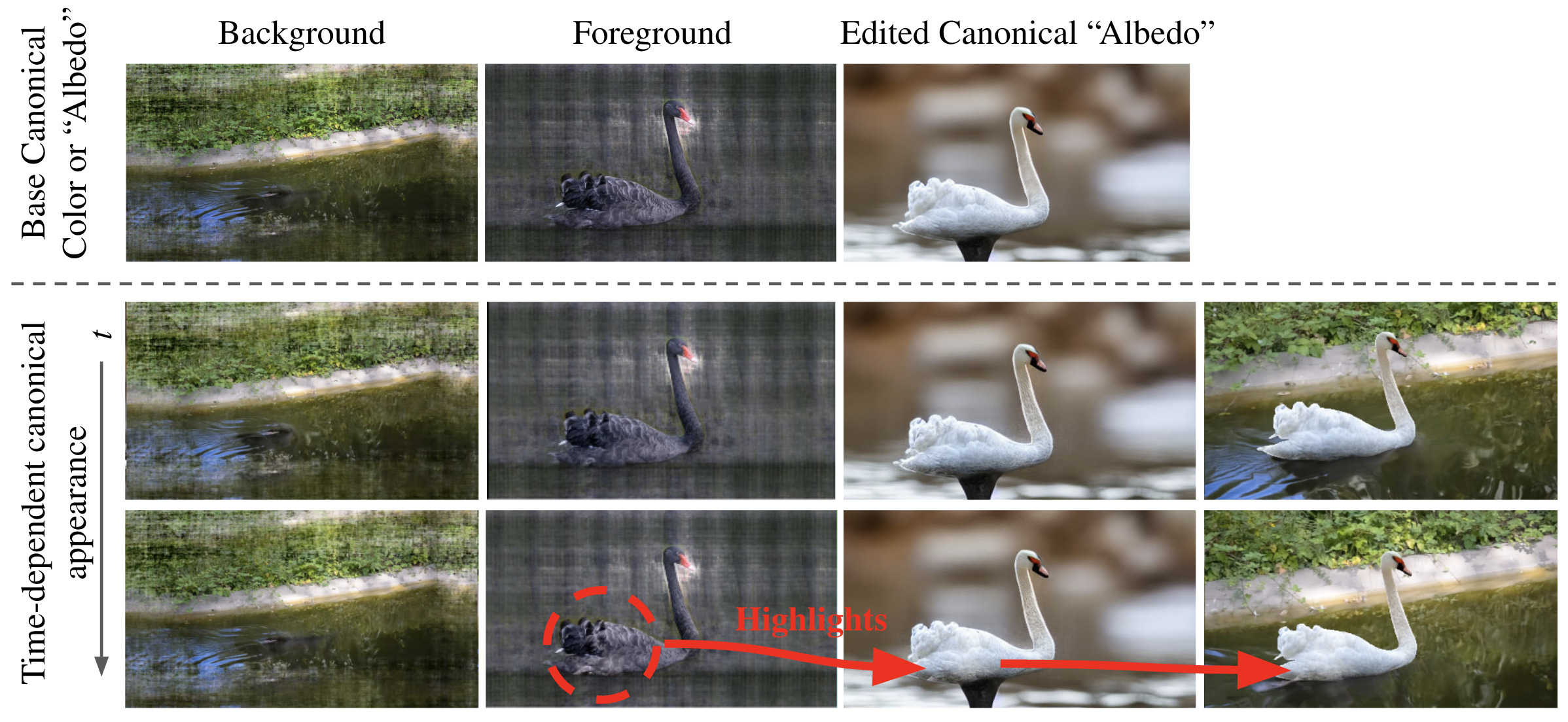}
  \vspace*{-4mm}
  \caption{Our \textit{VideoSPatS} modeling provides more naturally edited videos by disentangling occlusions, motion, and appearance. See supplement for videos.}
  \label{fig:color_deform}
\end{figure}

\begin{figure*}[t]
  \centering 
  \includegraphics[width=1.0\textwidth]{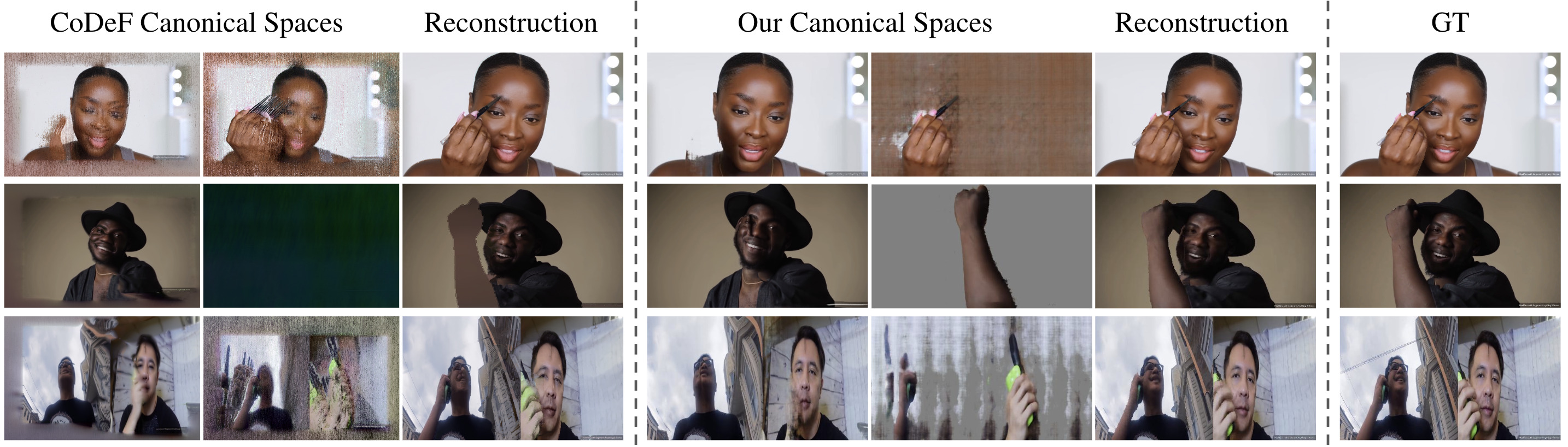}
  \vspace*{-5mm}
  \caption{Results on YT-in-the-wild videos (Creative Commons videos from YT, see supplemental for per-video details).}
  \label{fig:yt_results}
\end{figure*}
\subsection{Editing}
\label{subsec:editing}
Unlike previous works that predict single fixed canonical images \cite{kasten2021layered, ouyang2024codef}, our method predicts a \textit{continuous color deformation spline field}, which means that appearance disentangled from motion can be controlled by interpolating this color deformation field. For this reason, we propose a new approach to perform editing on the base colors $\vc_0$ and then apply the temporally varying appearance on top. This allows for better highlights, shadows, or other changes in intensity values that are not to be modeled as motion in the scene. Given editing algorithm $\mathcal{A}(\cdot)$, rendered base color image \xe{I_0}, and color deformation image $\Delta_{\xe{I}_t}$, the editing of our canonical space is given by
\begin{equation} \label{eq:editing}
\xe{I}^{C_{ed}}_t(\vx) = \sigma\left(\ln{\tfrac{\mathcal{A}(\xe{I}_0(\vx))}{(1 -\mathcal{A}(\xe{I}_0(\vx)))}} + \Delta_{\xe{I}_t}(\vx)\right),
\end{equation}
where images $\xe{I}_0$ and $\Delta_{\xe{I}_t}$ are obtained by sampling all $\vx$ within a window (and applying $S(\tP_c(\vx), t)$) in $f_{\theta_c}$. The consistent video propagation of the edited canonical space is then given by 
\begin{equation} \label{eq:propagation}
\xe{I}^{ed}_t(\vx) = \xe{I}^{C_{ed}}_t(S(\tP_s(\vx), t)),
\end{equation}
which is, in practice, approximated via bilinear interpolation using commonly available functions in deep learning libraries, such as \textit{grid\_sample}. This process is depicted in Fig. \ref{fig:color_deform}, where deep editing (by Stable Diffusion \cite{stable_difussion}) is applied to the canonical base or ``albedo'' color. \camready{For more conceptual details on motion and appearance decomposition for video editing, please see the supplementary material.}

\subsection{Initialization}
\label{subsec:init}
We implicitly encourage a canonical space that resembles the observed space by initializing the last fully connected layers of $f_{\theta_s}$, $f_{\theta_c}$, and $f_{\theta_\alpha}$ with zeros and no bias. This prevents large-valued mappings (e.g. far away control points) from appearing in early iterations.

\section{Experiments and Results}
\label{sec:results}
We present extensive experimental results and ablation evaluations in this section. Additional videos and extended experiments can be found in the supplemental materials.

\subsection{Implementation Details} 
We train our model, the \textit{VideoSPatS}, with a batch size of 10k random coordinates for up to 100k iterations using an initial learning rate of 1e-4, which is progressively halved at $50\%$, $70\%$, $80\%$, and $90\%$ of the training iterations. We train our models via Adam optimization with default betas ($\beta_1=0.9$, $\beta_2=0.999$). For optimal video fitting, we set the number of control points to the number of video frames. For a smoother, more regularized fitting, we fix the number of control points to be half of the number of frames, i.e., the curve described by 3 control points should satisfy the reconstruction of six frames. Note that we assign the same number of control points for both spatial and color deformation. Our models require about 4 GB of an A10G GPU for training and testing. Under this setup, a 50-frame length video with a 512$\times$288 image resolution is learned in 90 minutes. 

For all our main experiments, all our MLPs share the same architecture inspired by \cite{mildenhall2021nerf}, an 8 fully-connected-ReLu layer network with 256 channels and a skip connection at layer 4 with positional encoding at its inputs. The only difference among our MLPs is in the number of input and output channels. We also apply an optimization schedule to our color deformation model learning, using $\vc_0$ only for $50\%$ of the iterations in the reconstruction loss. Such a schedule helps prevent appearance and motion entanglement until the deformation field has warmed up.

\vspace*{1mm}
\noindent\textbf{YT-in-the-wild video dataset.} We collected a set of 12 short video clips from YouTube under the Creative Commons license. These videos each contain at least 50 consecutive frames (no cuts), showing people in motion and dynamic occlusions. We downscale these videos to 512$\times$288 for faster experimentation and run SAM 2~\cite{ravi2024sam2} to extract coarse foreground-background masks for training, which are eroded up to 11 pixels to simulate even coarser masks. For a 11-pixel eroded guidance mask, we set up the boundary mask $m_k$ with $k=31$ to test our alpha mask refinement objective functions.

\vspace*{1mm}
\noindent\textbf{DAVIS \cite{davis_data_Perazzi_CVPR_2016}}. This dataset comprises casual videos of subjects (usually a single one) being filmed. We use the provided segmentation masks and a resolution of $432 \times 768$ for training.

\vspace*{1mm}
\noindent\textbf{CoDeF dataset.} This public dataset, released with a state-of-the-art video representation model, CoDeF~\cite{ouyang2024codef}, contains short videos taken mostly from movies, featuring different characters and complex scenes. We use the guidance masks provided in the dataset's repository for training.\looseness=-1

\subsection{Evaluation}
We evaluate our method \textit{VideoSPatS} on the aforementioned three datasets and compare it against the state-of-the-art video modeling method, CoDeF~\cite{ouyang2024codef}. Our results show that our method generalizes well to both in-the-wild videos with complex occlusions and fast motion, as well as to general scenes, such as those from DAVIS and CoDeF dataset.

\vspace*{2mm}
\noindent\textbf{Result on YT-in-the-wild video dataset.}
Fig.~\ref{fig:yt_results} depicts canonical spaces for both the face background and the occluder foreground, along with composited rendering results for CoDeF and our method. As noted, although CoDeF is able to render a final composited image, it struggles to consistently model the canonical spaces of fast moving objects, leading to suboptimal foreground-background separation and motion disentanglement. This can be observed in the repeated instances in the first and second columns of Fig.~\ref{fig:yt_results}. In contrast, our method generates consistent canonical spaces, even in the challenging scenarios where multiple moving objects are present, such as in the case of the bottom row.

\begin{figure}[t]
  \centering 
  \includegraphics[width=0.49\textwidth]{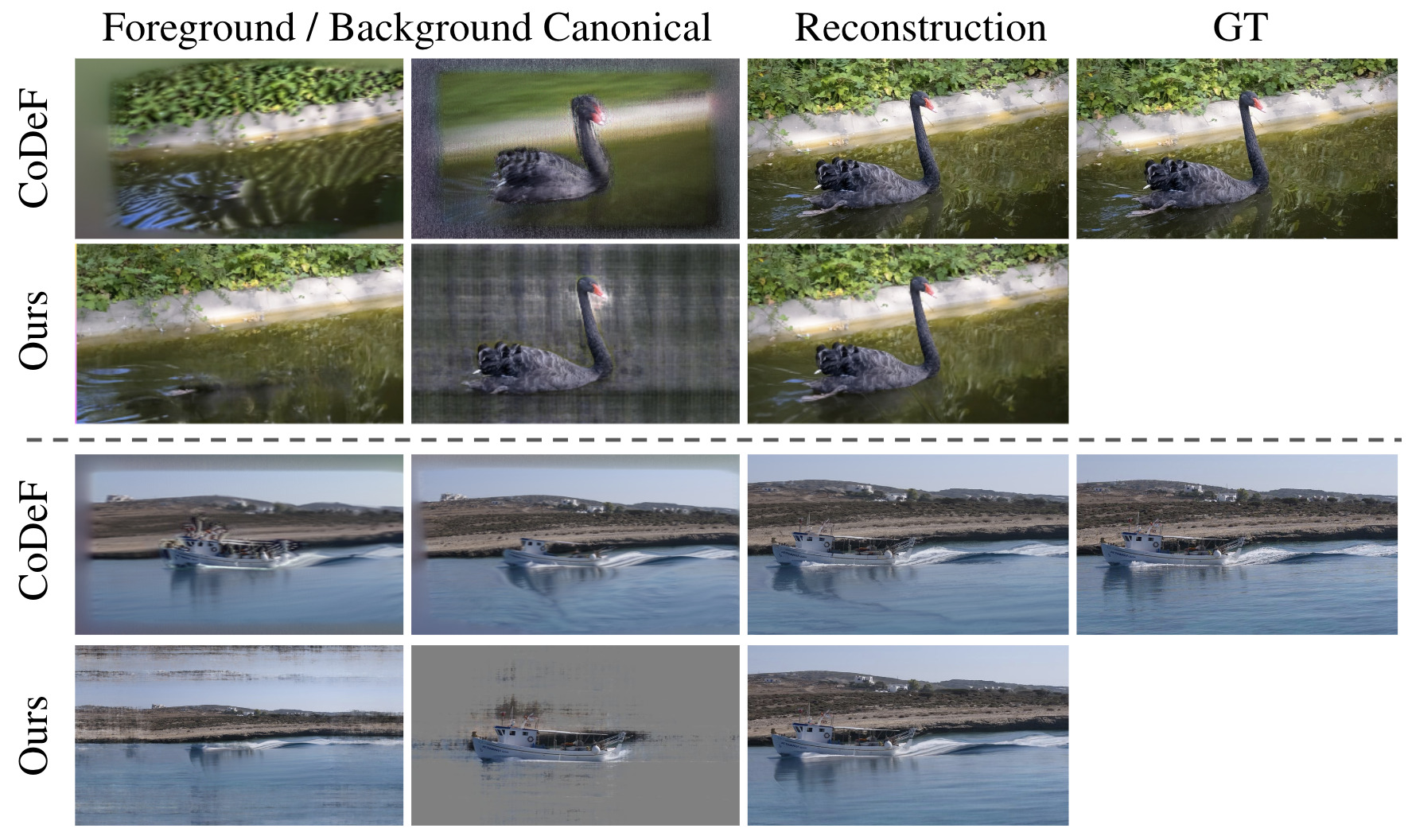}
  \vspace*{-4mm}
  \caption{Results on DAVIS \cite{davis_data_Perazzi_CVPR_2016}. Compared with CoDeF \cite{ouyang2024codef}, our method generates consistent and separated canonical spaces.}
  \label{fig:davis}
\end{figure}

\vspace*{2mm}
\noindent\textbf{Results on DAVIS.}\quad
Fig.~\ref{fig:davis} shows our fitting results on several scenes from the DAVIS dataset, which are compared with those of CoDeF. 
While CoDeF achieves slightly better reconstruction, it struggles to consistently model a semantically reasonable canonical space, as observed in the noisy black swan example of Fig.~\ref{fig:davis} (top row). We remark that obtaining an editable canonical space is much more critical than high reconstruction quality, as it enables propagating semantic-aware edits. 
We also highlight that our method separates foreground and background contents more reliably, as shown in the boat renderings of Fig.~\ref{fig:davis} (bottom row). \camready{We measured video editing quantitative results in terms of warping consistency between edited and warped-and-edited frames. We used RAFT \cite{teed2020raft} to obtain the original frames' optical flow to warp edited frames at $t$+$n$ into $t$. Ours outperforms CoDeF \cite{ouyang2024codef} and Def. Sprites \cite{ye2022deformable} by the considerable margins of \textbf{4.66dB} and \textbf{0.4dB}, respectively. See the supplemental for more details.}

\noindent\textbf{Editing results.} Fig.~\ref{fig:color_deform} illustrates the effectiveness of our editing method on the `blackswan' scene of the DAVIS dataset.
Thanks to our proposed color spline deformation fields, temporally varying appearance, such as highlights, can be propagated into the deep edited video. Please refer to the supplemental material for more editing results.

\begin{figure}[t]
  \centering 
  \includegraphics[width=0.48\textwidth]{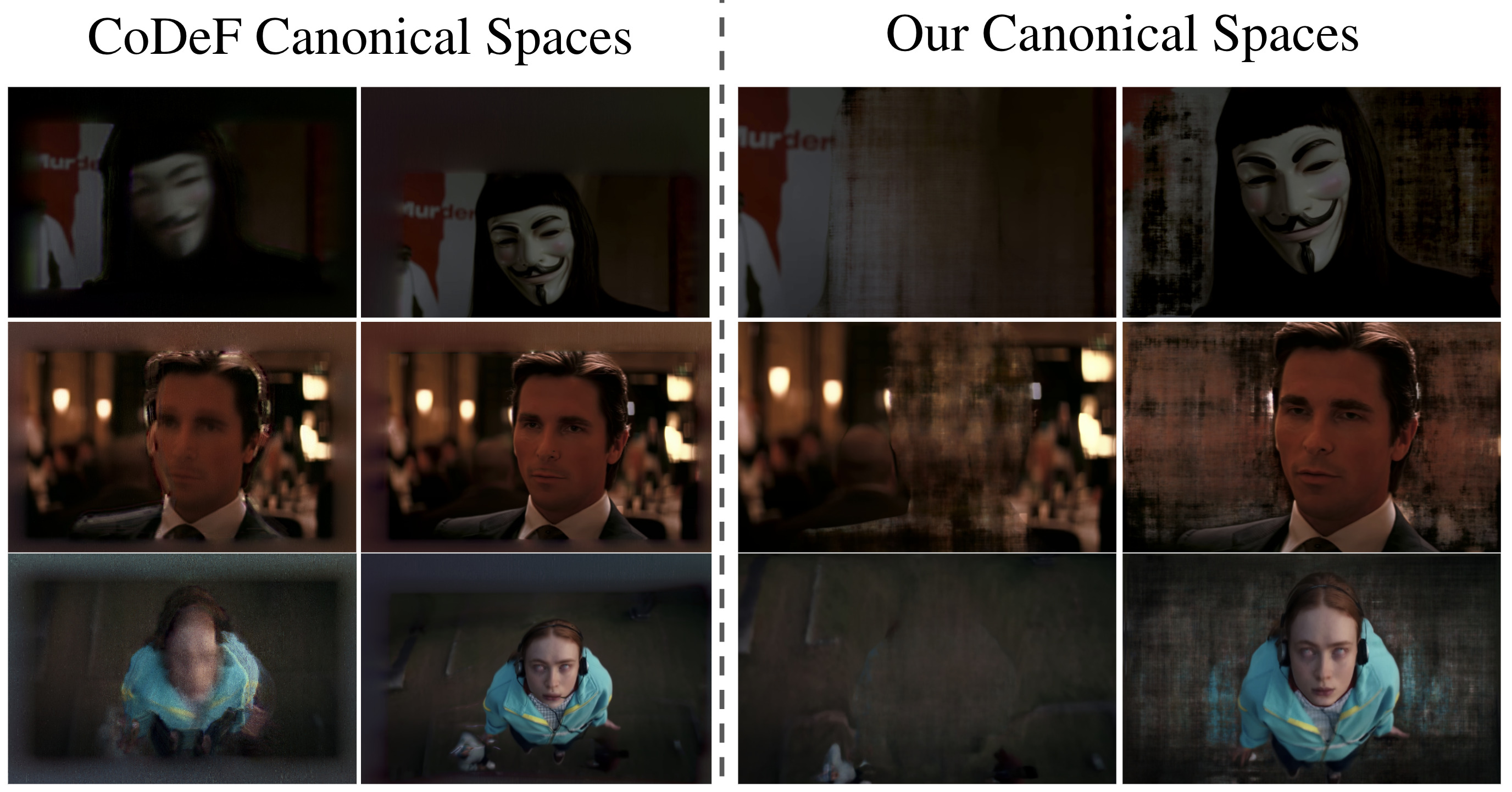}
  \vspace*{-4mm}
  \caption{Qualitative comparisons of inferred canonical spaces on the CoDeF Dataset \cite{ouyang2024codef}.}
  \label{fig:codef_data}
\end{figure}

\vspace*{2mm}
\noindent\textbf{Results on CoDeF dataset.}
Fig.~\ref{fig:codef_data} shows a qualitative comparison of the learned canonical spaces obtained by CoDeF and our proposed method. Due to the simplicity of the motion on the CoDeF dataset, the final composited rendering is not displayed here (see supplemental material). As can be noted, even when both approaches generate reasonable canonical spaces for the characters in the videos, only our method implicitly learns to reconstruct a clean background and maintains the canonical space of the foreground well aligned and ill-formed with respect to the input video.

\begin{figure*}[t]
  \centering 
  \includegraphics[trim={0 0 8cm 0}, clip, width=0.95\textwidth]{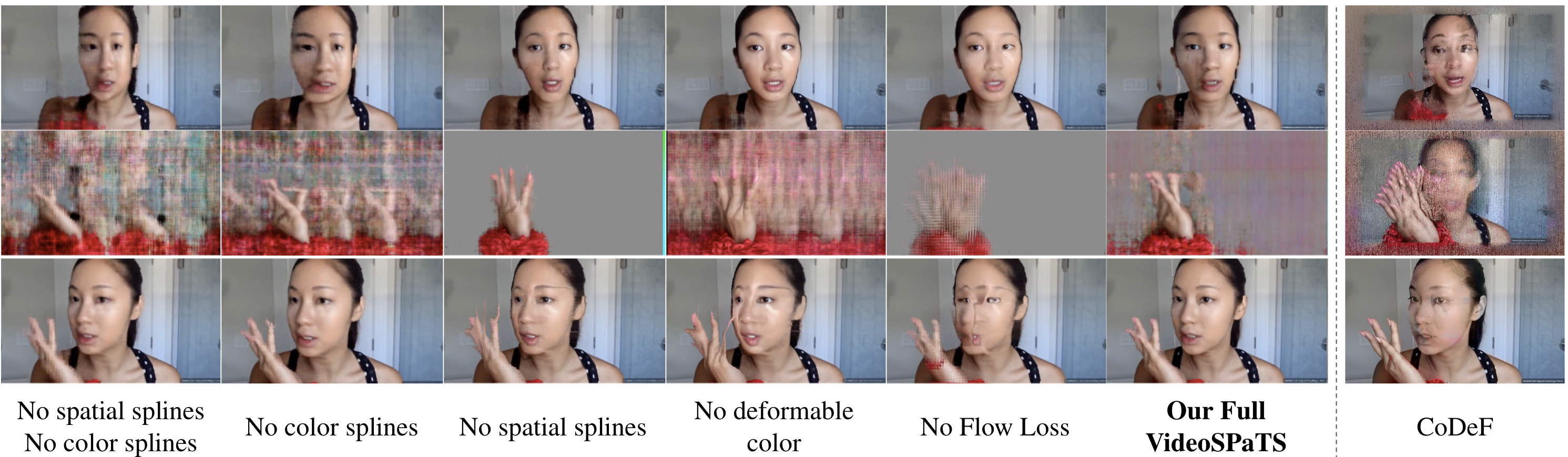}
  \vspace{-2mm}
  \caption{Ablations. From left to right, each feature in our \textit{VideoSPatS} produces better renderings and canonical spaces suitable for editing.}
  \label{fig:ablation_studies}
  \vspace{-4mm}
\end{figure*}

\subsection{Ablation Analysis}
We present extensive ablation analysis, illustrating the effects of each of our contributions and design choices, such as our proposed spatial and color deformation spline fields and different objective terms. Fig. \ref{fig:ablation_studies} shows the ablation studies that lead to an improved canonical space reconstruction and better motion and appearance disentanglement. 

In the \textbf{first column}, we show the effects of not incorporating our spline deformation fields, neither for spatial nor for color deformations. Directly predicting the deformation fields is prone to deform the canonical space. This is not the case for our proposed spline deformation fields. Ours contain a smooth interpolation inductive bias, and MLPs are conditioned on a lower dimensionality input $(u,c)$ instead of $(u,v,t)$, which is common in models without spline controlled deformations. In the \textbf{second column}, we only utilize splines for $\mathcal{D}_s$ (not for $\mathcal{D}_c$). While this alleviates the distortion in the canonical space (Fig.~\ref{fig:ablation_studies}, top row), the color deformation is still not fully disentangled from spatial deformations. In the \textbf{third column}, we do not use spatial splines. While employing color splines improves the resulting canonical space, the aspect ratio and final rendered quality are still far from suboptimal.

In the \textbf{fourth column}, we use splines for $\mathcal{D}_s$ but no deformable color. As can be seen, the canonical space is reasonable, but the rendered frame is unrealistic. Besides, the face of the girl wrongly displays a rather uniform color. In contrast, the models that use deformable colors tend to better contrast the shading on the left and right side (dark vs bright colors of the target frame as shown at the bottom row).
Due to the inability to handle the temporally varying color, the results from this version show massive warping artifacts on the left side of the girl's face. In the \textbf{fifth column}, we show the effect of removing the optical flow loss, which proves that flow-based regularization plays a crucial role in controlling motion under large deformations.

Finally, the \textbf{sixth column} illustrates our full-blown model with a reasonable canonical space that reflects the darkening/shading of the left side of the girl's face and a cleaner composited rendering on the bottom. Note that the ablated models in columns 1-5 struggle to generate a consistent canonical space for highly deformable dynamic objects (in this case the hand), while our full model is more consistent. 

\begin{figure}[t]
  \centering 
  \includegraphics[width=0.47\textwidth]{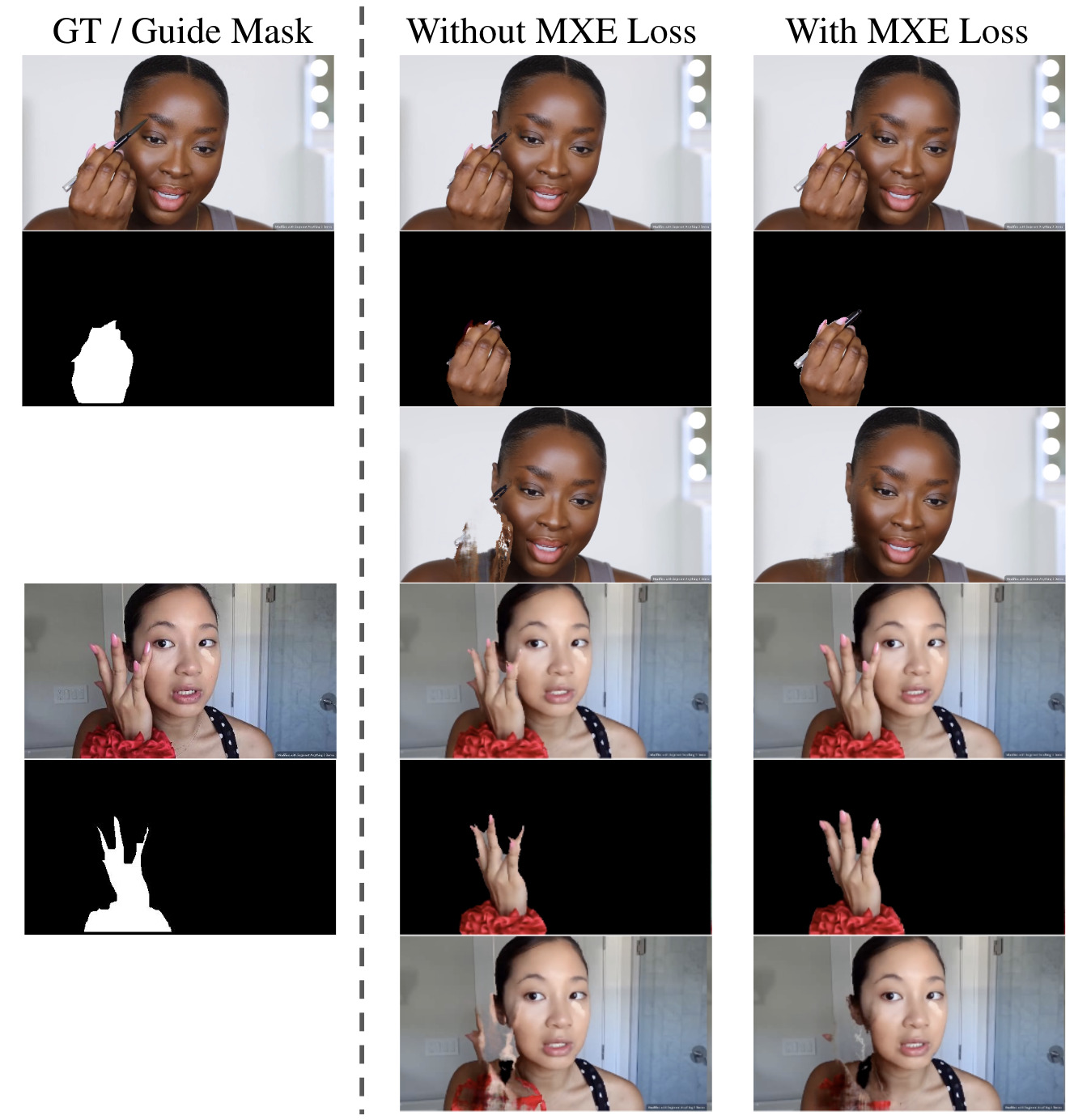}
  \vspace*{-2mm}
  \caption{Effects of the proposed error max loss in Eq. (\ref{eq:mxe_loss}).}
  \label{fig:mxe}
  \vspace{-4mm}
\end{figure}

Fig. \ref{fig:mxe} additionally illustrates the effectiveness of our layer separation objective function $l_{sep}$. Our model without the proposed MXE loss fails to refine the alpha mask, resulting in poor occlusion disentanglement. For instance, hand artifacts appear in the face image background and incomplete fingers in the occluder foreground.

\subsection{Limitations}
\label{sec:limitations}

Even though our method has increased the range of motions that an implicit video representation can handle, it is still restricted by the extent of deformation and self-occlusions in the input videos. Interestingly, our method is less so limited by the length of the videos, as fewer control points can be employed (see \textit{supplemental material}). 

\section{Conclusion}
\label{sec:conclusion}

In this work, we have presented a method to disentangle occlusions, appearance, and motion from videos through two different deformation fields, controlled by implicit neural spline functions. We show that these spline-based spatial deformation fields can represent complex motion in a dynamic scene while still preserving the semantic features in the inferred canonical spaces. We also showed that spline-powered color deformation fields are robust for motion and appearance disentanglement, generating base or ``albedo'' canonical spaces that allow propagation of time-dependent effects in the edited images. Such a property enhances the plausibility of these methods \textit{w.r.t.} previous state-of-the-art methods. When used in conjunction with the spatial deformation spline fields, our color deformation spline fields consistently yield appropriate canonical spaces suitable for color editing. It also allows for high-quality rendering of the modified content.
We remark that our approach is generic and flexible enough for arbitrary content separation and editing. However, its particular suitability for facial video processing opens up exciting future directions, such as semantic edits of expressions, facial identity, and material and lighting properties.

\clearpage
\appendix
\setcounter{figure}{8}

\twocolumn[{
\renewcommand\twocolumn[1][]{#1}
\maketitlesupplementary
\begin{center}\centering
  \includegraphics[width=0.99\textwidth]{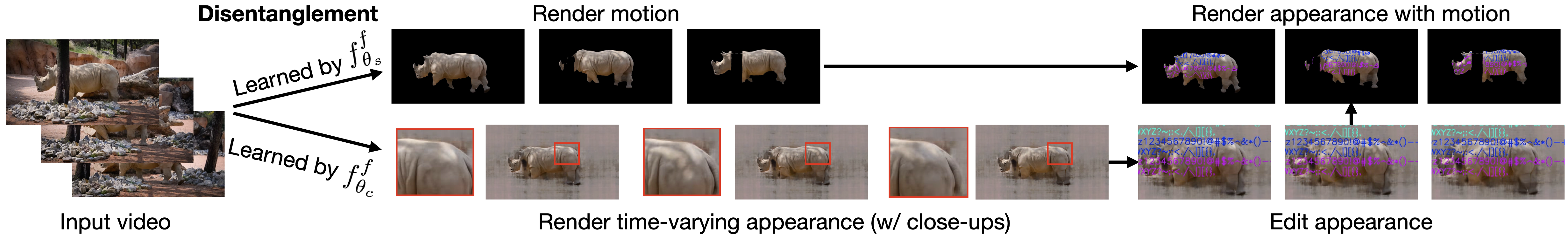}
  \captionof{figure}{Disentangling motion and appearance.}
  \label{fig:r_disentangle}
\end{center}
}]

In this supplemental material, we provide additional implementation details and experimental results. We present further comparisons with previous methods~\cite{kasten2021layered,ye2022deformable}, as well as additional results on texture editing, motion editing, and training on longer sequences. We also include additional ablation studies on loss terms, guidance masks, and control points. Finally, we discuss some failure cases and the ethical implications of our method and datasets. Furthermore, we provide several \textit{videos} in in our project website \url{https://juanluisg-flwls.github.io/videospats-website/}, which are critical for visualizing time-dependent appearance and temporally consistent reconstructions and edits produced by our method. \textit{We strongly encourage readers to visit our site for the best viewing experience.}

\section{Additional details on disentangling motion and appearance.}
\label{sec:supp_details_disen}

\camready{For the sake of better conceptualization, we provide the additional Fig. \ref{fig:r_disentangle}. As shown in Fig. \ref{fig:r_disentangle} for the foreground image (rhino), $f_{\theta_s}^f$ learns to model motion (a.k.a. spatial deformation field), while $f_{\theta_c}^f$ learns to model time-dependent appearance (a.k.a. color deformation field). Note that such disentanglement of motion and appearance allows us to perform editing on the rhino that is not distorted by the time-dependent appearance (e.g. shadows). In addition, as we learn a base color and deformation color splines, we can seamlessly blend appearance changes with color edits, as shown in the right-hand side of Fig. \ref{fig:r_disentangle}.}

\section{Additional implementation details}
\label{sec:supp_details}

Our \textit{VideoSPaTS} takes 90 minutes to fit a 512$\times$288, 50 frames video. 
However, the training time could be reduced with additional engineering efforts, such as replacing MLPs with optimized embedders like those in tiny-cuda-nn~\cite{tiny-cuda-nn}.
This can potentially provide between 2$\times$ and 10$\times$ training speed-ups. In addition, our method does not require running the models during inference / editing for every single frame, since a single run suffices to obtain the deformation and color control points, providing further speedups during inference and editing.

We employed periodical positional encoding \cite{attentionallyouneed} for our deformation models. 

The weights in Eq.(16) of the main paper are empirically set to balance their respective terms with respect to $l_{rec}$. Specifically:
\begin{itemize}
\setlength\itemsep{0em}

\item $\lambda_{fl}=100$ is set with a relatively high value as the coordinate error has very small magnitudes in comparison with the color errors in $l_{rec}$. 

\item $\lambda_{\mathcal{D}_s}=0.1$ is set to slightly regularize deformations. See Section \ref{sec:sup_lds_effects} for more details.

\item $\lambda_{\mathcal{D}_c}=0.001$ is set to a relatively low value to regularize color deformation while still allowing it to learn, as such color deformation is enabled after 50\% of the training.
\end{itemize}

\begin{figure}[h]
    \centering
    \small
    \renewcommand{\arraystretch}{1.5}
\setlength{\tabcolsep}{4pt}
\begin{tabular}{lccc}
\hline
Scene & CoDeF & Deformable Sprites & Ours \\ 
\hline
Bear  & 27.52/0.84 & \textbf{30.94/0.96} & 30.82/0.95 \\
Train & 21.53/0.87 & 27.08/0.94 & \textbf{27.68/0.92} \\
Rhino & 24.66/0.81 & 28.60/0.94 & \textbf{29.35/0.94} \\
\hline
Average & 24.57/0.84 & 28.87/\textbf{0.95} & \textbf{29.23}/0.94 \\
\hline
\end{tabular}
    \vspace{-3mm}
    \captionof{table}{Video editing quantiative results \textsc{psnr/ssim}}
    \label{tab:editing}
\end{figure}

\section{Additional results}
\label{sec:supp_results}
We show additional results in this section. Please refer to our site \url{https://juanluisg-flwls.github.io/videospats-website/} for additional video visualizations. 

\begin{figure*}[h]
  \centering 
  \renewcommand{\arraystretch}{0.5}
\setlength{\tabcolsep}{1.2pt}
\begin{tabular}{cc|cc|cc}
\multicolumn{2}{c}{a) Deformable Sprites \cite{ye2022deformable}} & \multicolumn{2}{c}{a) CoDeF \cite{ouyang2024codef}} & \multicolumn{2}{c}{b) Our VideoSPatS} \\
\includegraphics[width = 1.11in]{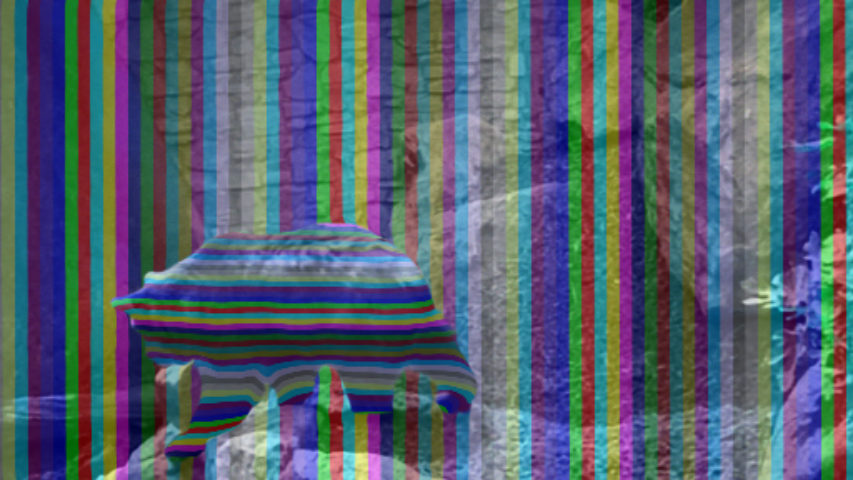} &
\includegraphics[width = 1.11in]{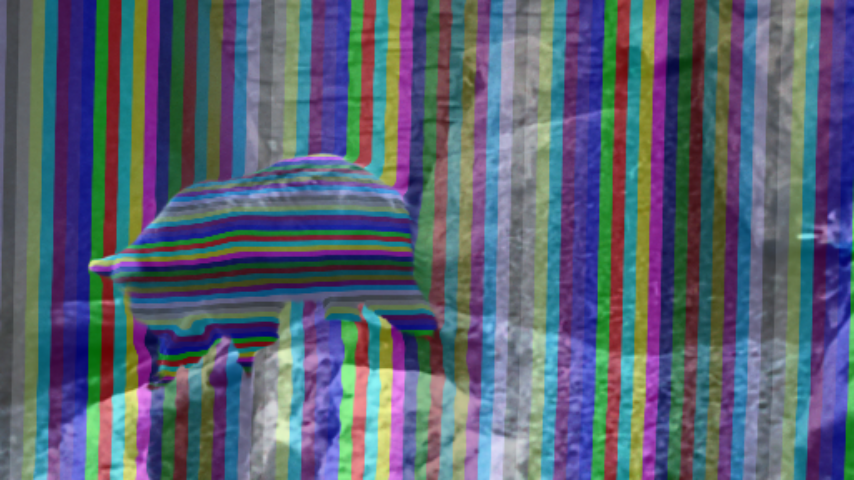} &
\includegraphics[width = 1.11in]{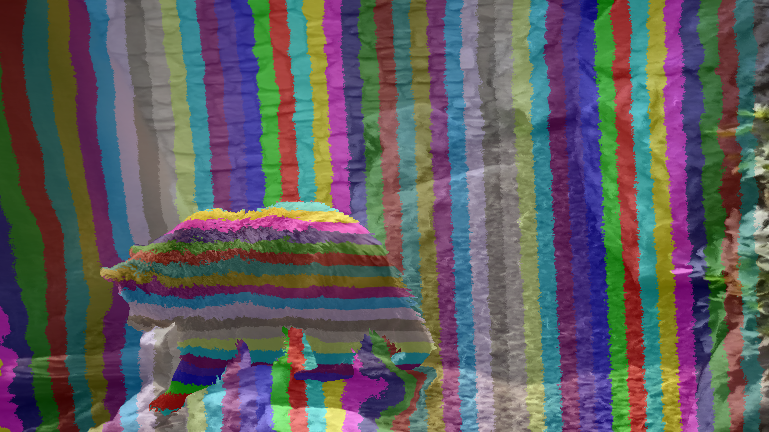} &
\includegraphics[width = 1.11in]{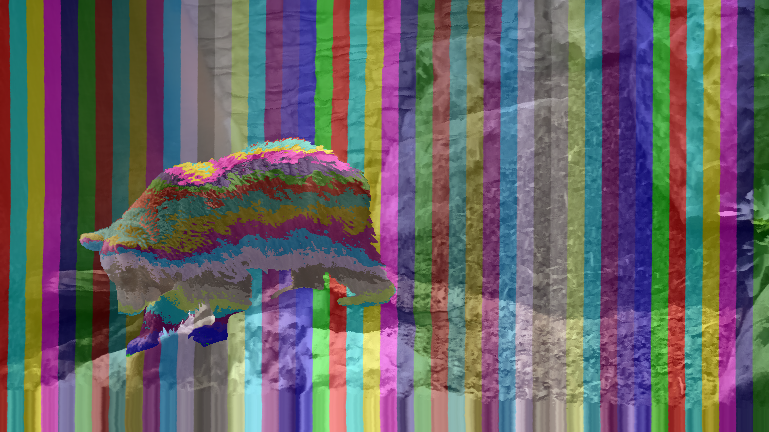} &
\includegraphics[width = 1.11in]{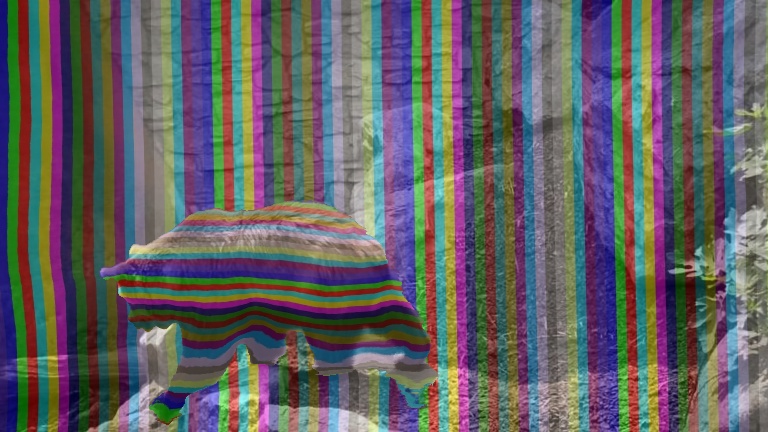} &
\includegraphics[width = 1.11in]{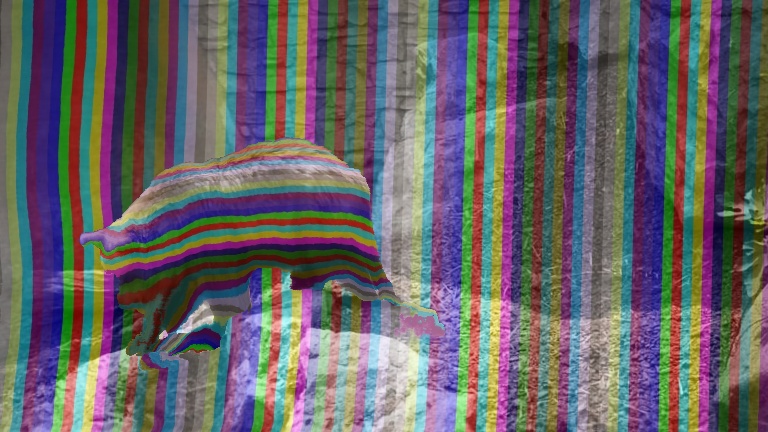} \\

\includegraphics[width = 1.11in]{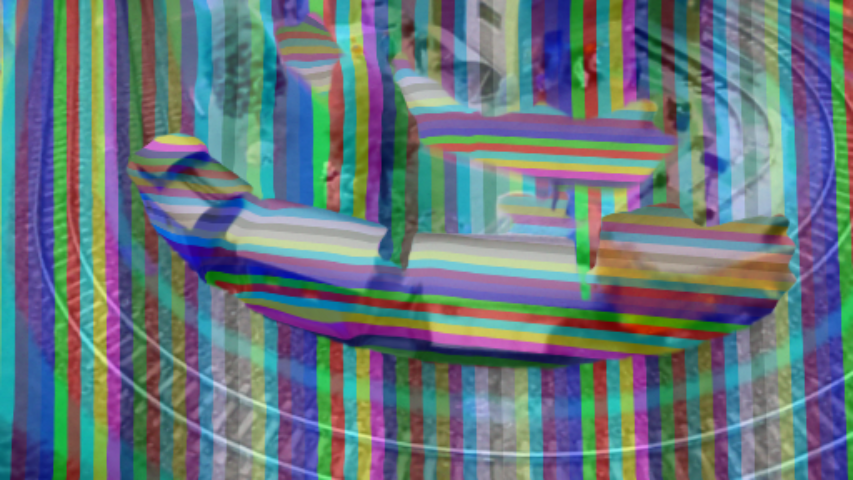} &
\includegraphics[width = 1.11in]{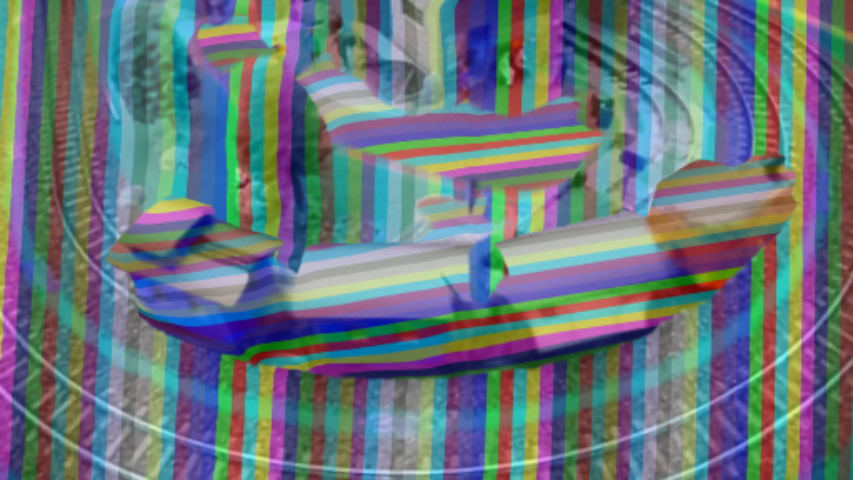} &
\includegraphics[width = 1.11in]{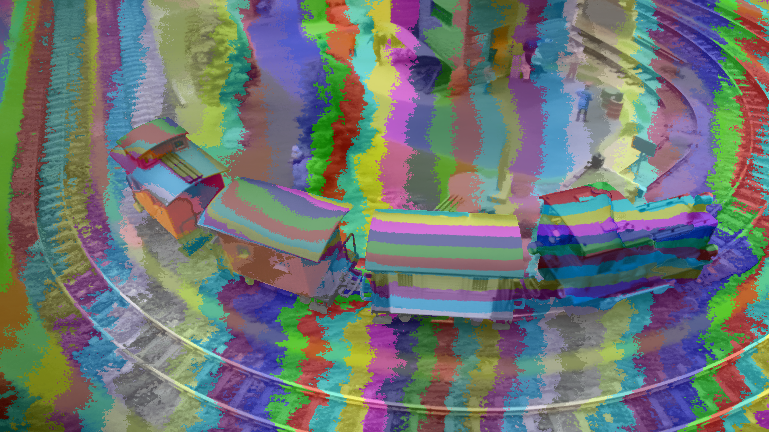} &
\includegraphics[width = 1.11in]{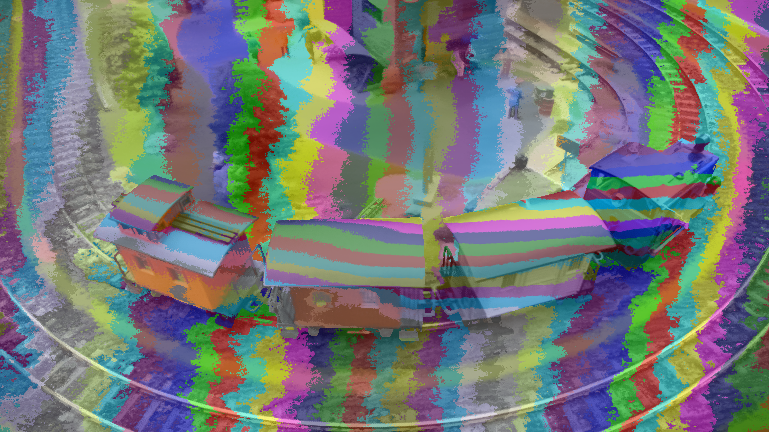} &
\includegraphics[width = 1.11in]{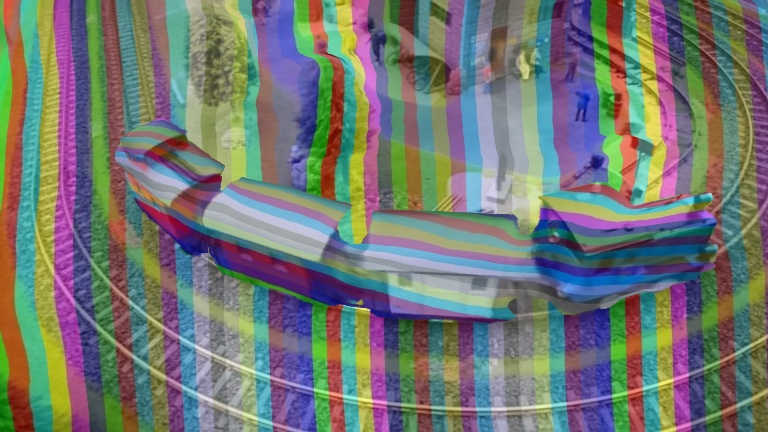} &
\includegraphics[width = 1.11in]{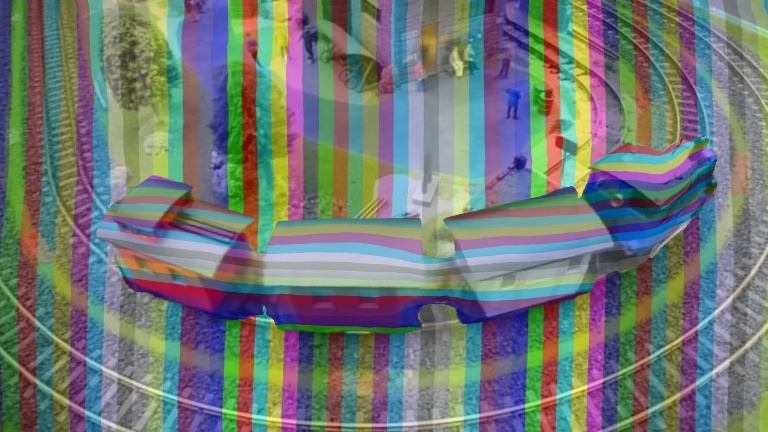} \\

\end{tabular}
  \caption{Editing results. Note inconsistencies in (b) as deformation fields incorrectly model time-varying appearance in the bear's fur.}
  \label{fig:r_comp_editing}
\end{figure*}

\begin{figure*}[h]
  \centering 
  \setlength{\tabcolsep}{1pt}
\renewcommand{\arraystretch}{0.6}
\small
\begin{tabular}{c|cccc}
Ground Truth & Neural Layered Atlases & Deformable Sprites & CoDeF & Our \textit{VideoSPatS} 
\\

\includegraphics[width=0.195\linewidth]{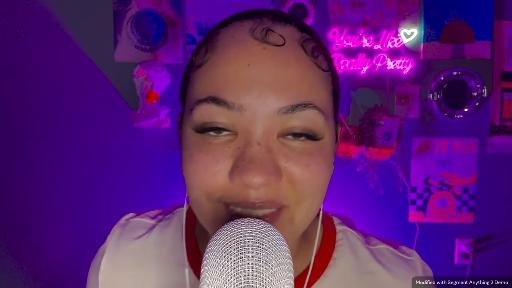} & 
\includegraphics[width=0.195\linewidth]{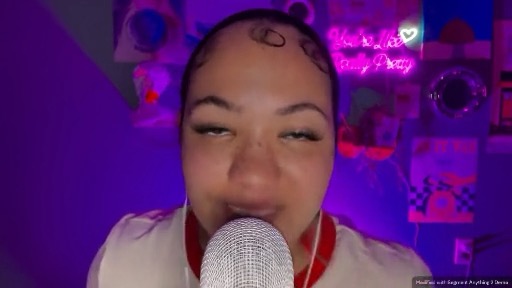} & 
\includegraphics[width=0.195\linewidth]{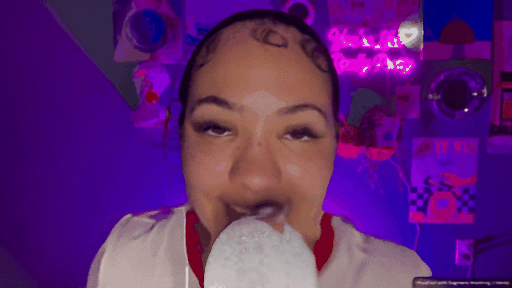} & 
\includegraphics[width=0.195\linewidth]{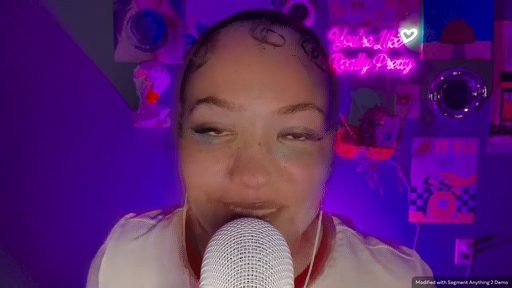} & 
\includegraphics[width=0.195\linewidth]{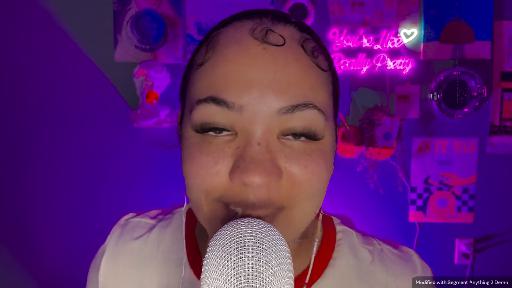} 
\\

\includegraphics[width=0.195\linewidth]{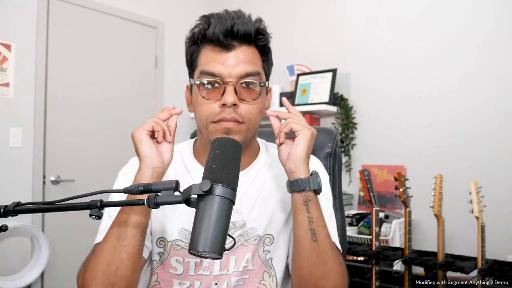} & 
\includegraphics[width=0.195\linewidth]{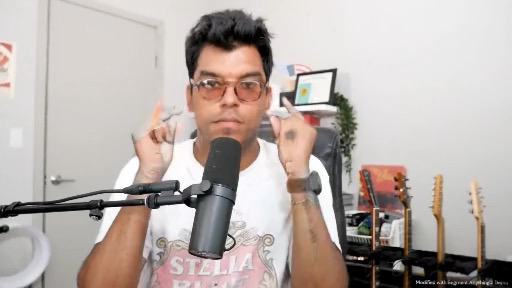} & 
\includegraphics[width=0.195\linewidth]{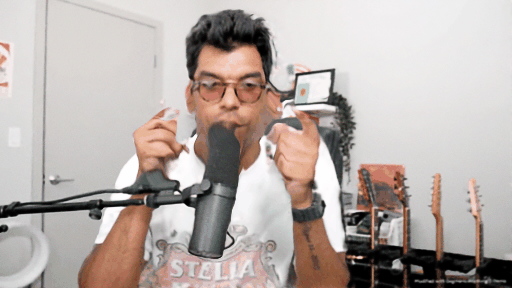} & 
\includegraphics[width=0.195\linewidth]{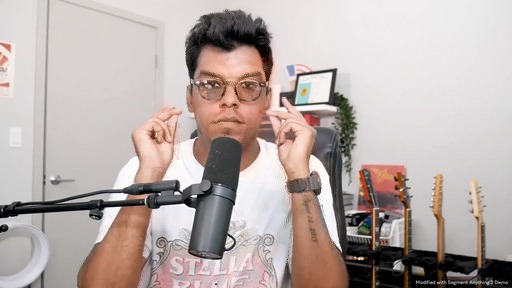} & 
\includegraphics[width=0.195\linewidth]{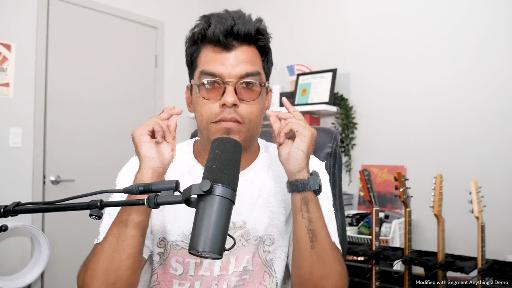} 
\\

\includegraphics[width=0.195\linewidth]{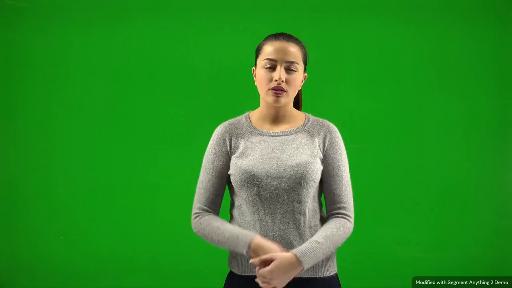} & 
\includegraphics[width=0.195\linewidth]{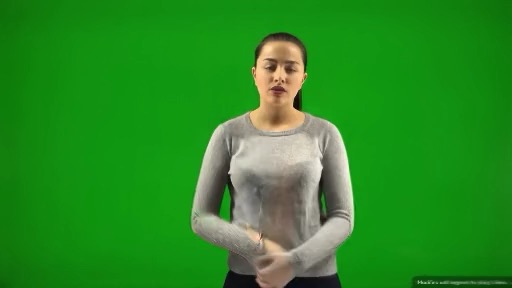} & 
\includegraphics[width=0.195\linewidth]{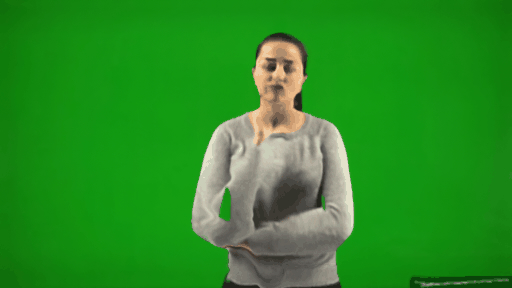} & 
\includegraphics[width=0.195\linewidth]{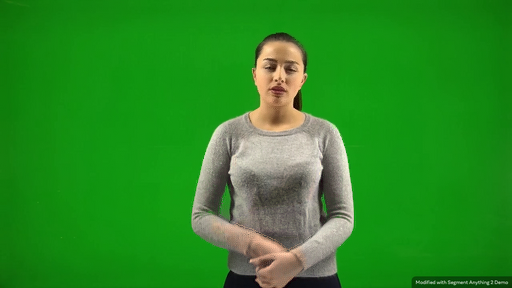} & 
\includegraphics[width=0.195\linewidth]{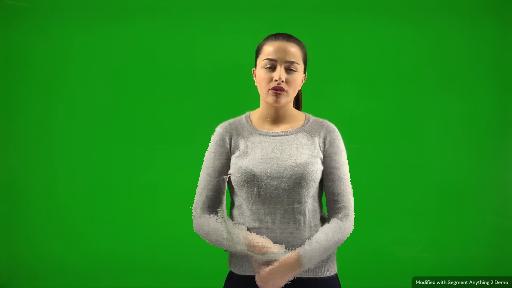} 
\\

\includegraphics[width=0.195\linewidth]{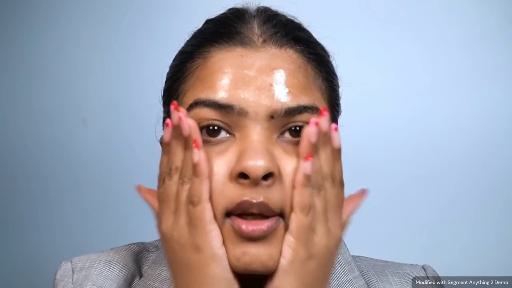} & 
\includegraphics[width=0.195\linewidth]{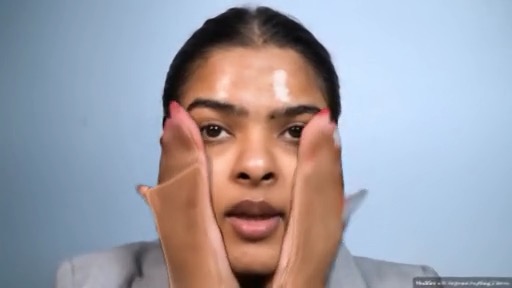} & 
\includegraphics[width=0.195\linewidth]{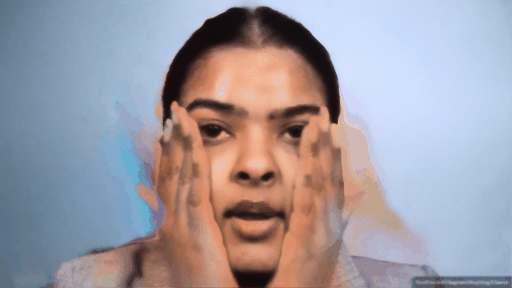} & 
\includegraphics[width=0.195\linewidth]{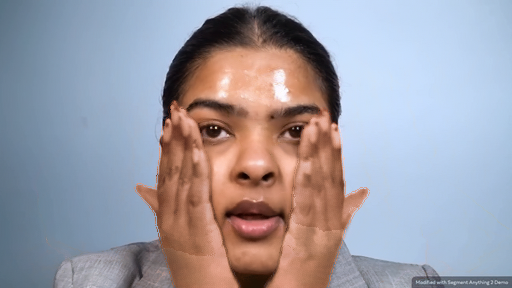} & 
\includegraphics[width=0.195\linewidth]{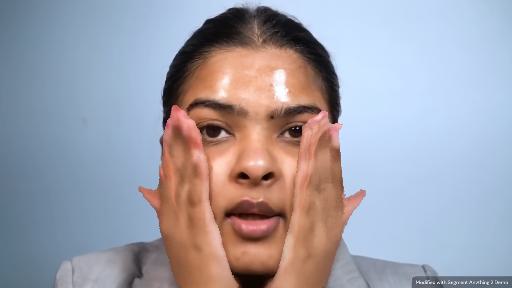} 
\\
\end{tabular}

  \vspace*{-3mm}
  \caption{Video reconstruction comparisons with other methods. Our method consistently implicitly reconstructs videos, while the other methods fail on one case (CoDeF) or multiple cases (Layered Atlases, Deformable Sprites).}
  \label{fig:sup_other_comp_rec}
\end{figure*}

\begin{figure*}[h]
  \centering 
  \setlength{\tabcolsep}{1pt}
\renewcommand{\arraystretch}{0.5}
\small
\begin{tabular}{cccc}

Layered Atlasses &
Deformable Sprites &
CoDeF &
Our \textit{VideoSPatS}
\\
\begin{tabular}{@{}c@{}}
\includegraphics[height=0.14\linewidth]{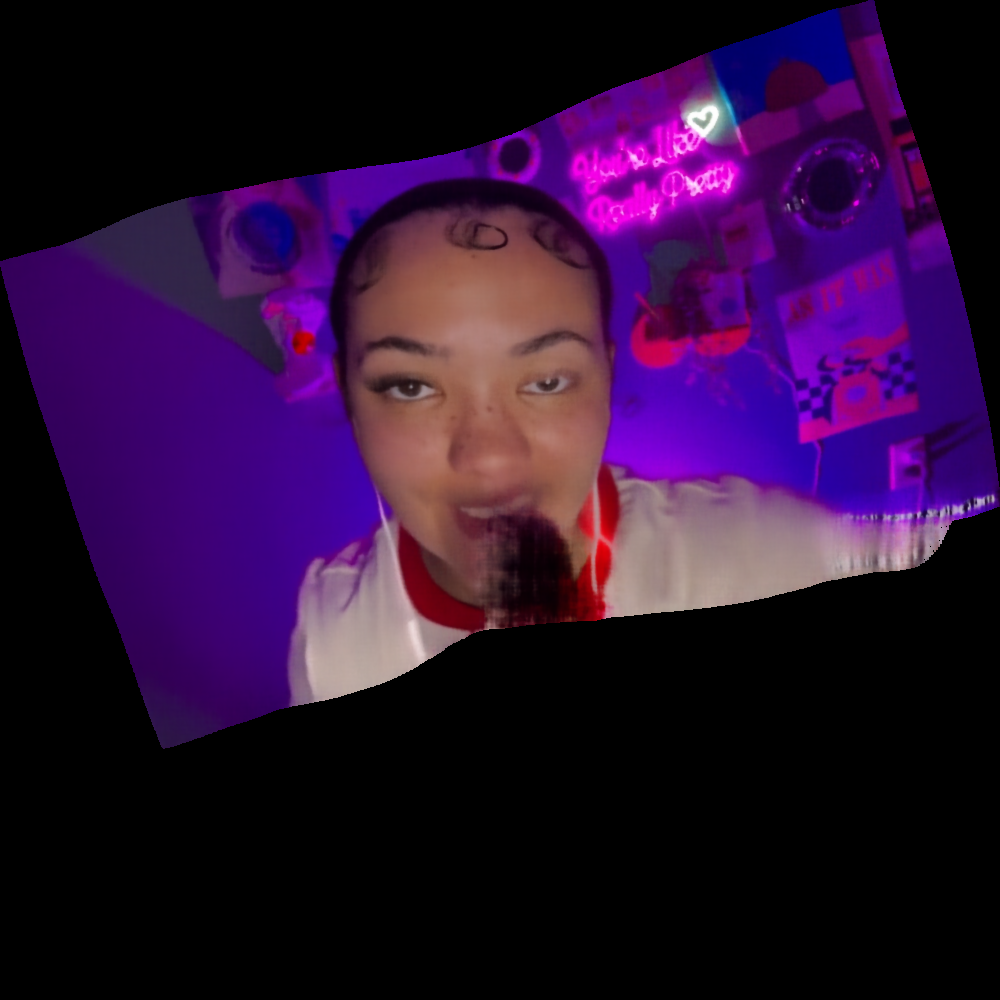} \\ \includegraphics[height=0.14\linewidth]{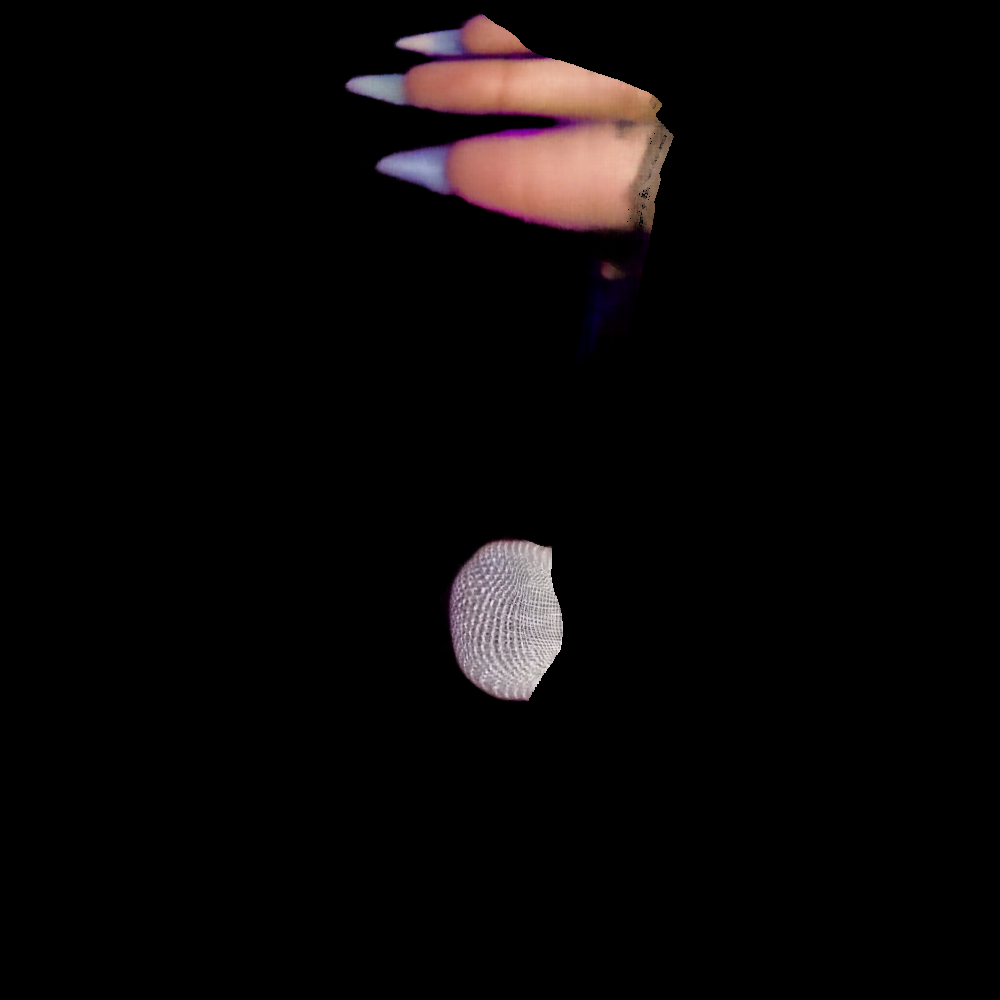}
\end{tabular} &
\begin{tabular}{@{}c@{}}
\includegraphics[height=0.14\linewidth]{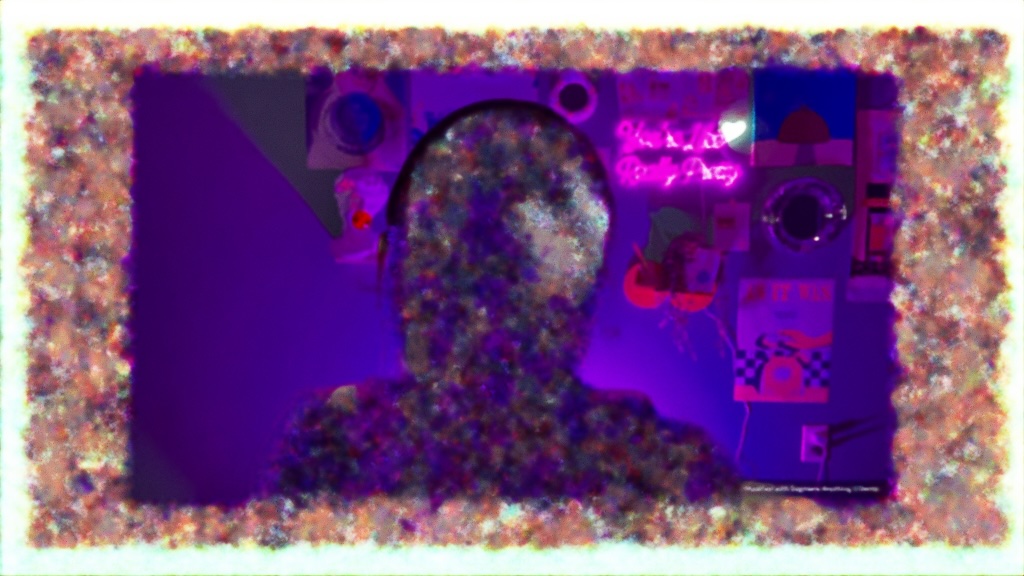} \\ \includegraphics[height=0.14\linewidth]{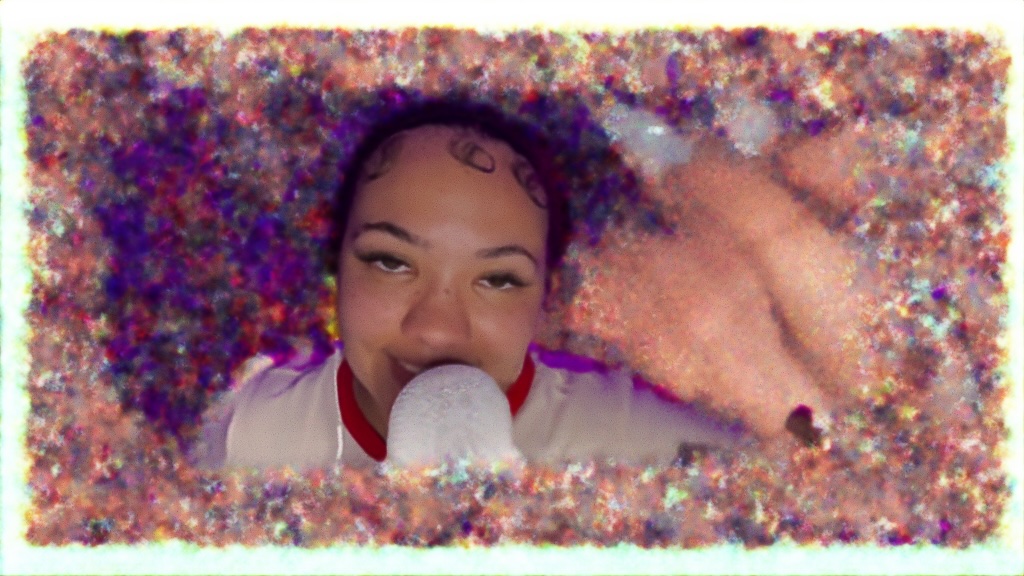}
\end{tabular} &
\begin{tabular}{@{}c@{}}
\includegraphics[height=0.14\linewidth]{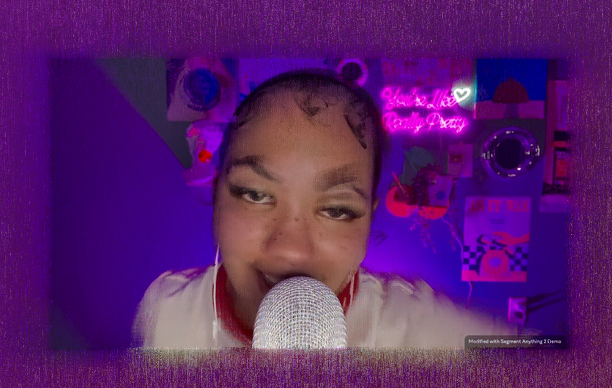} \\ 
\includegraphics[height=0.14\linewidth]{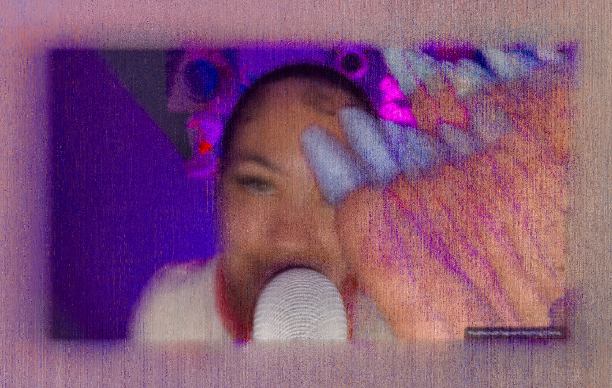}
\end{tabular} &
\begin{tabular}{@{}c@{}}
\includegraphics[height=0.14\linewidth]{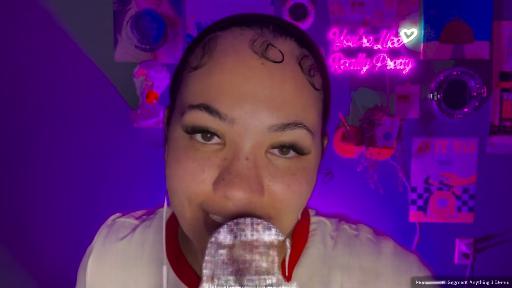} \\ 
\includegraphics[height=0.14\linewidth]{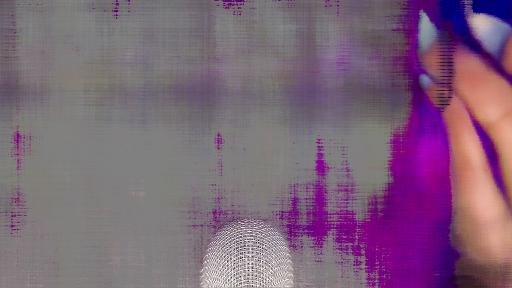}
\end{tabular} 
\\

\begin{tabular}{@{}c@{}}
\includegraphics[height=0.14\linewidth]{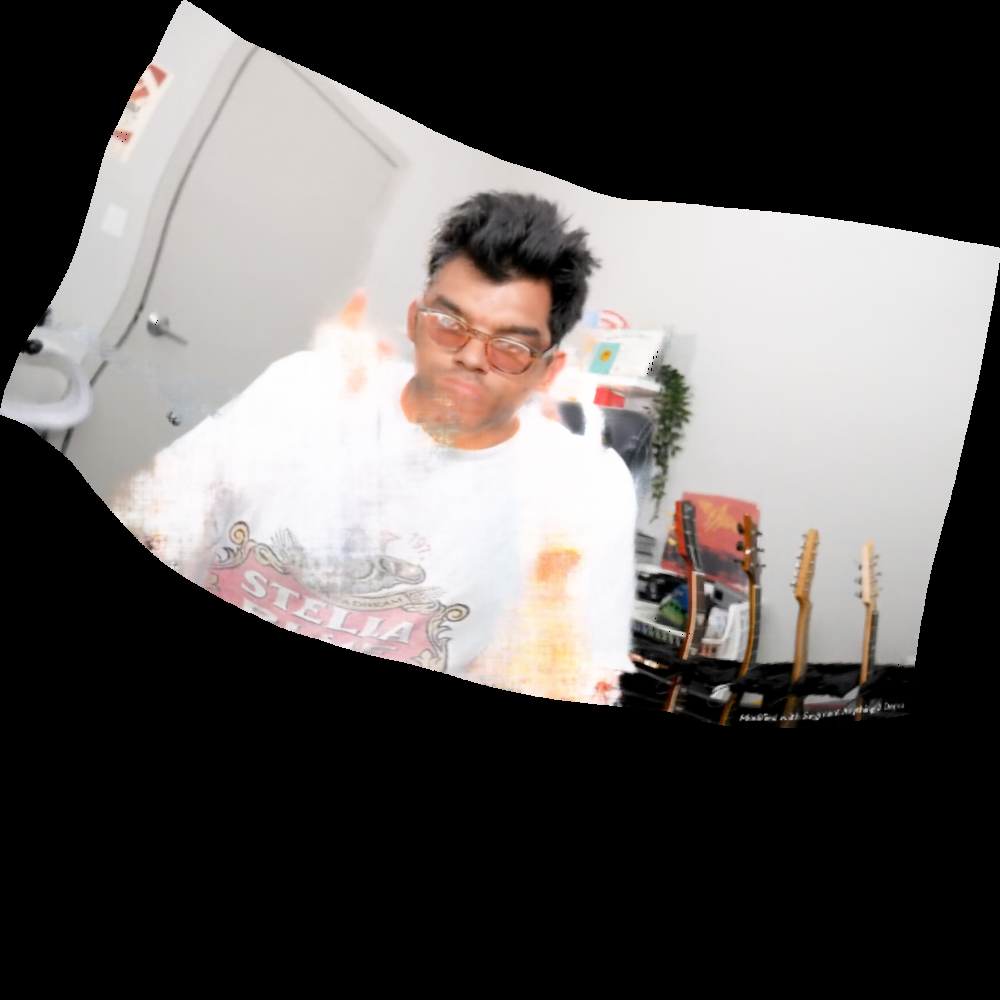} \\ \includegraphics[height=0.14\linewidth]{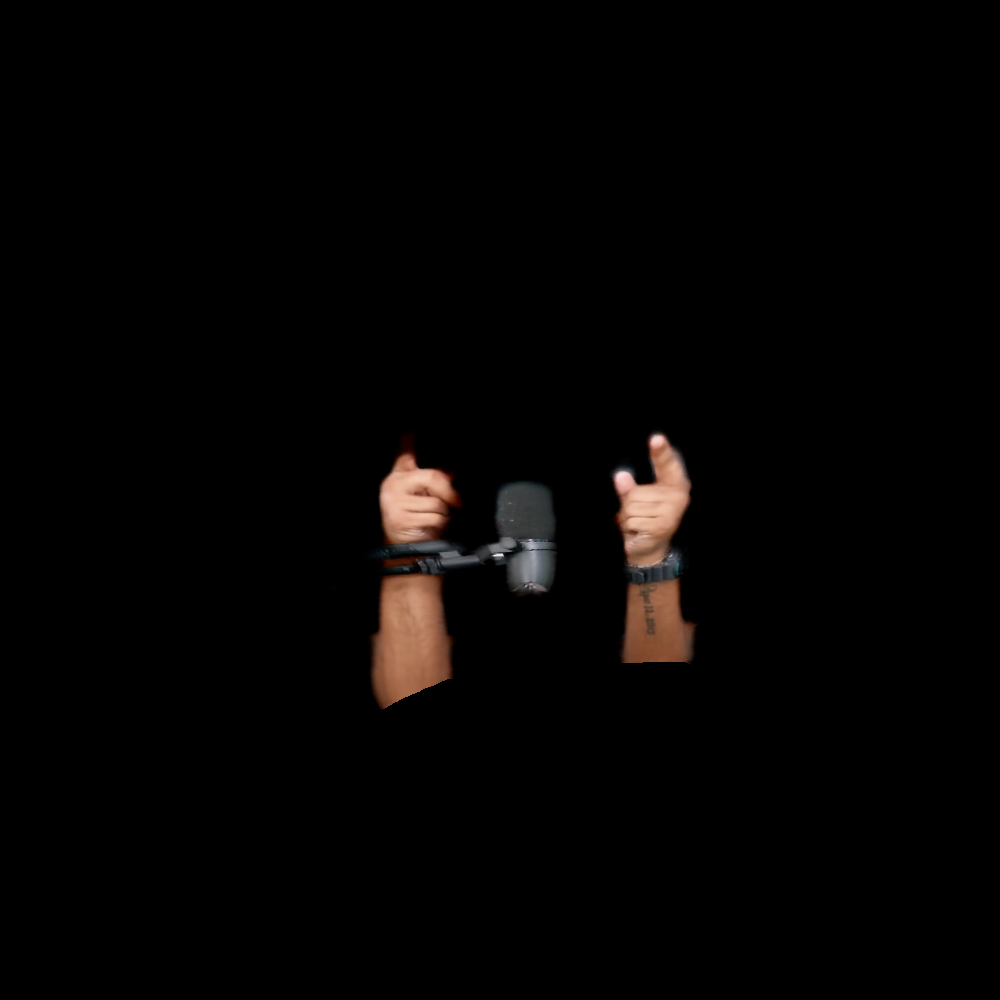}
\end{tabular} &
\begin{tabular}{@{}c@{}}
\includegraphics[height=0.14\linewidth]{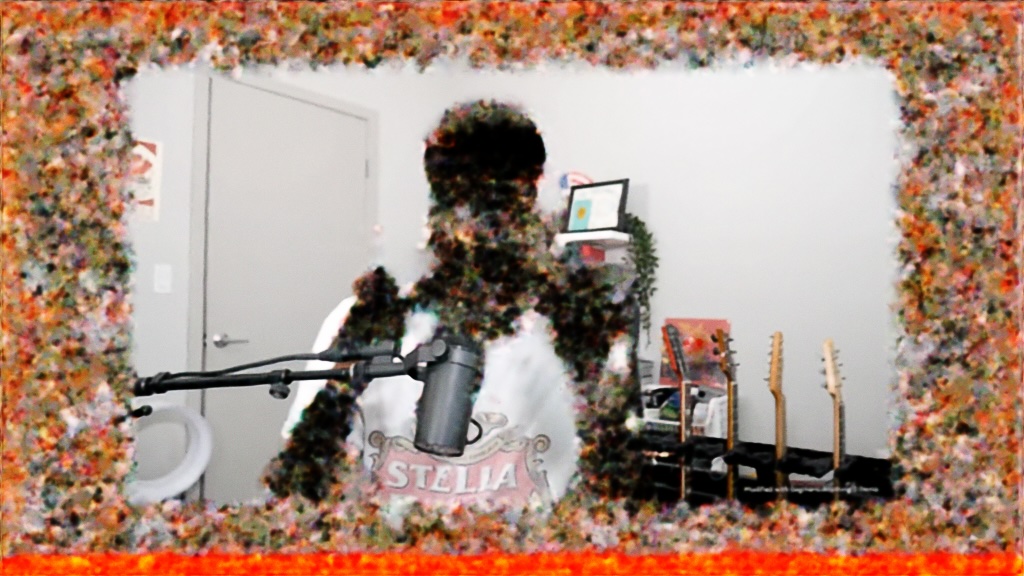} \\ \includegraphics[height=0.14\linewidth]{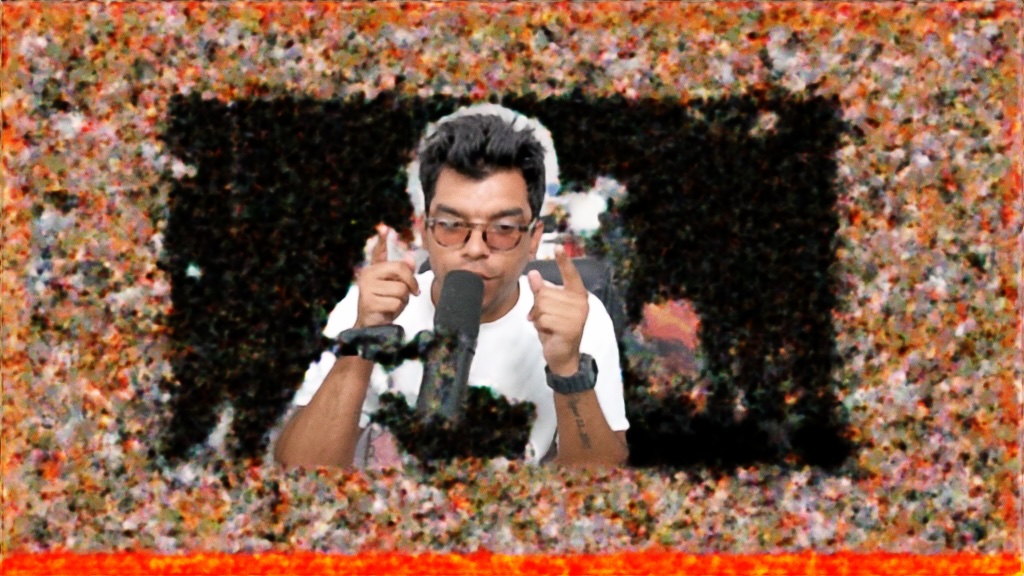}
\end{tabular} &
\begin{tabular}{@{}c@{}}
\includegraphics[height=0.14\linewidth]{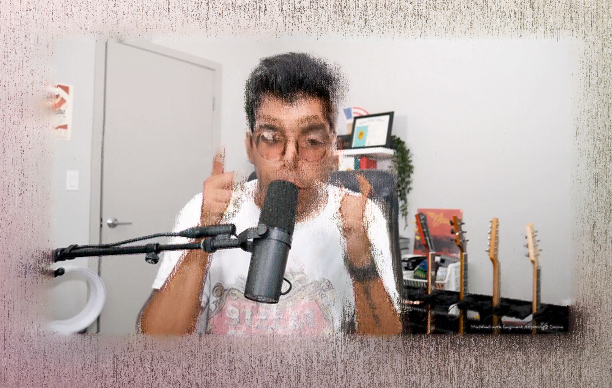} \\ 
\includegraphics[height=0.14\linewidth]{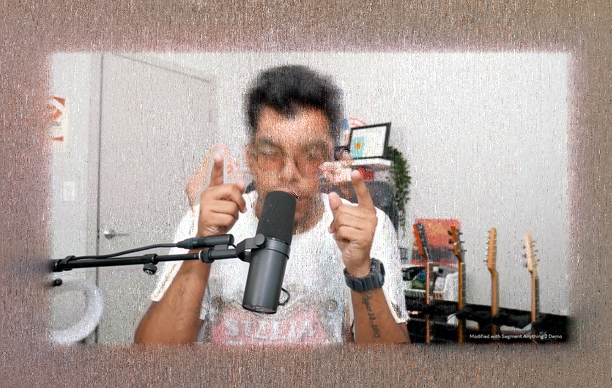}
\end{tabular} &
\begin{tabular}{@{}c@{}}
\includegraphics[height=0.14\linewidth]{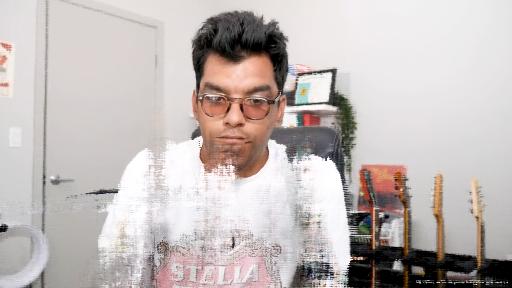} \\ 
\includegraphics[height=0.14\linewidth]{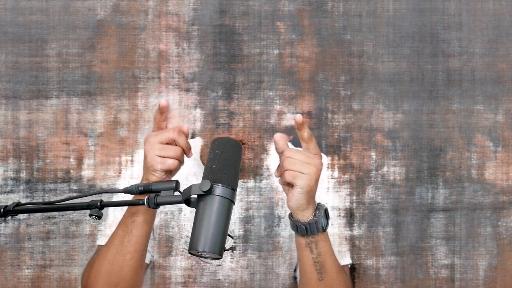}
\end{tabular} 
\\

\begin{tabular}{@{}c@{}}
\includegraphics[height=0.14\linewidth]{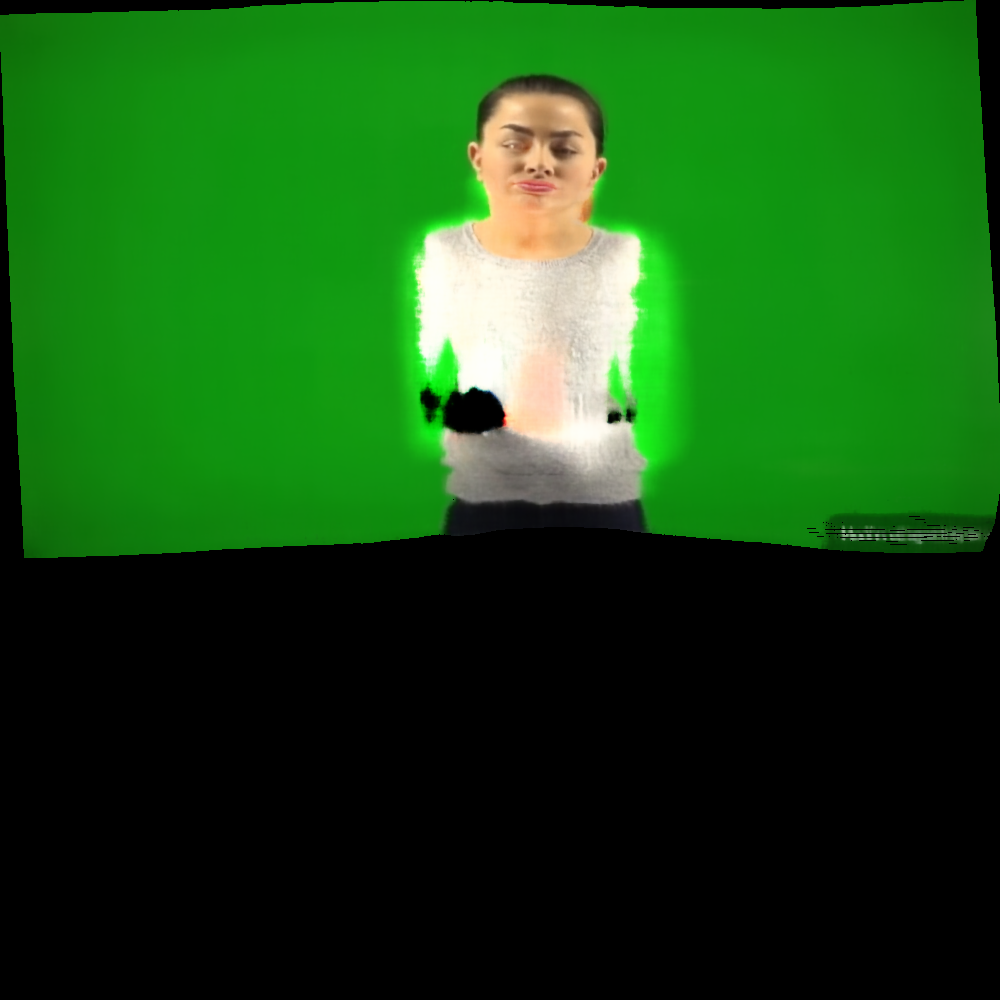} \\ \includegraphics[height=0.14\linewidth]{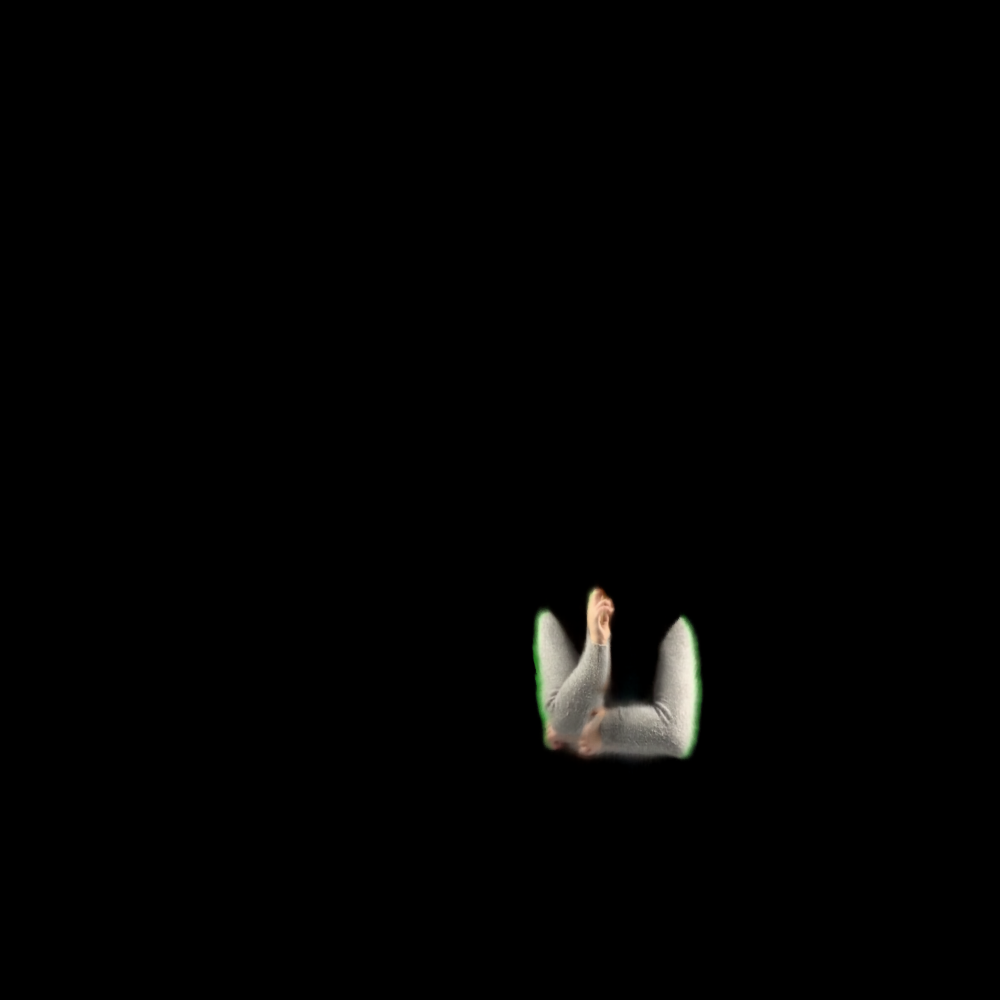}
\end{tabular} &
\begin{tabular}{@{}c@{}}
\includegraphics[height=0.14\linewidth]{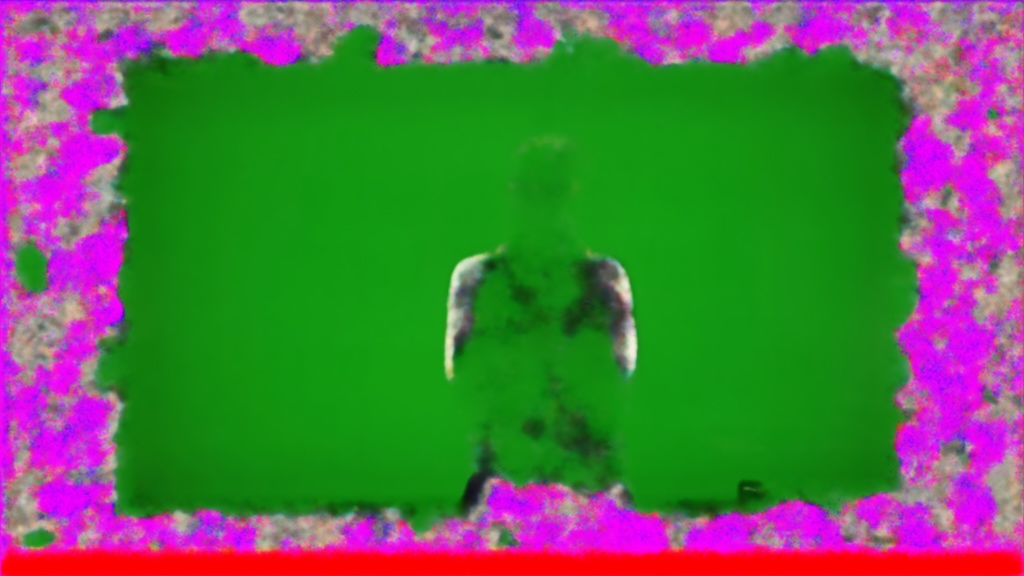} \\ \includegraphics[height=0.14\linewidth]{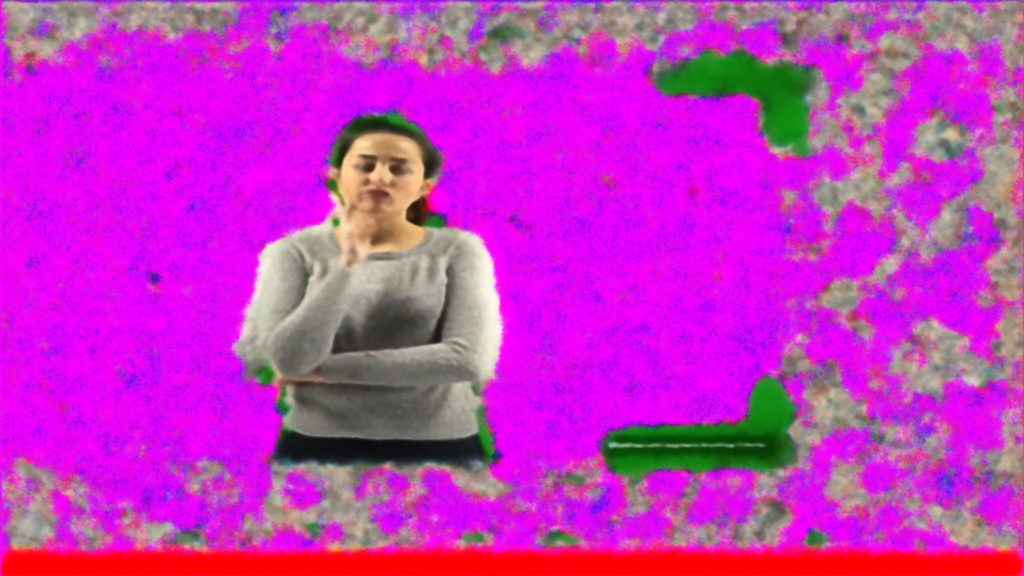}
\end{tabular} &
\begin{tabular}{@{}c@{}}
\includegraphics[height=0.14\linewidth]{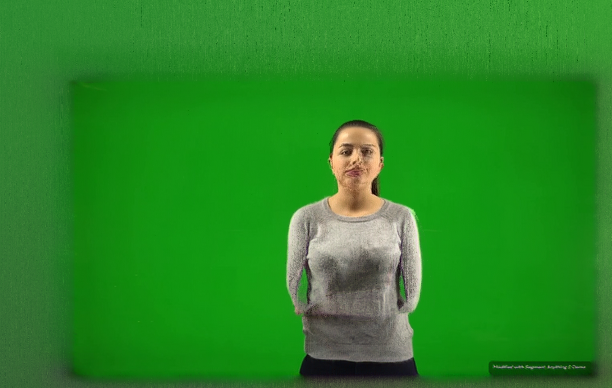} \\ 
\includegraphics[height=0.14\linewidth]{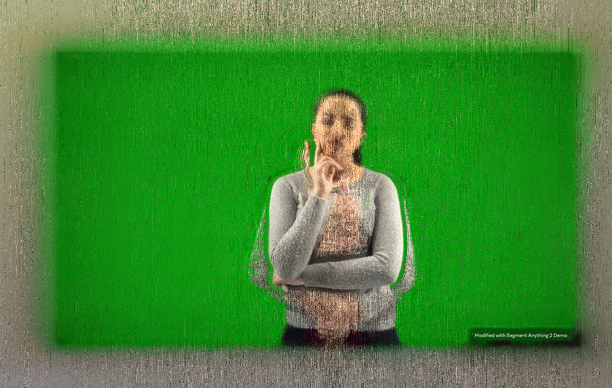}
\end{tabular} &
\begin{tabular}{@{}c@{}}
\includegraphics[height=0.14\linewidth]{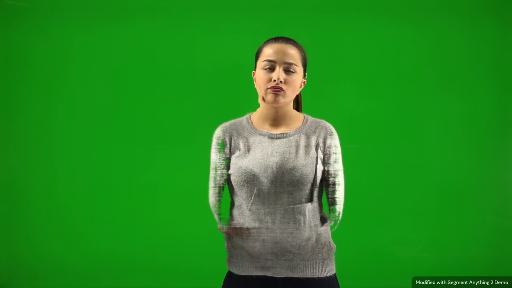} \\ 
\includegraphics[height=0.14\linewidth]{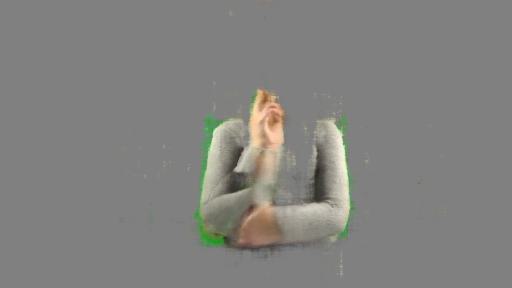}
\end{tabular} 
\\

\begin{tabular}{@{}c@{}}
\includegraphics[height=0.14\linewidth]{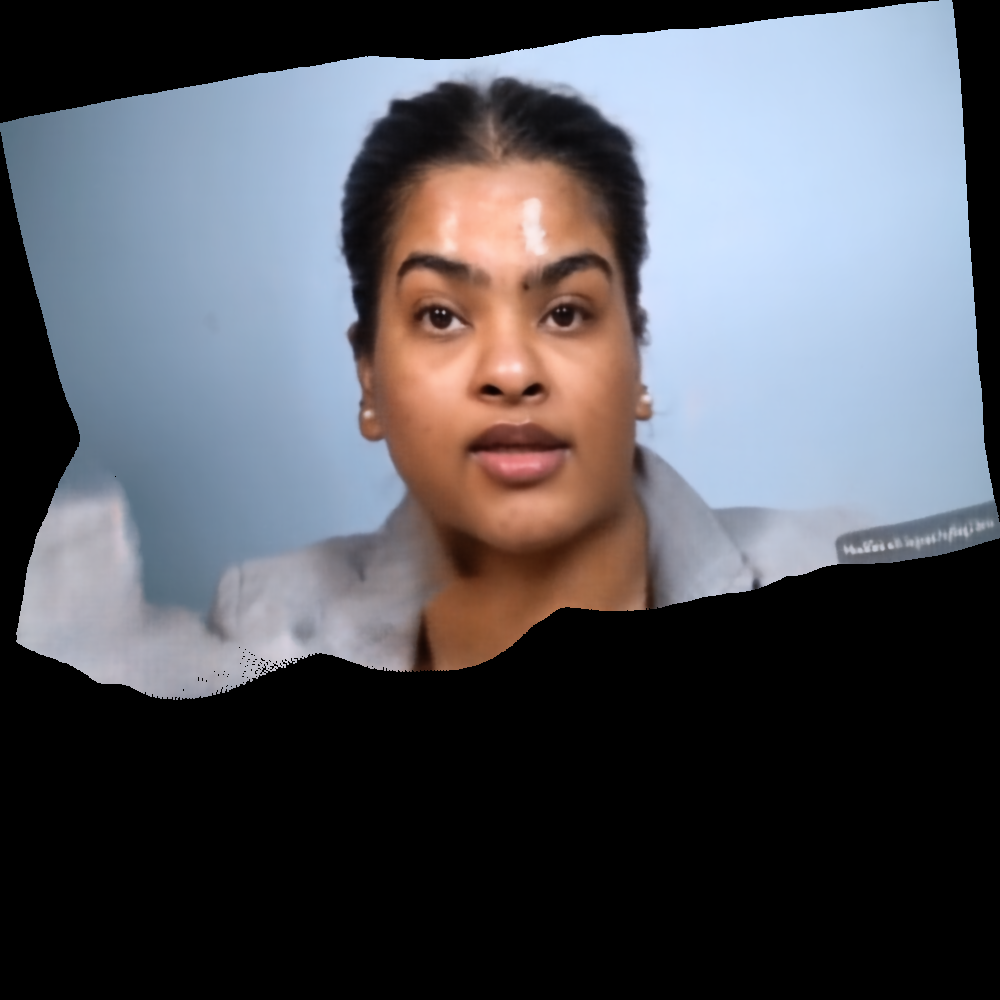} \\ \includegraphics[height=0.14\linewidth]{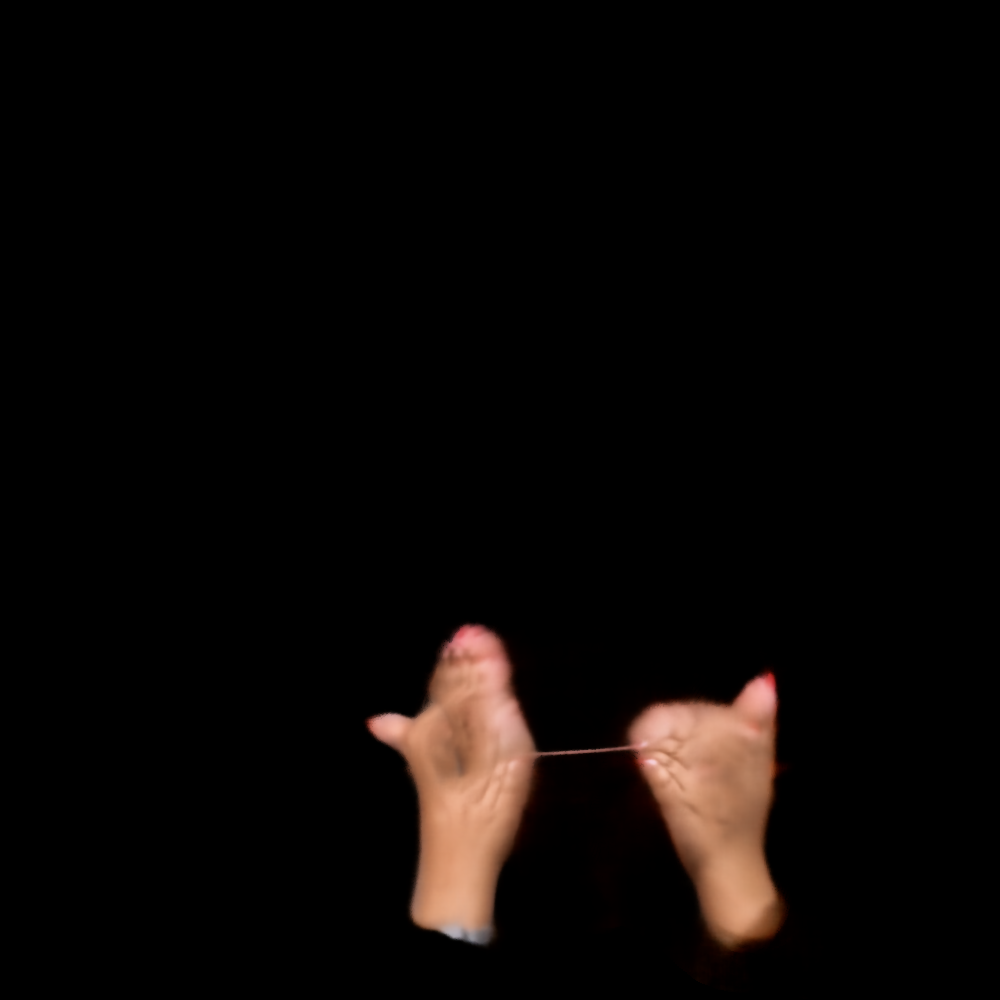}
\end{tabular} &
\begin{tabular}{@{}c@{}}
\includegraphics[height=0.14\linewidth]{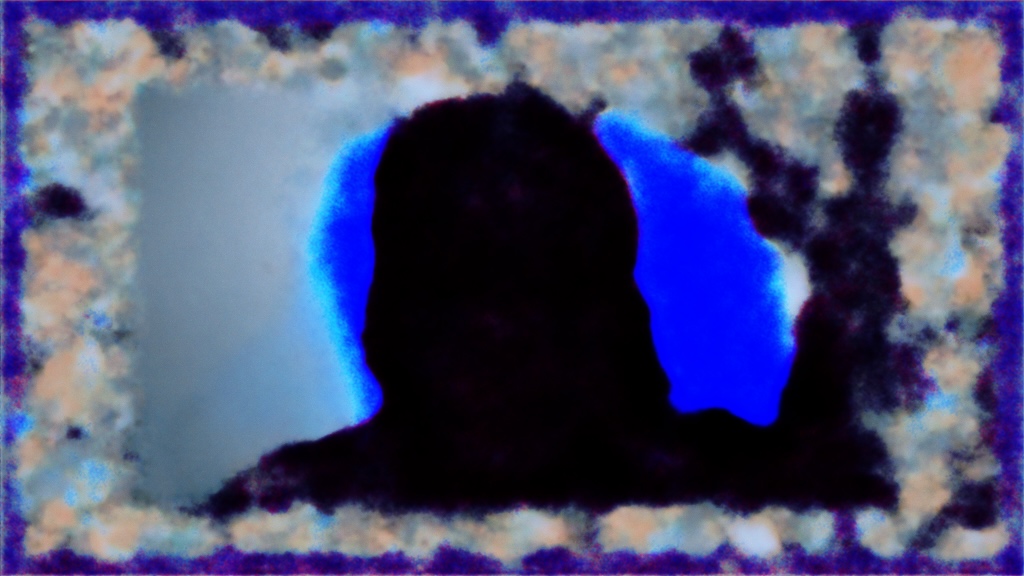} \\ \includegraphics[height=0.14\linewidth]{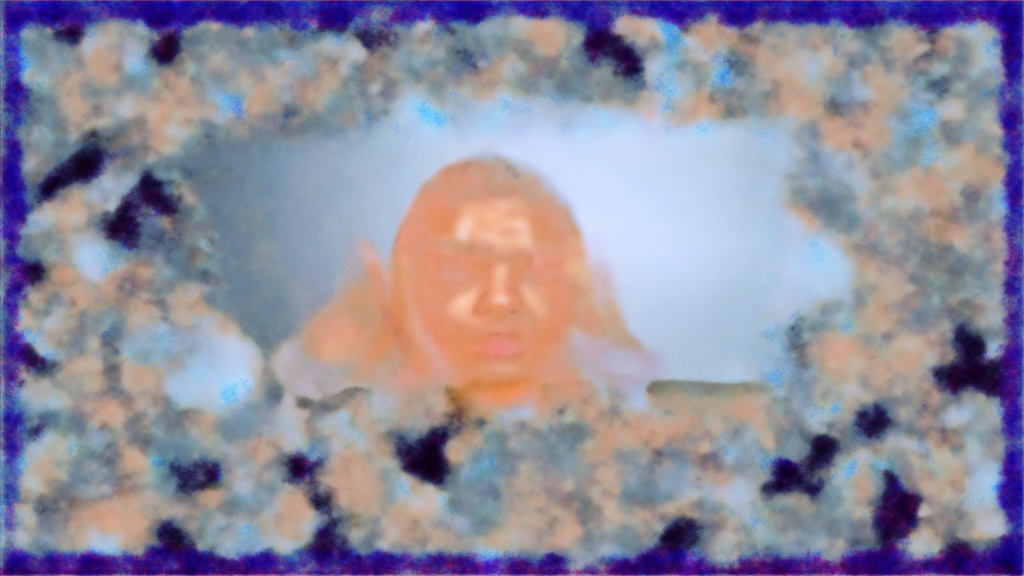}
\end{tabular} &
\begin{tabular}{@{}c@{}}
\includegraphics[height=0.14\linewidth]{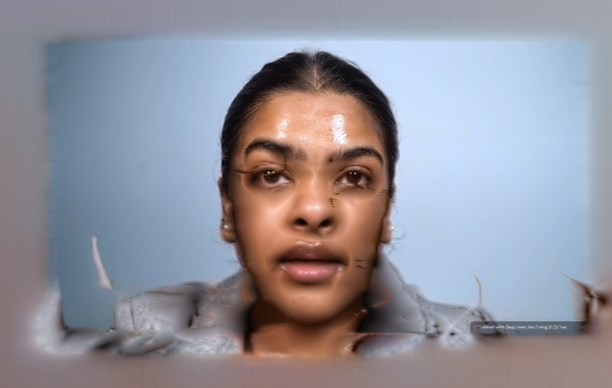} \\ 
\includegraphics[height=0.14\linewidth]{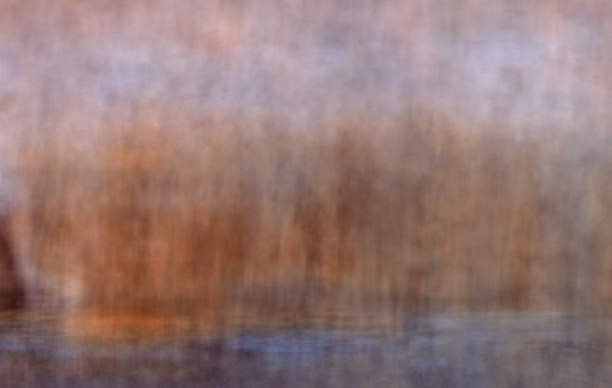}
\end{tabular} &
\begin{tabular}{@{}c@{}}
\includegraphics[height=0.14\linewidth]{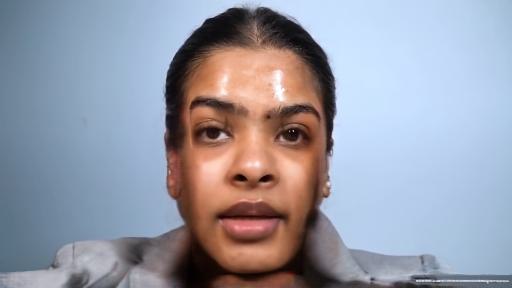} \\ 
\includegraphics[height=0.14\linewidth]{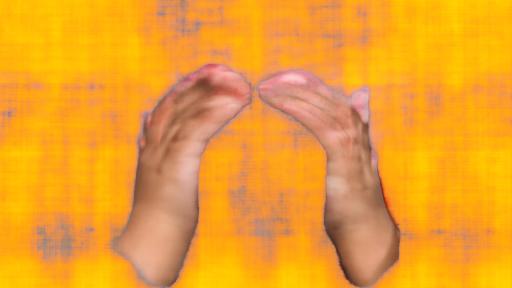}
\end{tabular} 
\\

\end{tabular}
  \vspace*{-3mm}
  \caption{Comparisons of the obtained canonical spaces with other methods. For every two rows, the top row corresponds to the background canonical space, and the bottom row corresponds to the foreground canonical space. Our \textit{VideoSPaTS} consistently yields more editing-intuitive and feasible canonical spaces.}
  \label{fig:sup_other_comp}
\end{figure*}

\subsection{Quantitative results}
\camready{As mere video reconstruction metrics are not indicative of editing performance, we show video editing quantitative results in Table \ref{tab:editing} in terms of warping consistency, measured in avgerage PSNR and SSIM between edited and warped edited frames. We use RAFT \cite{teed2020raft} to obtain the original frames' optical flow to warp edited frames at $t$+$n$ into $t$. We set $n$=3 for a significant difference in terms of scene optical flow. Ours outperforms CoDeF\cite{ouyang2024codef} by a large margin in terms of PSNR and SSIM, corresponding well to the visuals in Fig. \ref{fig:r_comp_editing}. With respect to Deformable Sprites \cite{ye2022deformable}, our method outperforms it by \textbf{0.4dB} in terms of PSNR, but more importantly, our VideoSpatS can model the time-dependent appearance (e.g. shadows on bear's fur), yielding a more realistic and disentangled reconstruction and editing than the fixed colors in Deformable Sprites \cite{ye2022deformable}.}

\subsection{Canonical spaces and reconstruction}
We present additional qualitative results and comparisons with previous methods, including Neural Layered Atlases~\cite{kasten2021layered}, Deformable Sprites~\cite{ye2022deformable} and Codef~\cite{ouyang2024codef},
in terms of video reconstruction and canonical space estimation, as shown in Fig.~\ref{fig:sup_other_comp_rec} and Fig.~\ref{fig:sup_other_comp}, respectively. 

Fig. \ref{fig:sup_other_comp_rec} shows that, unlike Neural Layered Atlases and Deformable Sprites, our method consistently yields more detailed reconstructions. Although CoDeF generates very detailed renderings, its canonical spaces are not suitable for editing, as shown in Fig. \ref{fig:sup_other_comp}. In contrast, our method generates intuitive canonical spaces that are well-suited for editing.

\begin{figure*}[h]
  \centering 
  \setlength{\tabcolsep}{0.5pt}
\begin{tabular}{ccccc}
Reference & \multicolumn{2}{c}{CoDeF Editing (Canonic and Render)} & \multicolumn{2}{c}{\textit{VideoSPatS} Editing (Canonic and Render)} \\

\includegraphics[width = 1.3in, height = 0.73in]{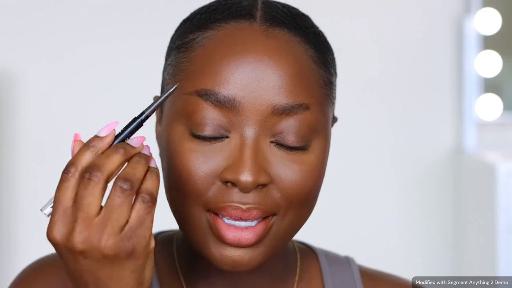} &
\includegraphics[width = 1.3in, height = 0.73in]{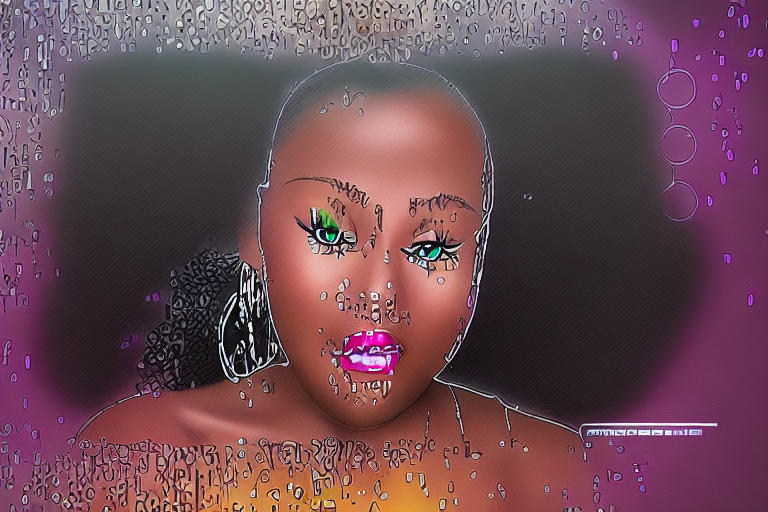} &
\includegraphics[width = 1.3in, height = 0.73in]{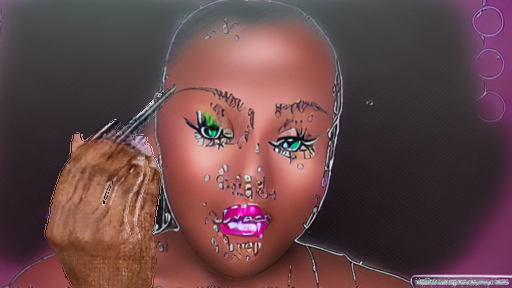} &
\includegraphics[width = 1.3in, height = 0.73in]{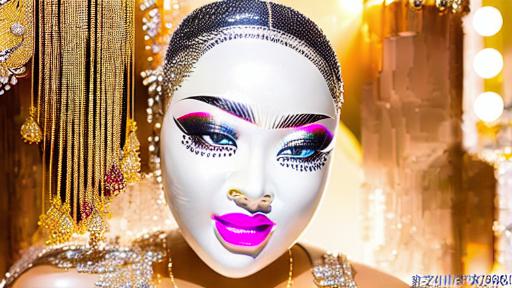} &
\includegraphics[width = 1.3in, height = 0.73in]{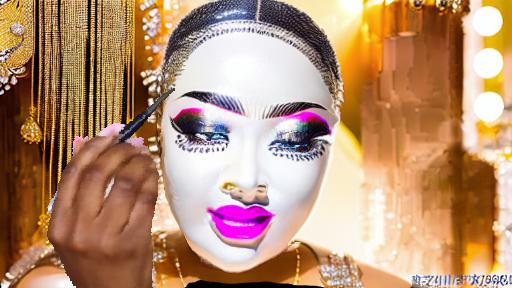} \\

\includegraphics[width = 1.3in, height = 0.73in]{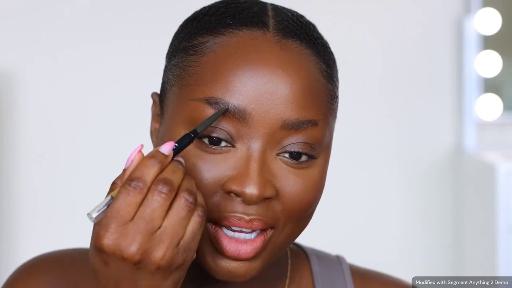} &
\includegraphics[width = 1.3in, height = 0.73in]{figures/sup_res_editing/canonic_1_codef_make.png} &
\includegraphics[width = 1.3in, height = 0.73in]{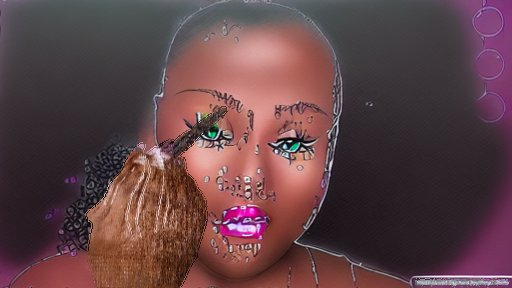} &
\includegraphics[width = 1.3in, height = 0.73in]{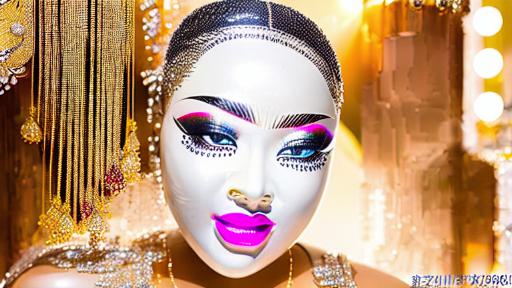} &
\includegraphics[width = 1.3in, height = 0.73in]{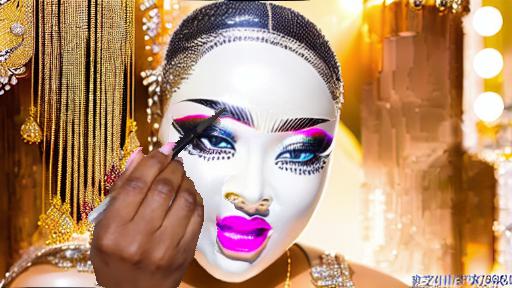} \\

\includegraphics[width = 1.3in]{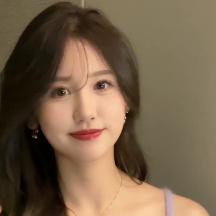} &
\includegraphics[width = 1.3in]{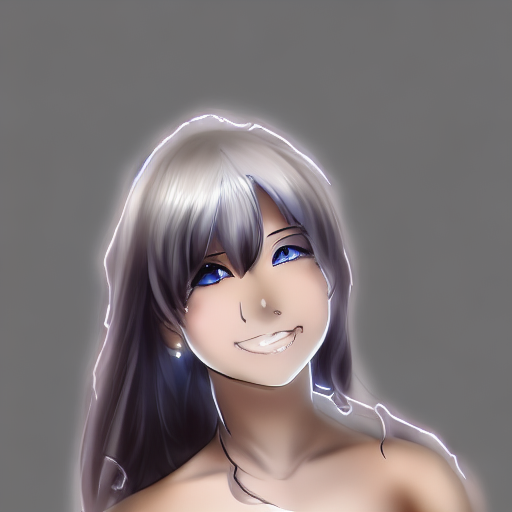} &
\includegraphics[width = 1.3in]{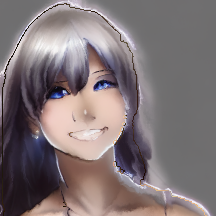} &
\includegraphics[width = 1.3in]{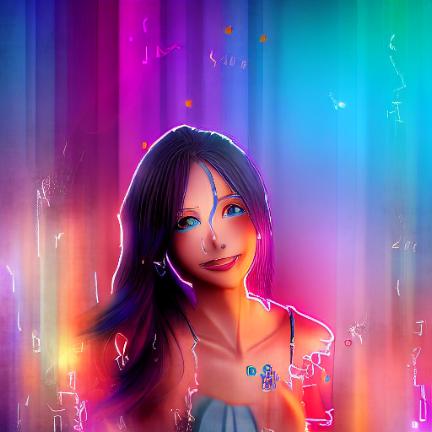} &
\includegraphics[width = 1.3in]{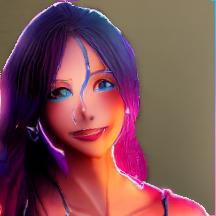} \\

\includegraphics[width = 1.3in]{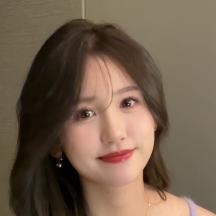} &
\includegraphics[width = 1.3in]{figures/sup_res_editing/canonic_codef_beauty.png} &
\includegraphics[width = 1.3in]{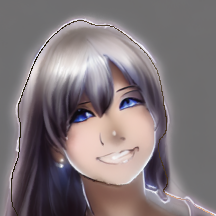} &
\includegraphics[width = 1.3in]{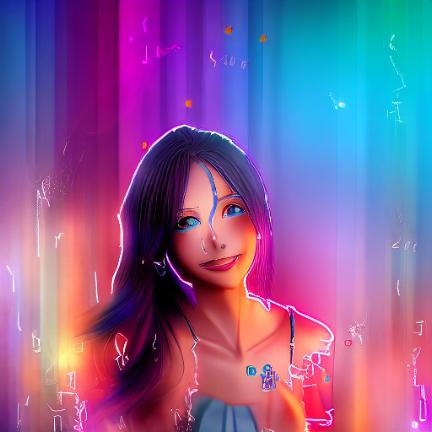} &
\includegraphics[width = 1.3in]{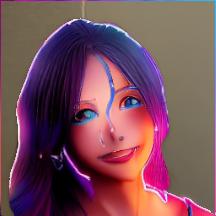} \\

\includegraphics[width = 1.3in, height = 0.73in]{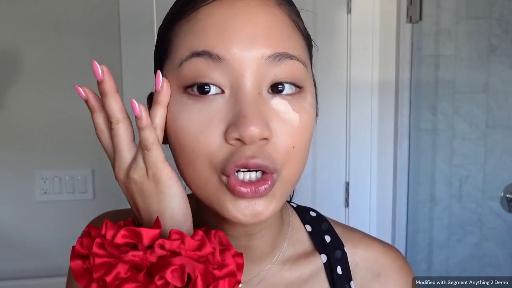} &
\includegraphics[width = 1.3in, height = 0.73in]{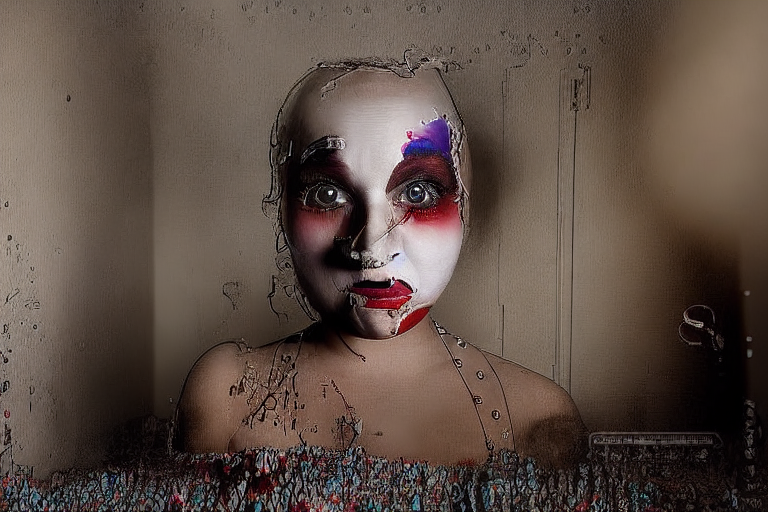} &
\includegraphics[width = 1.3in, height = 0.73in]{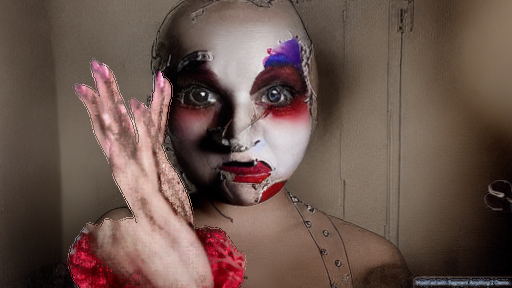} &
\includegraphics[width = 1.3in, height = 0.73in]{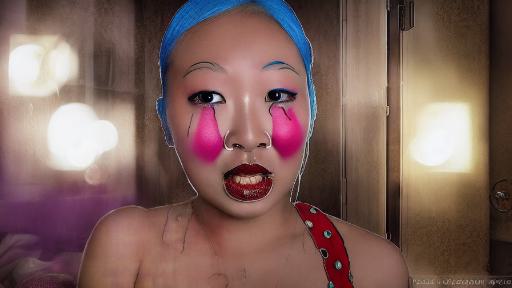} &
\includegraphics[width = 1.3in, height = 0.73in]{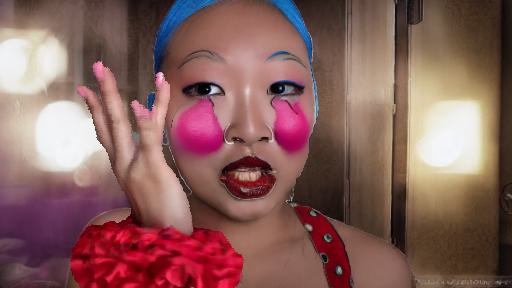} \\

\includegraphics[width = 1.3in, height = 0.73in]{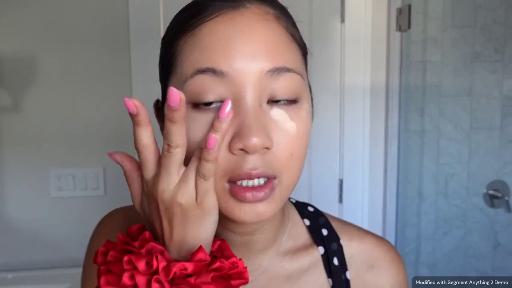} &
\includegraphics[width = 1.3in, height = 0.73in]{figures/sup_res_editing/image_codef_clown_canonic.png} &
\includegraphics[width = 1.3in, height = 0.73in]{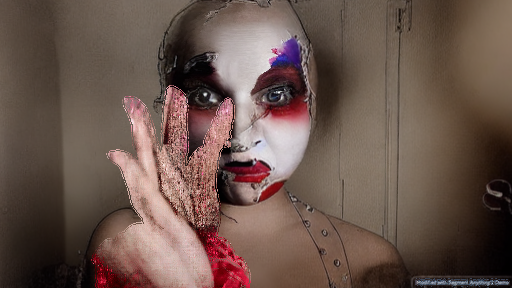} &
\includegraphics[width = 1.3in, height = 0.73in]{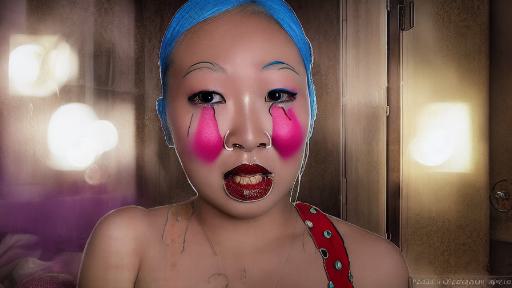} &
\includegraphics[width = 1.3in, height = 0.73in]{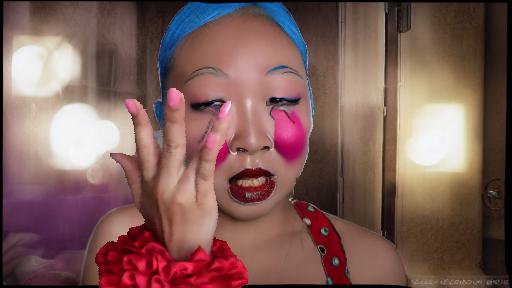} \\

\end{tabular}
  \caption{Additional editing results. The consistency of our canonical spaces allows for better deep editing than that of CoDeF.}
  \label{fig:add_editing}
\end{figure*}

\subsection{Comparisons to diffusion-based methods.} 
\begin{figure}[h]
  \centering 
  \scriptsize
  \renewcommand{\arraystretch}{1.1}
\setlength{\tabcolsep}{1pt}
\begin{tabular}{c|c|c}
Original & ReVideo & Ours \\
\includegraphics[width = 1.1in]{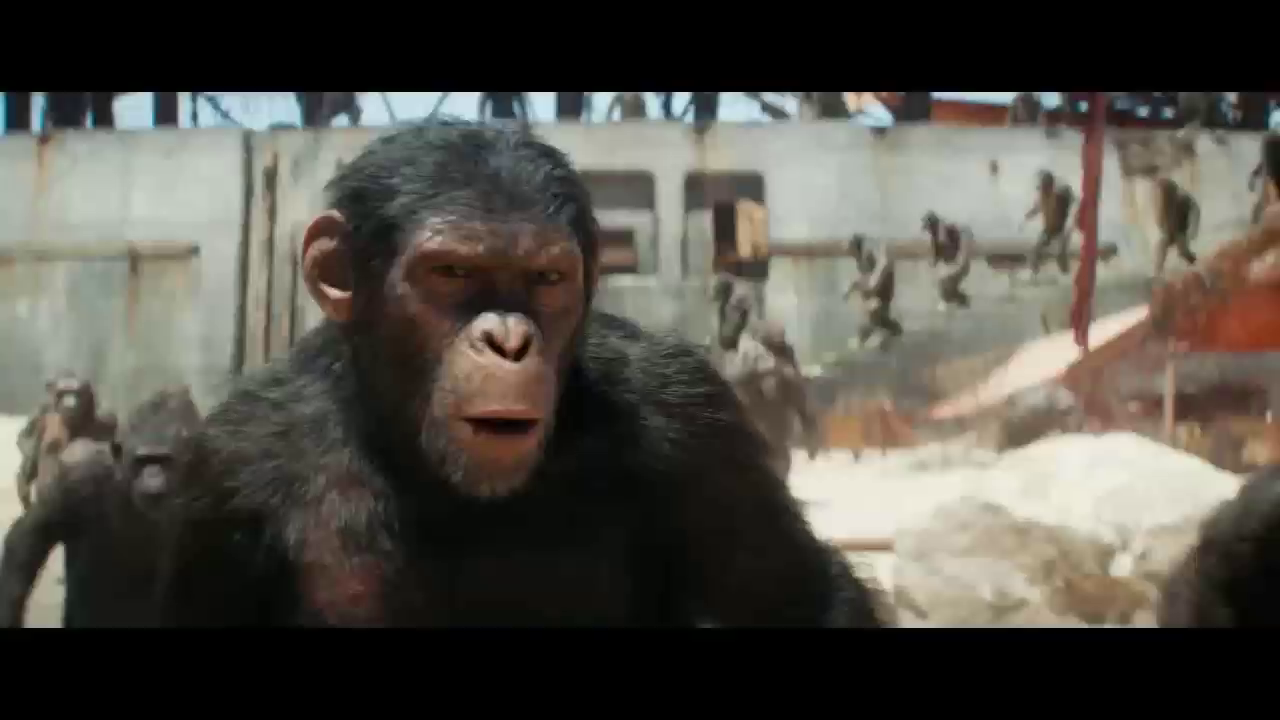} &
\includegraphics[width = 1.1in]{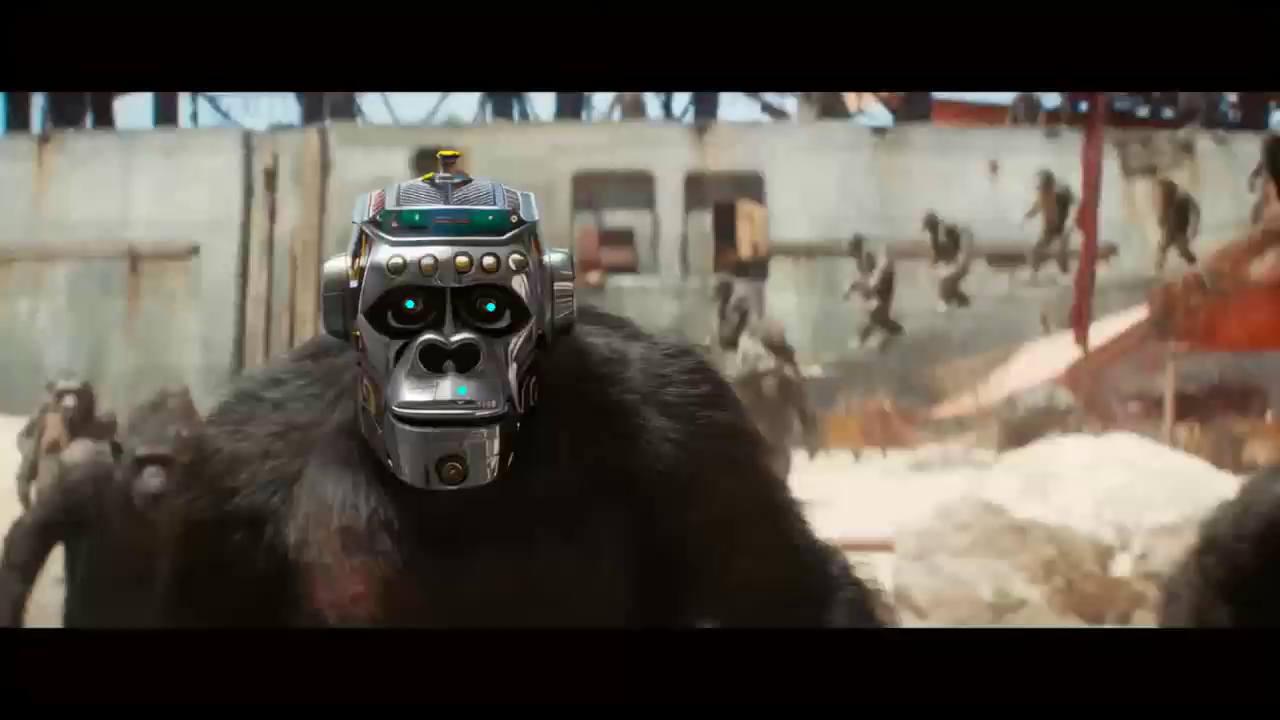} &
\includegraphics[width = 1.1in]{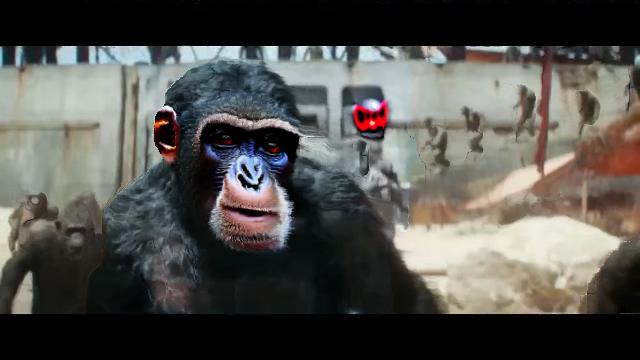} \\

\includegraphics[width = 1.1in]{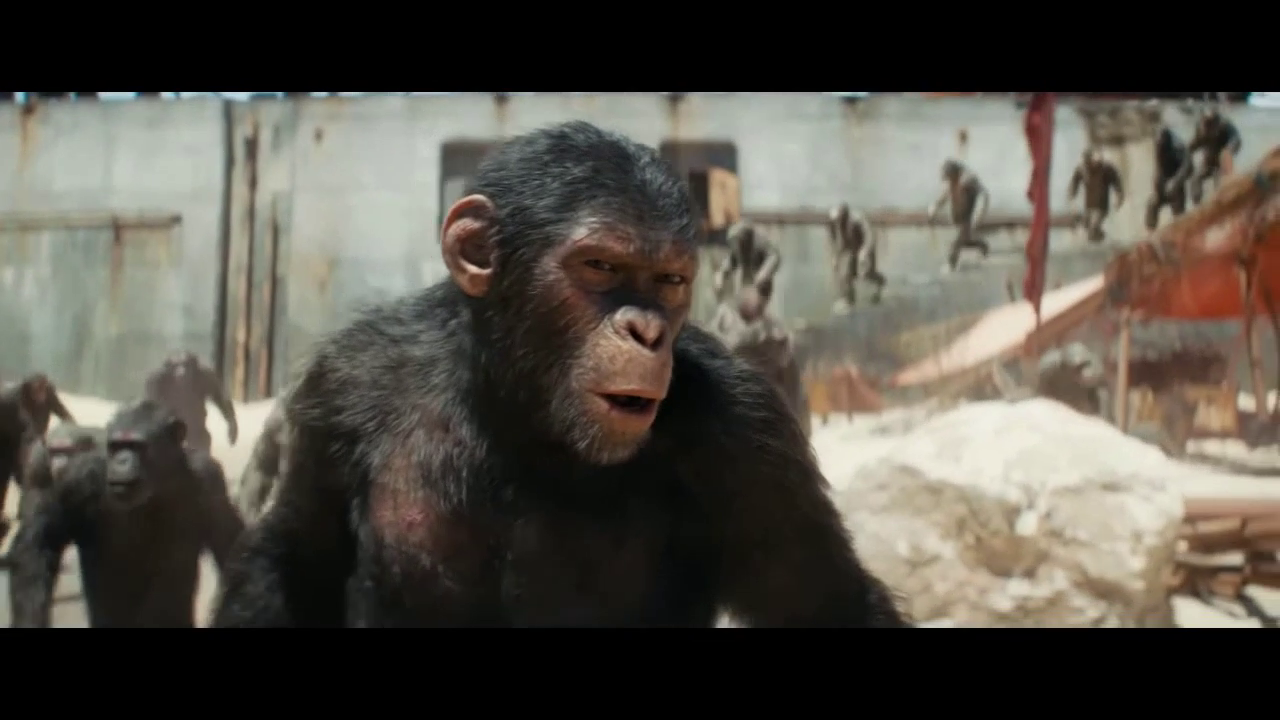} &
\includegraphics[width = 1.1in]{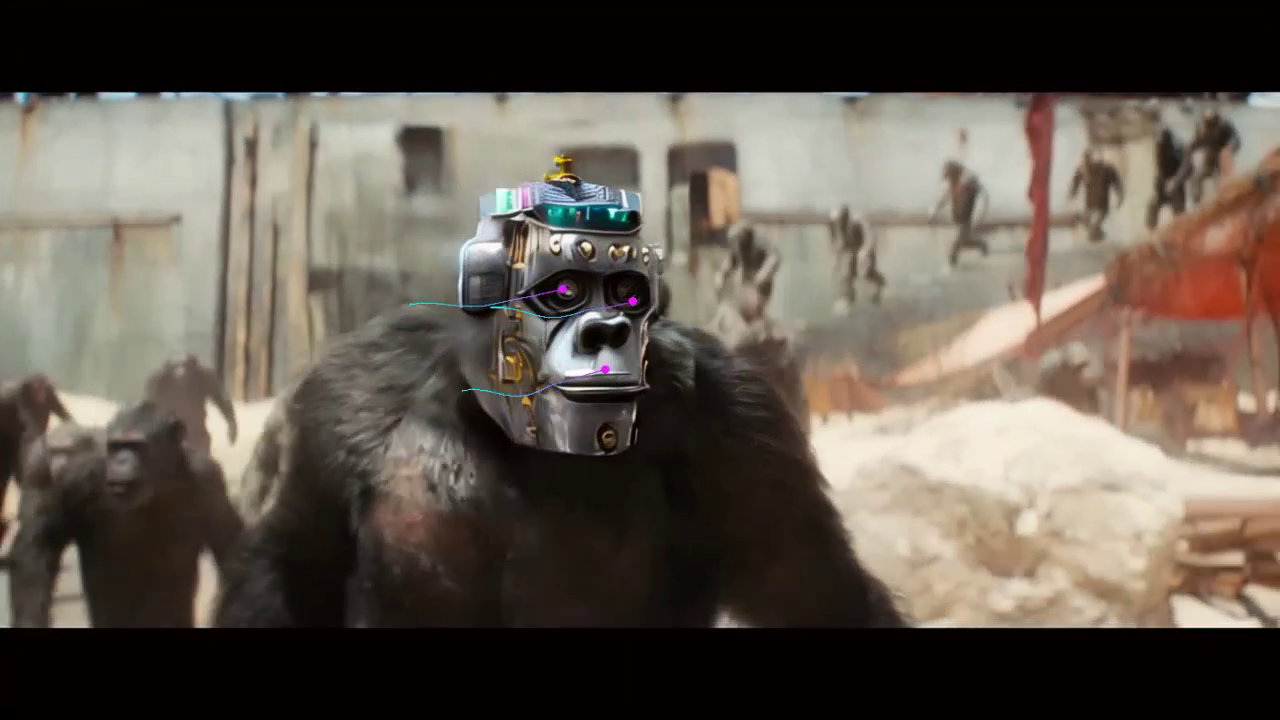} &
\includegraphics[width = 1.1in]{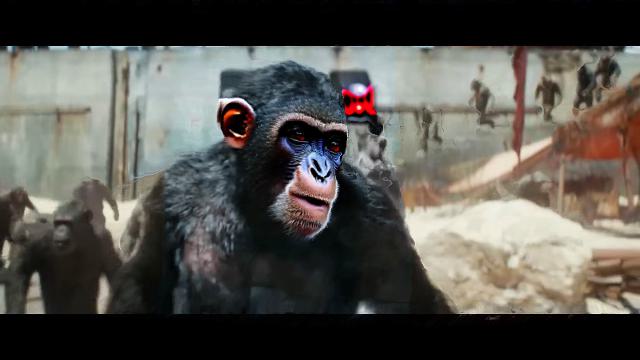} \\

\end{tabular}
  \caption{Comparison to ReVideo \cite{revideo}.}
  \label{fig:r_vs_revideo}
\end{figure}

\camready{Although flexible for semantic video editing, diffusion-based methods such as \cite{I2VEdit, revideo} are not designed for time-dependent appearance editing or do not support motion editing. Our method, closer to warping-based video modeling, focuses on modeling motion, appearance, and occlusions, so we did not compare to general video editing approaches in the main paper. For completeness, we provide an additional comparison to ReVideo \cite{revideo} in Fig. \ref{fig:r_vs_revideo}. Ours keeps original head poses and temporal consistency, and ReVideo changes semantics.}

\begin{figure*}[ht]
  \centering 
  \setlength{\tabcolsep}{1.5pt}
\begin{tabular}{cccccc}
Amplified by $\times 1.5$ & Amplified $\times 0.5$ & $v$ off every 2 frames & $u, v$ off every 3, 2 frames & GT \\

\includegraphics[width = 1.3in]{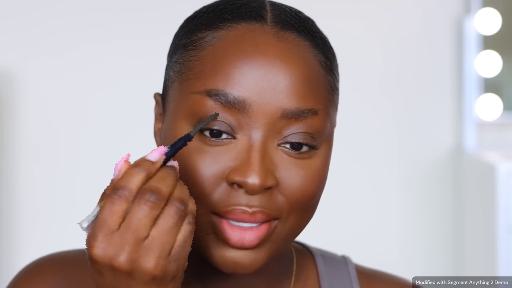} &
\includegraphics[width = 1.3in]{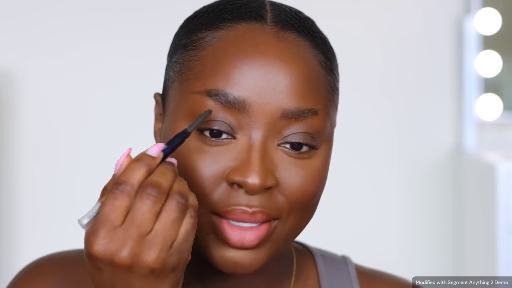} &
\includegraphics[width = 1.3in]{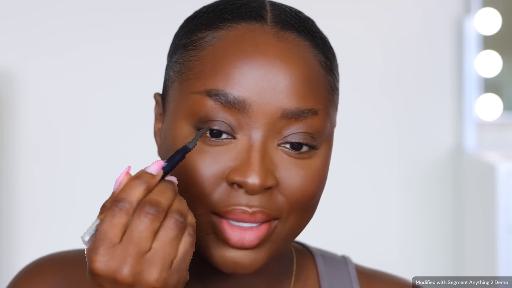} &
\includegraphics[width = 1.3in]{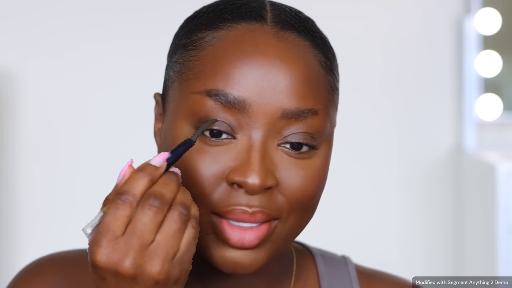} &
\includegraphics[width = 1.3in]{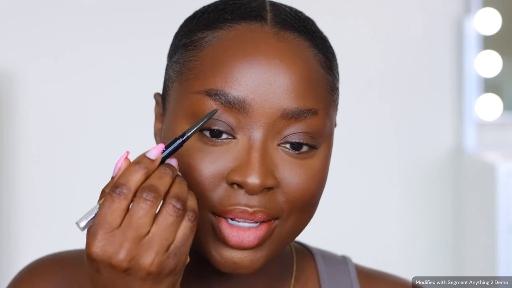} \\

\includegraphics[width = 1.3in]{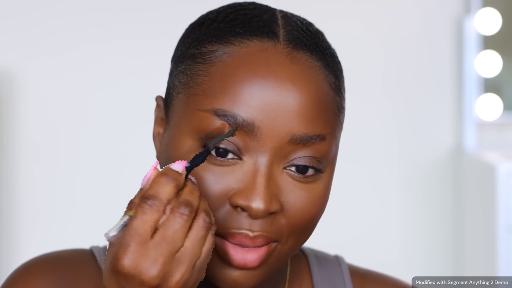} &
\includegraphics[width = 1.3in]{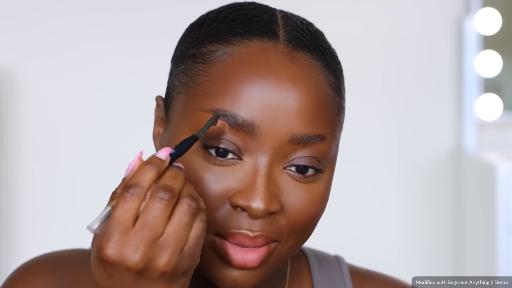} &
\includegraphics[width = 1.3in]{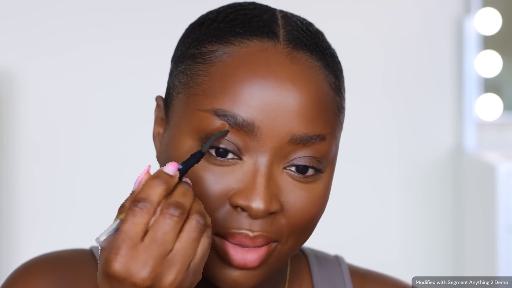} &
\includegraphics[width = 1.3in]{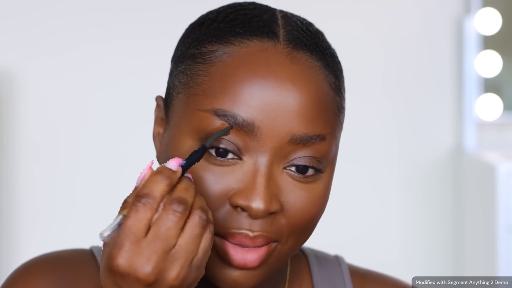} &
\includegraphics[width = 1.3in]{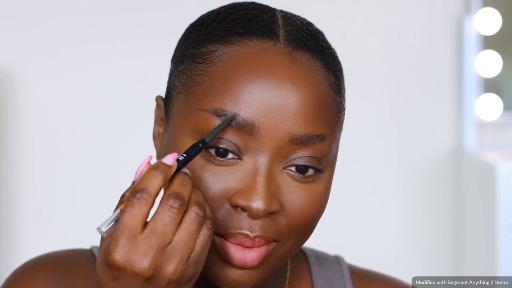} \\

\includegraphics[width = 1.3in]{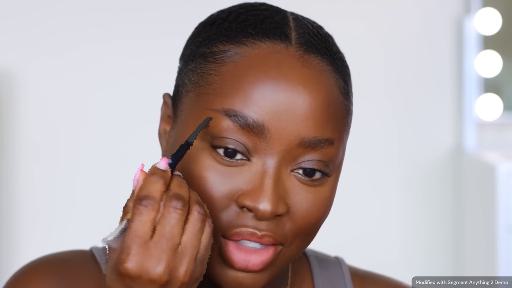} &
\includegraphics[width = 1.3in]{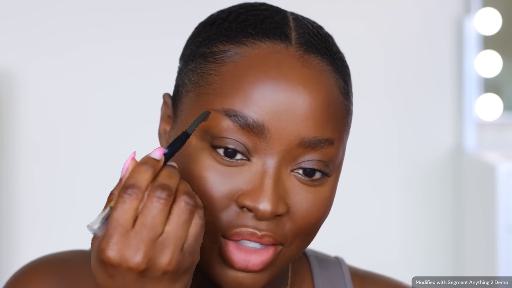} &
\includegraphics[width = 1.3in]{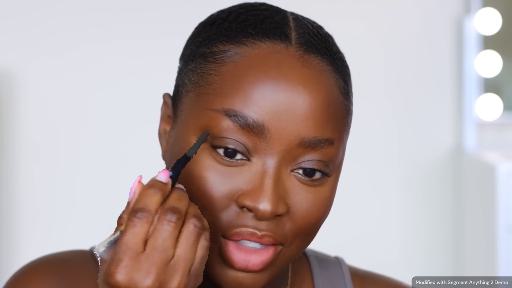} &
\includegraphics[width = 1.3in]{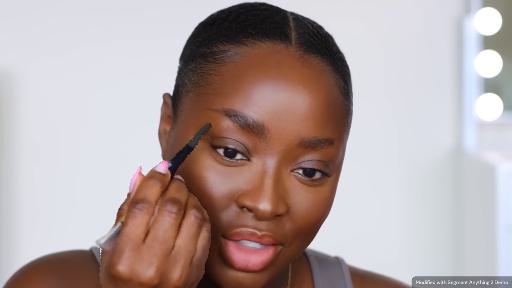} &
\includegraphics[width = 1.3in]{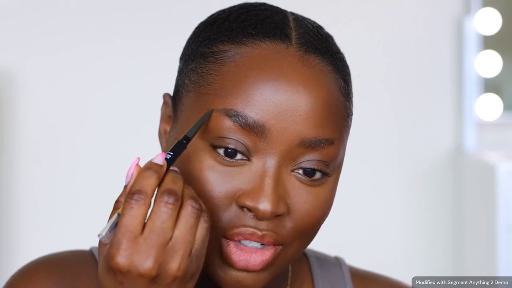} \\





\end{tabular}
  \caption{Additional results on motion editing by control points. See our videos for a better visualization.}
  \label{fig:motion_editing}
\end{figure*}

\subsection{Texture editing}
We provide additional editing results in Fig~\ref{fig:add_editing}. We use ControlNet~\cite{zhang2023adding} to apply editing on the canonical space. Note that the inconsistencies of canonical spaces in CoDeF prevent ControlNet from generating a high quality edit, as shown in the first and last rows of Fig~\ref{fig:add_editing}. In contrast, our method generates more edit-friendly canonical spaces that are translated into higher-quality, temporally-consistent images.

\subsection{Motion editing}
By modifying the precomputed control points, we can smoothly perform motion editing. For instance, we can select every $m$ control point of each foreground pixels and apply a vertical offset. Thanks to the spline nature of our deformation fields, we can smoothly transfer this new motion into the rendered video. Thanks to our spline deformation fields, instead of rendering frames where the foreground is instantly ``teleporting'' to the offset location, our motion-edited frames are smoothly rendered without discontinuities. Additional motion edits, such as amplification and diminishing of motion, are shown in the attached videos as well as in Fig.~\ref{fig:motion_editing}.

\subsection{Experiments on long sequences}
\begin{figure*}[h]
  \centering 
  \setlength{\tabcolsep}{0.5pt}
\begin{tabular}{cccccc}
0s & 1s & 2s & 3s & 6s & 10s \\

\includegraphics[width = 1.1in]{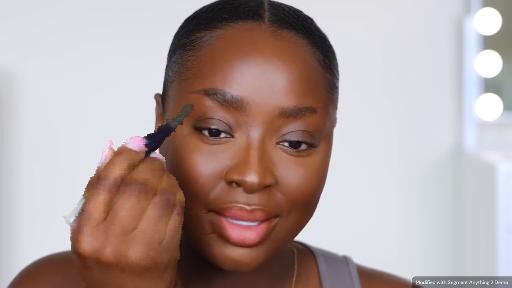} &
\includegraphics[width = 1.1in]{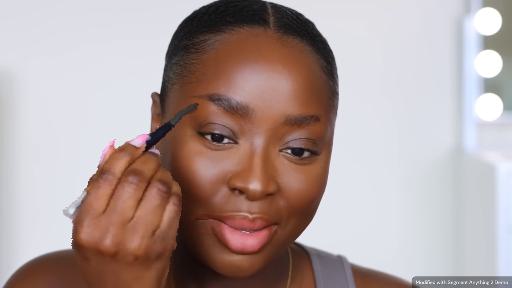} &
\includegraphics[width = 1.1in]{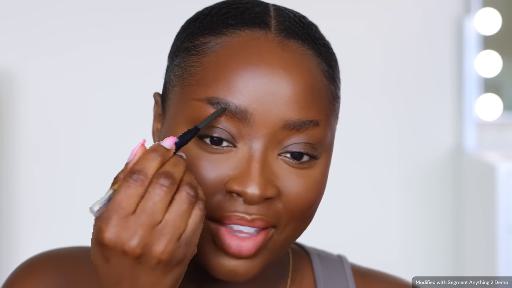} &
\includegraphics[width = 1.1in]{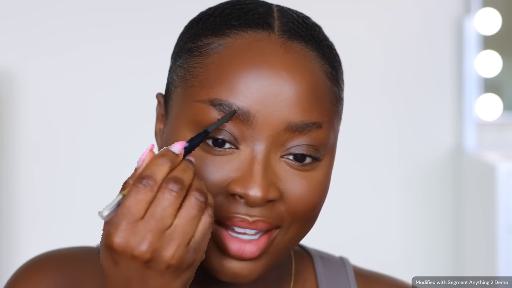} &
\includegraphics[width = 1.1in]{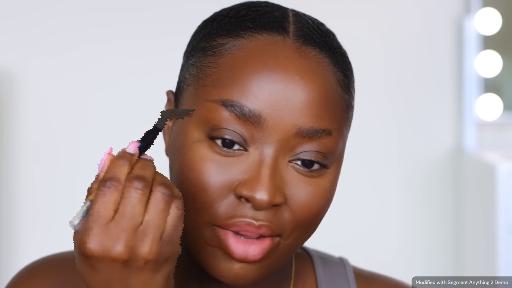} &
\includegraphics[width = 1.1in]{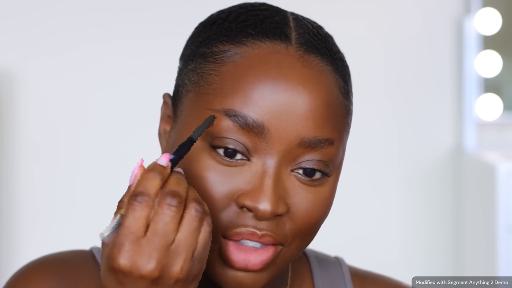} \\

\includegraphics[width = 1.1in]{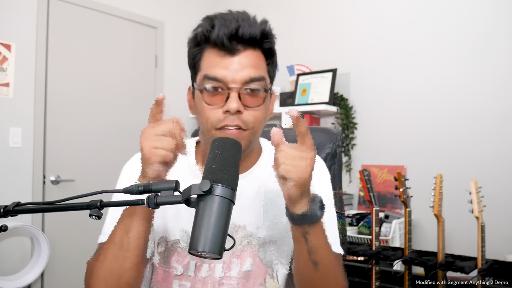} &
\includegraphics[width = 1.1in]{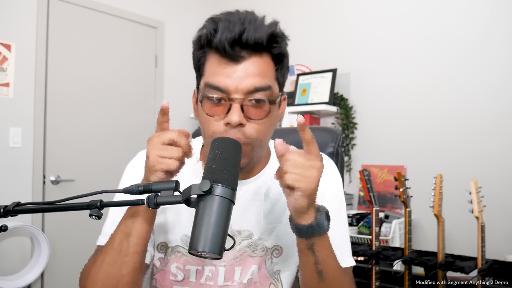} &
\includegraphics[width = 1.1in]{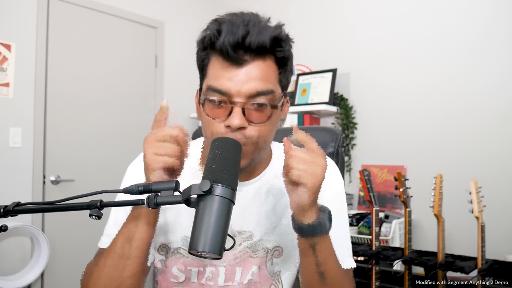} &
\includegraphics[width = 1.1in]{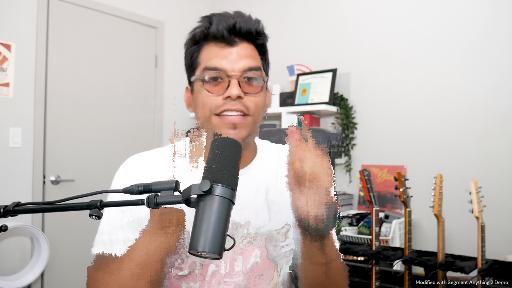} &
\includegraphics[width = 1.1in]{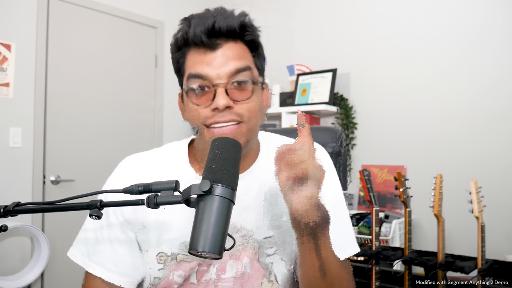} &
\includegraphics[width = 1.1in]{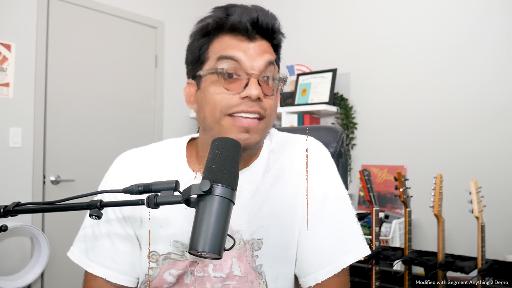} \\

\includegraphics[width = 1.1in]{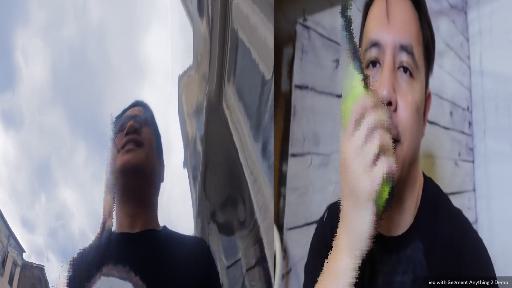} &
\includegraphics[width = 1.1in]{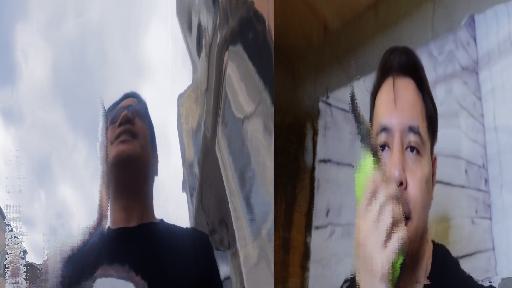} &
\includegraphics[width = 1.1in]{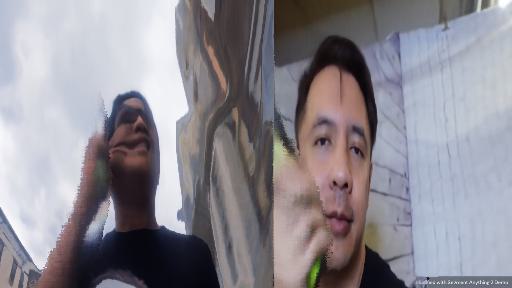} &
\includegraphics[width = 1.1in]{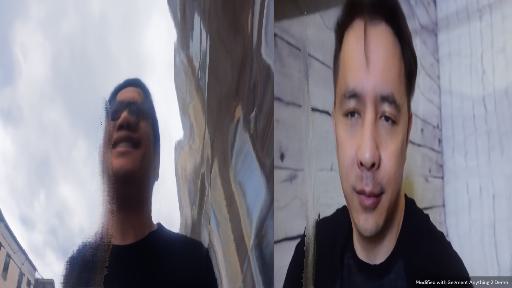} &
\includegraphics[width = 1.1in]{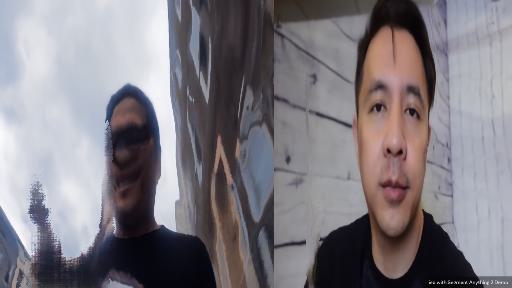} &
\includegraphics[width = 1.1in]{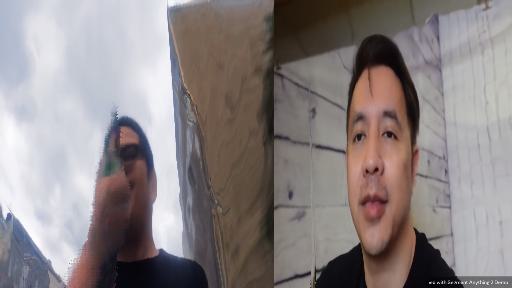} \\

\end{tabular}
  \caption{Additional results on long sequences. Our method can capture long-range relationships in long video sequences (10s). }
  \label{fig:add_long}
\end{figure*}

While most of the experiments mentioned above were conducted with videos of 50 frames, our method also performs well on longer sequences. Fig.~\ref{fig:add_long} presents additional results on sequences of 10 seconds. Our method is capable of capturing the long-range correspondences in longer videos.

\section{Additional ablation studies}
\label{sec:supp_ablation}

\subsection{Spatial regularization loss}
\label{sec:sup_lds_effects}

\begin{figure}[h]
  \centering 
  \setlength{\tabcolsep}{2pt}
\begin{tabular}{c|cc}
Ground Truth & With $l_{\mathcal{D}_s}$ & Without $l_{\mathcal{D}_s}$ \\

\includegraphics[width = 1.05in]{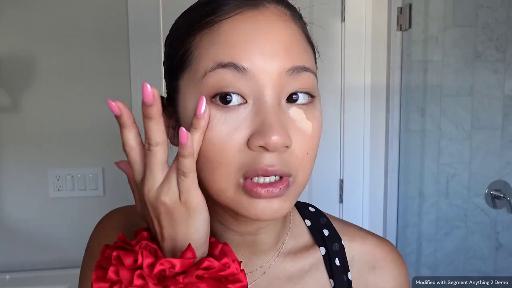} &
\includegraphics[width = 1.05in]{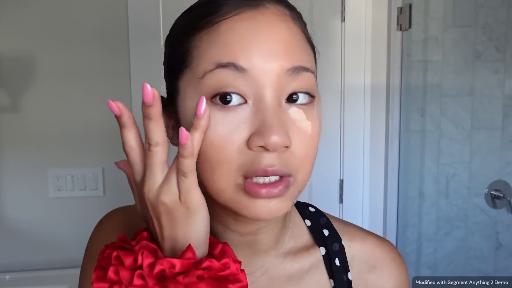} &
\includegraphics[width = 1.05in]{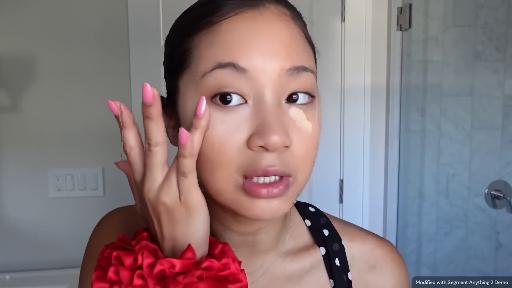} \\

&
\includegraphics[width = 1.05in]{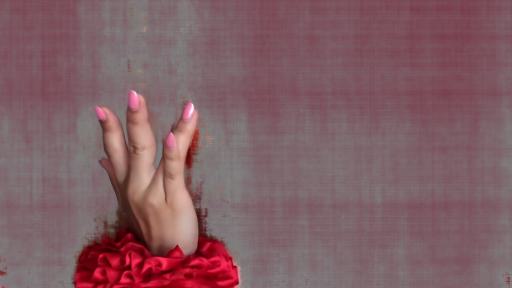} &
\includegraphics[width = 1.05in]{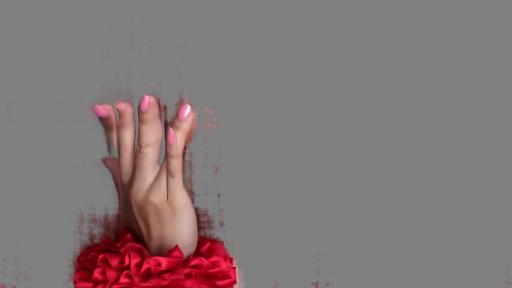} \\

&
\includegraphics[width = 1.05in]{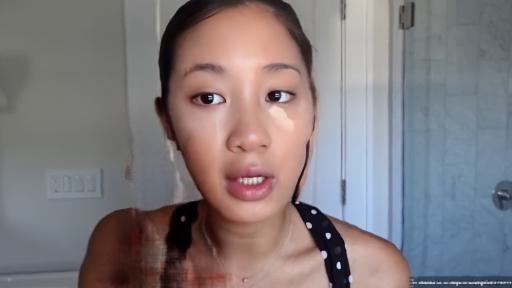} &
\includegraphics[width = 1.05in]{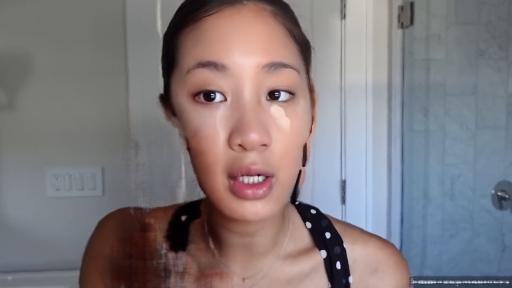} \\

\includegraphics[width = 1.05in]{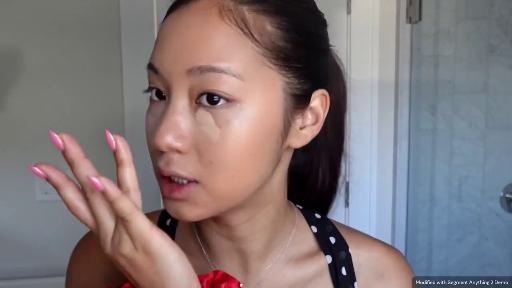} &
\includegraphics[width = 1.05in]{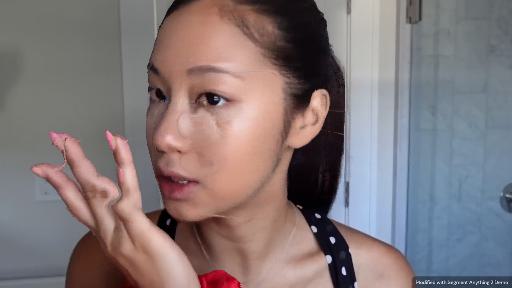} &
\includegraphics[width = 1.05in]{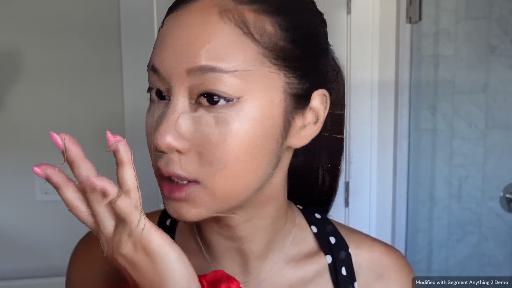} \\

&
\includegraphics[width = 1.05in]{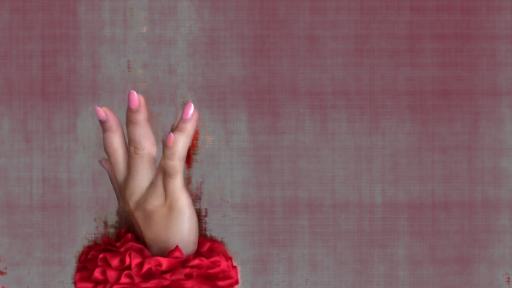} &
\includegraphics[width = 1.05in]{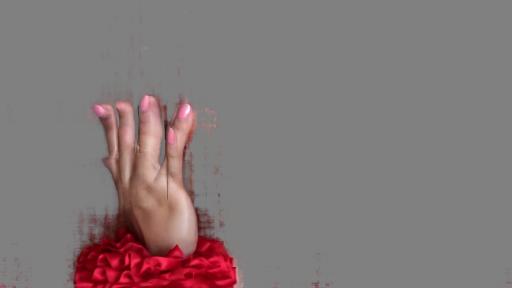} \\

&
\includegraphics[width = 1.05in]{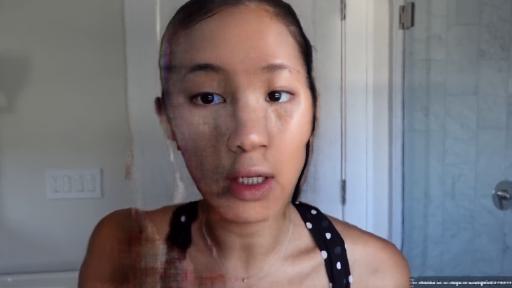} &
\includegraphics[width = 1.05in]{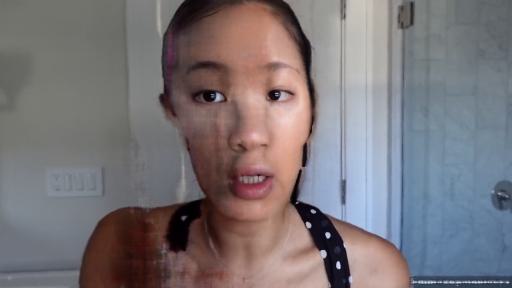} \\

\end{tabular}
  \caption{Effects of $l_{\mathcal{D}_s}$. \textit{From top to bottom:} Composited images, foreground occluder canonical spaces, and background face canonical spaces. Our model without $l_{\mathcal{D}_s}$ yields a slightly squeezed canonical space, with respect to the observed frames and our model with $l_{\mathcal{D}_s}$.}
  \label{fig:add_ablation_lds}
\end{figure}

We show the effects of the Spatial Splines Deformation Regularization loss, $l_{\mathcal{D}_s}$, in Fig. \ref{fig:add_ablation_lds}. Although the contribution of the regularization loss is minimal to the canonical space and final reconstruction, it still helps maintain a better aspect ratio between the canonical space and the observed space. This is because it encourages similar deformations between neighboring pixel locations, preventing the ``squeeze'' of the canonical space, as observed in the ``without $l_{\mathcal{D}_s}$'' column of Fig.~\ref{fig:add_ablation_lds}.

\subsection{Color regularization loss}
\label{sec:sup_ldc_effects}
\begin{figure}[h]
  \centering 
  \small
  \renewcommand{\arraystretch}{1.0}
\setlength{\tabcolsep}{1.0pt}
\begin{tabular}{cc|cc}
\multicolumn{2}{c}{Canonical Foreground w/ $D_c$} & \multicolumn{2}{c}{Canonical Foreground w/o $D_c$} \\
$t=0$ & $t=1$ & $t=0$ & $t=1$ \\
\includegraphics[width = 0.8in]{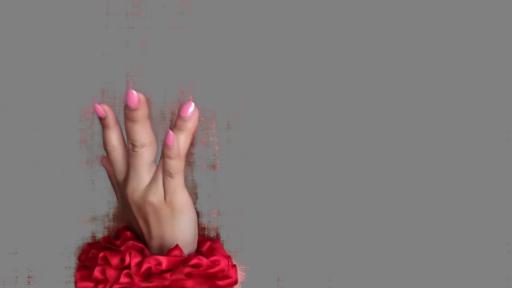} &
\includegraphics[width = 0.8in]{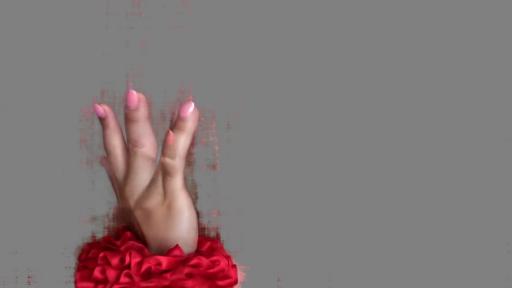} &
\includegraphics[width = 0.8in]{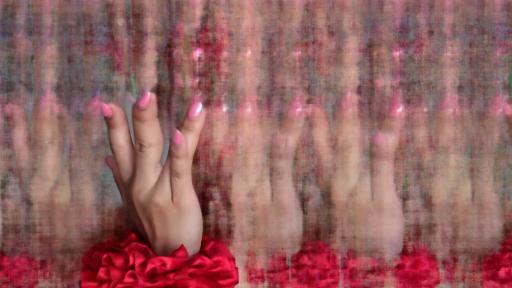} &
\includegraphics[width = 0.8in]{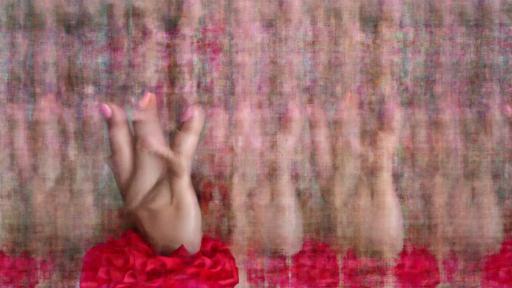} \\
\end{tabular}
  \vspace*{-2mm}
  \caption{Effects of $l_{D_c}$. $t=0$: start, $t=1$: end of the video.}
  \label{fig:r_dc}
\end{figure}

Fig.~\ref{fig:r_dc} depicts the effects of the Color Deformation Regularization loss, $l_{D_c}$, showing that not regularizing $P_c$ can lead to potential entanglement between motion and appearance in the canonical space, as shown in the bent finger on the rightmost image.

\begin{figure}[h]
    \centering
    \small
    \renewcommand{\arraystretch}{1}
\setlength{\tabcolsep}{2pt}
\begin{tabular}{ccc}
\includegraphics[width = 1.05in]{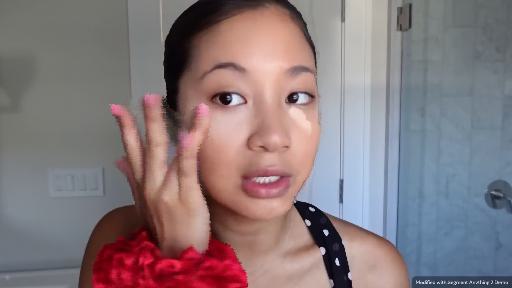} &
\includegraphics[width = 1.05in]{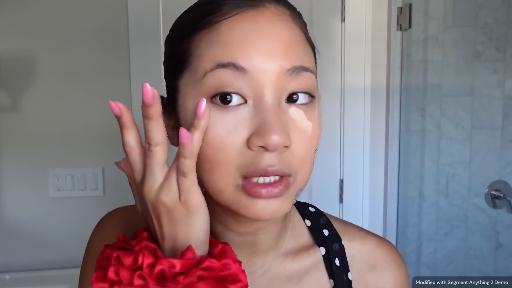} &
\includegraphics[width = 1.05in]{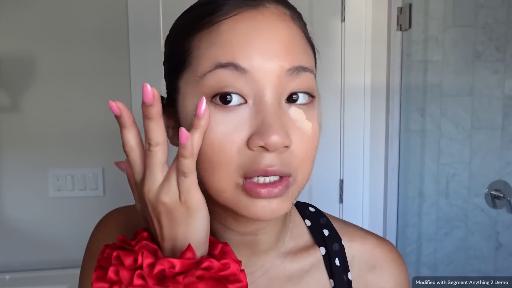} \\

10k-8.88min & 30k-27.23min & 100k-88.6min \\
30.67dB & 36.16dB & 37.21dB \\
\end{tabular}
    \vspace*{-2mm}
    \captionof{figure}{Effects of Iteration \# on reconstruction PSNR.}
    \label{fig:r_iterations}
\end{figure}

\subsection{Levels of guidance mask}
\begin{figure*}[h]
  \centering 
  \setlength{\tabcolsep}{0.5pt}
\begin{tabular}{cccc}
Guidance Mask & Foreground and alpha & Background & Composited \\
\includegraphics[width = 1.7in]{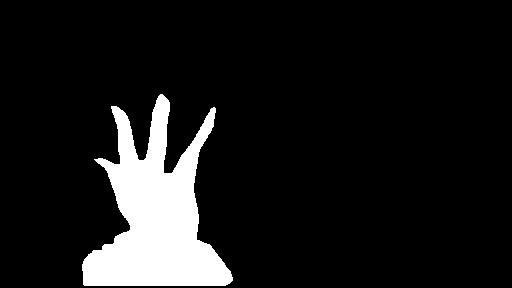} &
\includegraphics[width = 1.7in]{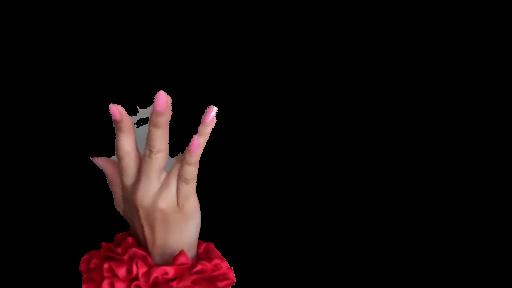} &
\includegraphics[width = 1.7in]{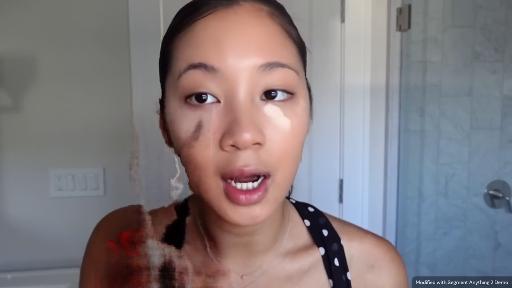} &
\includegraphics[width = 1.7in]{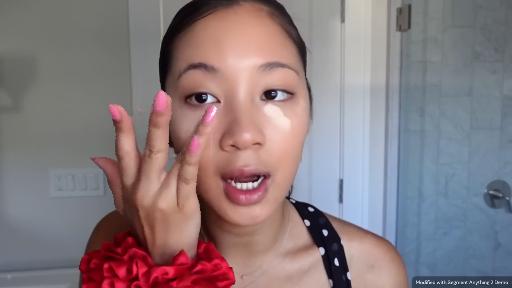}\\

\includegraphics[width = 1.7in]{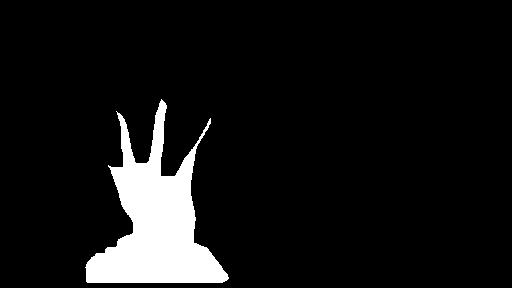} &
\includegraphics[width = 1.7in]{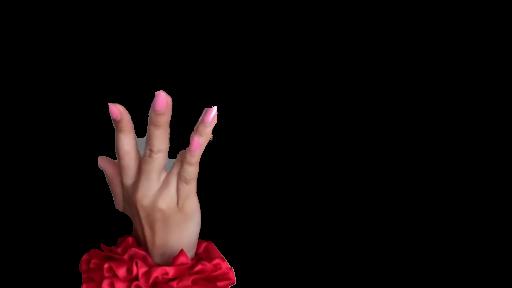} &
\includegraphics[width = 1.7in]{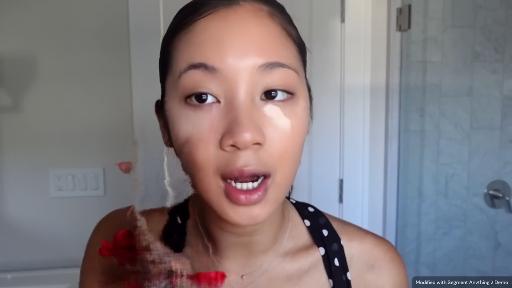} &
\includegraphics[width = 1.7in]{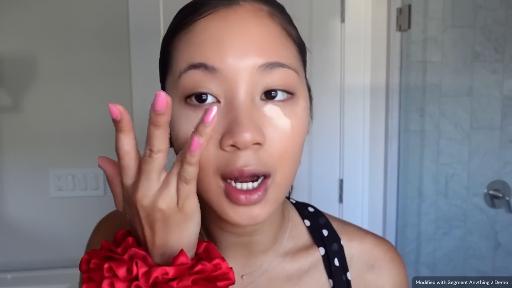}\\

\includegraphics[width = 1.7in]{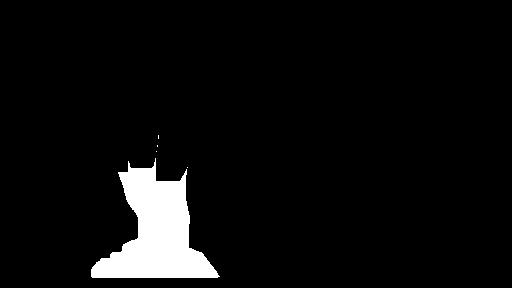} &
\includegraphics[width = 1.7in]{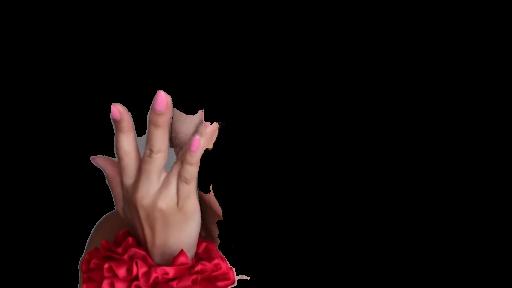} &
\includegraphics[width = 1.7in]{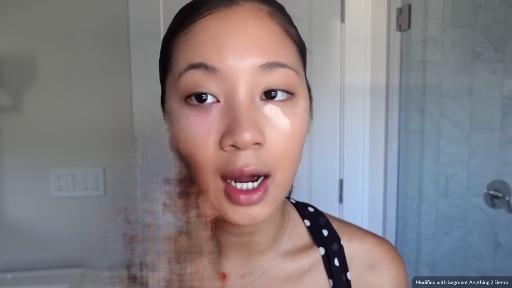} &
\includegraphics[width = 1.7in]{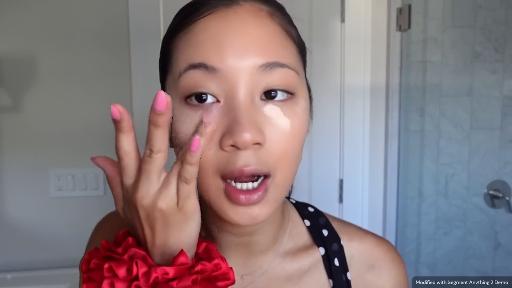}\\

\includegraphics[width = 1.7in]{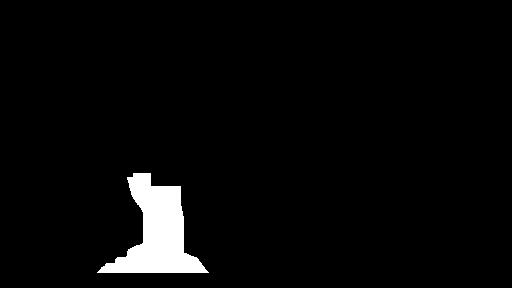} &
\includegraphics[width = 1.7in]{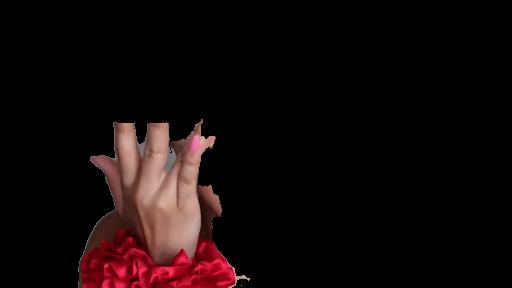} &
\includegraphics[width = 1.7in]{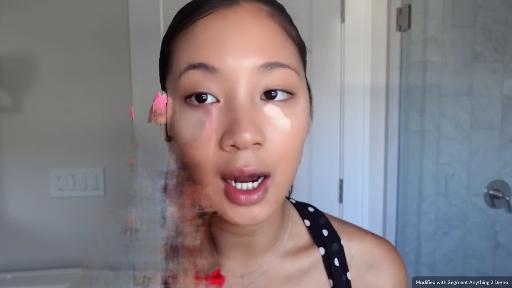} &
\includegraphics[width = 1.7in]{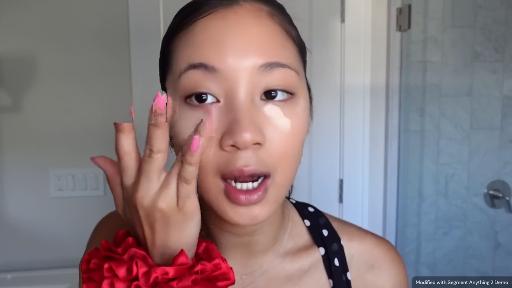}\\

\includegraphics[width = 1.7in]{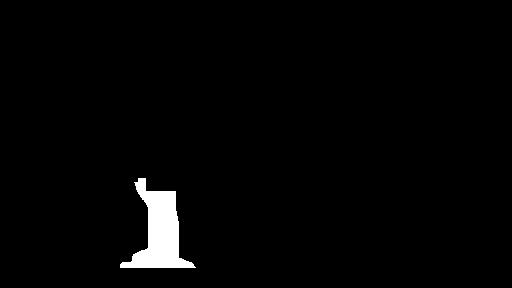} &
\includegraphics[width = 1.7in]{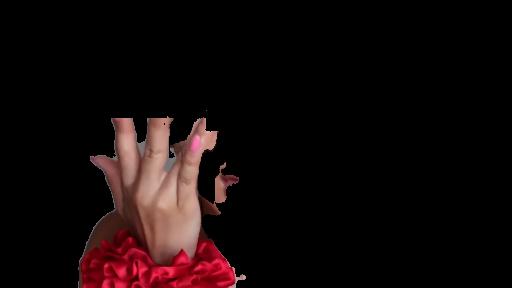} &
\includegraphics[width = 1.7in]{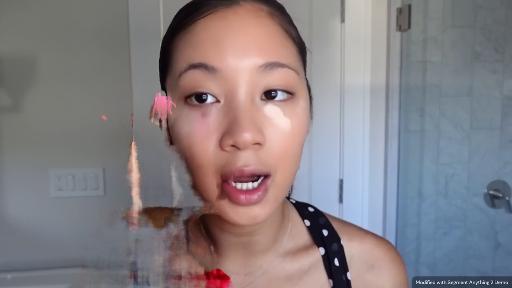} &
\includegraphics[width = 1.7in]{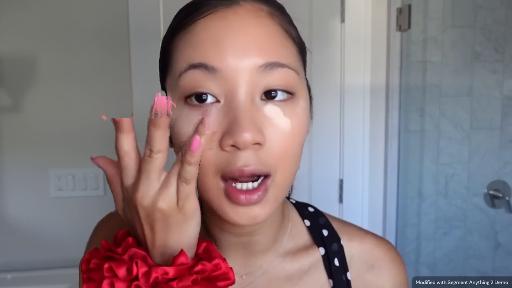}\\
\end{tabular}
  \caption{Additional ablation studies on guidance mask. Original guidance mask from SAM2\cite{ravi2024sam2} is eroded by 5, 11, 21, 31, and 41 pixels. Even under extreme erosion, our method can still reasonably separate the occluder foreground and the face background.}
  \label{fig:add_ablation_guide}
\end{figure*}

In the main paper, we show that our method can refine the guidance mask. Fig.~\ref{fig:add_ablation_guide} provides additional results on different levels of degradation of the guidance mask. 
\camready{In this supplemental study, our motivation is to show the robustness of the proposed method when the guidance mask is imperfect. As shown in Fig.~\ref{fig:add_ablation_guide} our proposed model can capture the foreground motion even with a rough mask.
Although our method cannot recover the mask when it is too heavily degraded (last row in Fig.~\ref{fig:add_ablation_guide}), it still succeeds with smaller degradation levels, supporting our design choices in Section \ref{subsec:optim}.}

\subsection{Number of control points}
\begin{figure*}[h]
  \centering 
  \setlength{\tabcolsep}{0.5pt}
\begin{tabular}{ccccc}
Canonical Foreground & Canonical Background & Foreground & Background & Composited \\

\includegraphics[width = 1.3in]{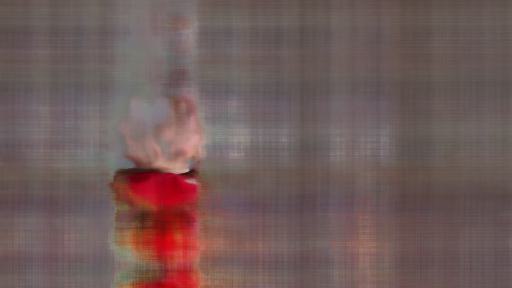} &
\includegraphics[width = 1.3in]{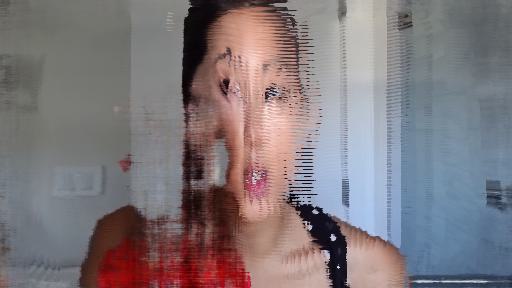} &
\includegraphics[width = 1.3in]{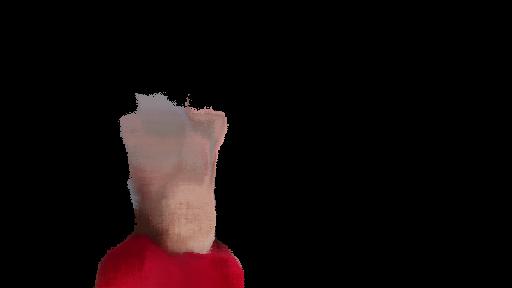} &
\includegraphics[width = 1.3in]{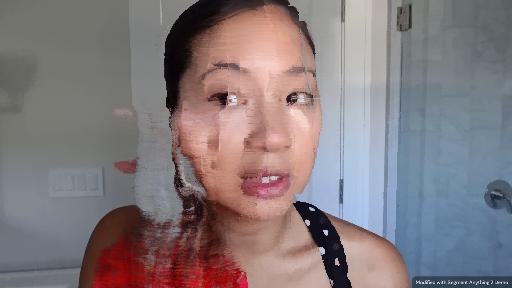} &
\includegraphics[width = 1.3in]{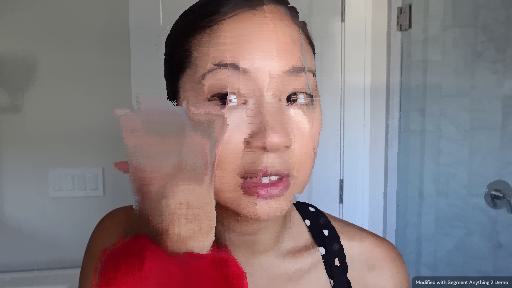} \\

\includegraphics[width = 1.3in]{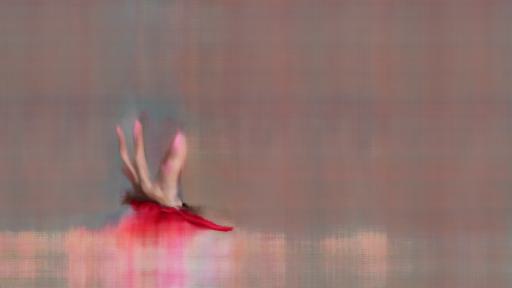} &
\includegraphics[width = 1.3in]{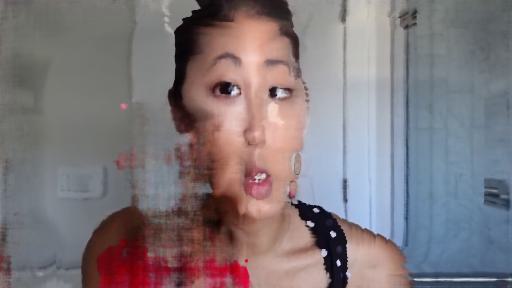} &
\includegraphics[width = 1.3in]{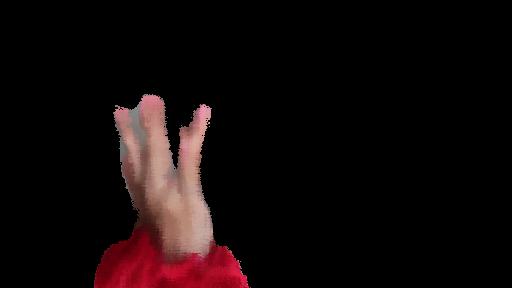} &
\includegraphics[width = 1.3in]{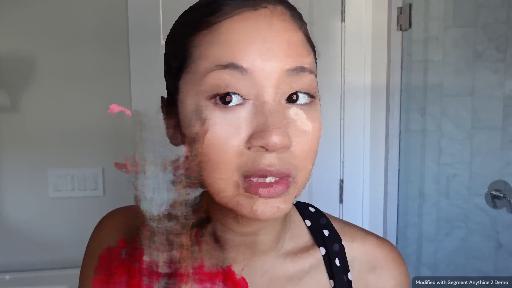} &
\includegraphics[width = 1.3in]{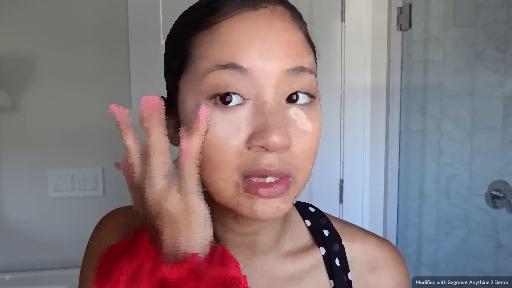} \\

\includegraphics[width = 1.3in]{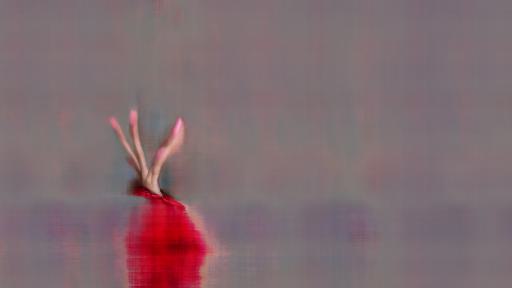} &
\includegraphics[width = 1.3in]{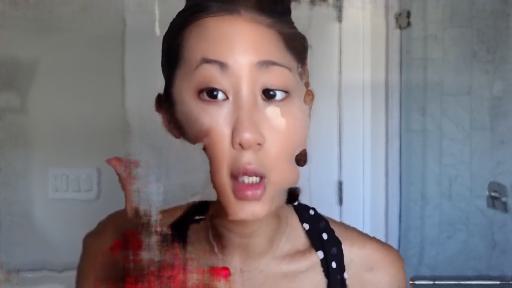} &
\includegraphics[width = 1.3in]{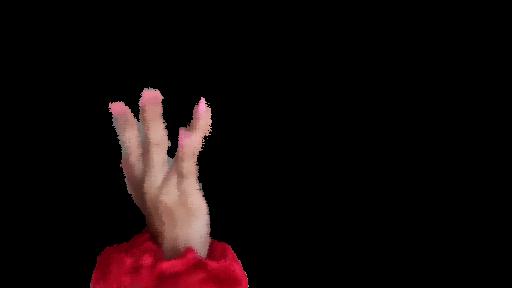} &
\includegraphics[width = 1.3in]{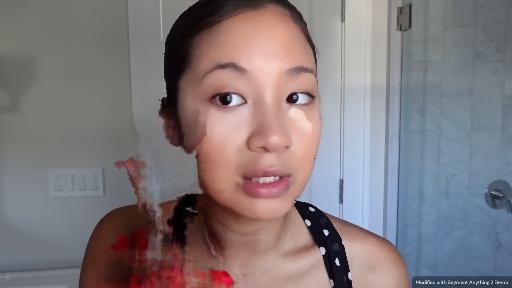} &
\includegraphics[width = 1.3in]{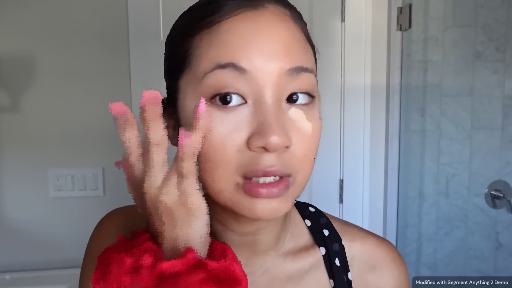} \\

\includegraphics[width = 1.3in]{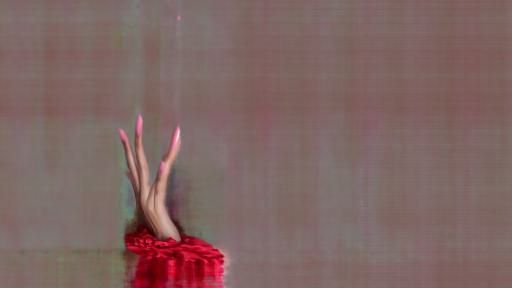} &
\includegraphics[width = 1.3in]{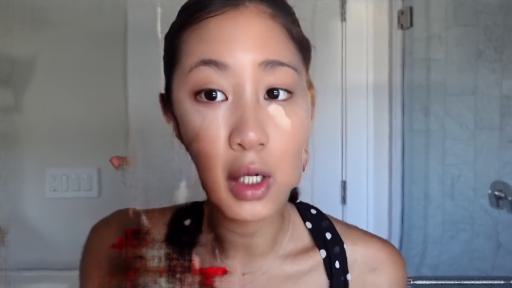} &
\includegraphics[width = 1.3in]{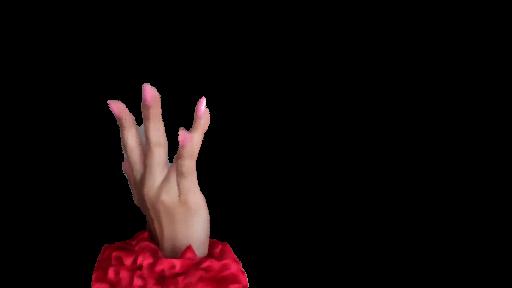} &
\includegraphics[width = 1.3in]{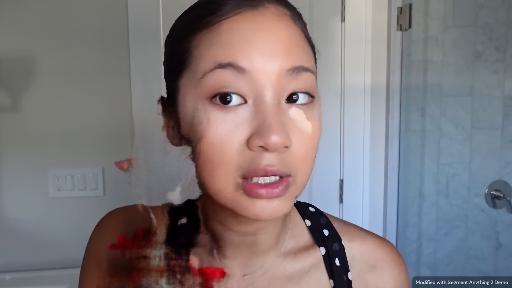} &
\includegraphics[width = 1.3in]{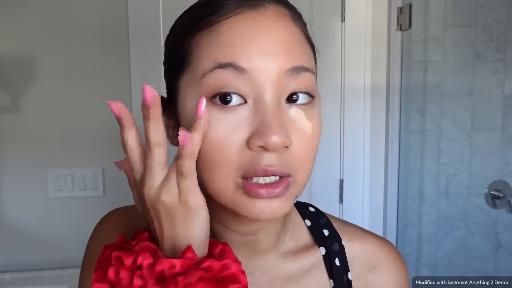} \\

\includegraphics[width = 1.3in]{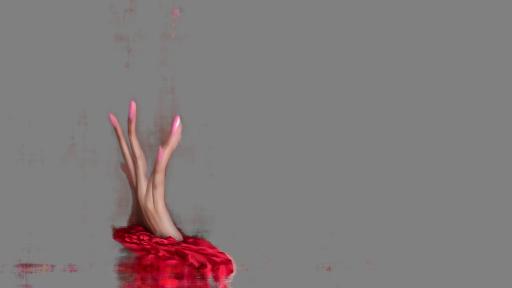} &
\includegraphics[width = 1.3in]{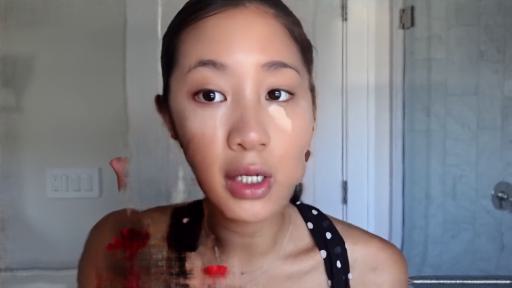} &
\includegraphics[width = 1.3in]{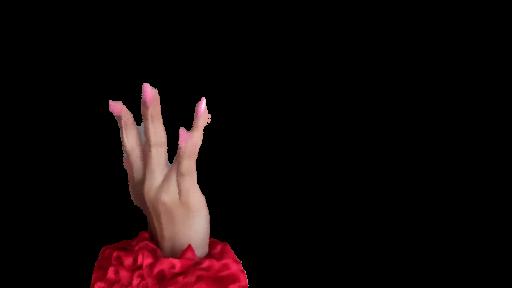} &
\includegraphics[width = 1.3in]{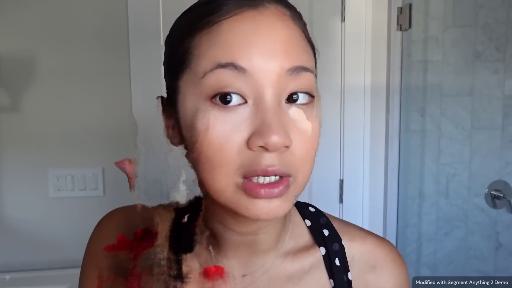} &
\includegraphics[width = 1.3in]{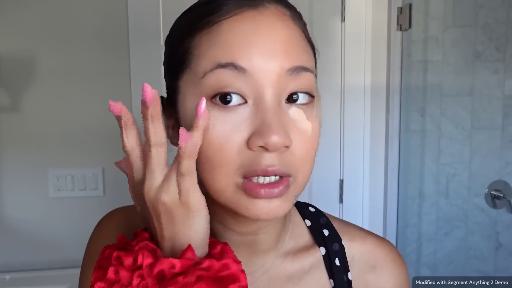} \\

\includegraphics[width = 1.3in]{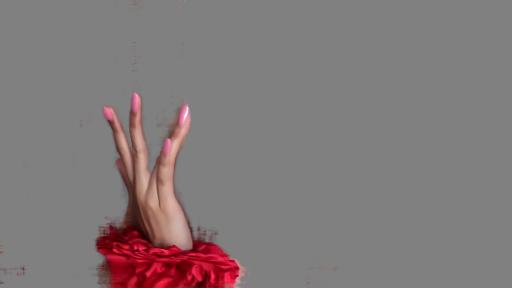} &
\includegraphics[width = 1.3in]{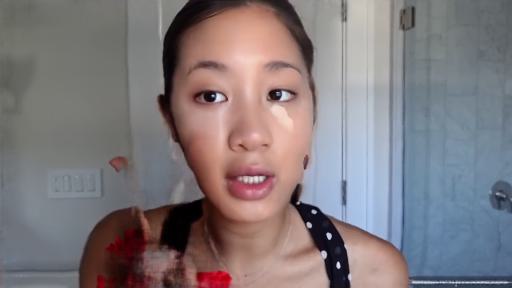} &
\includegraphics[width = 1.3in]{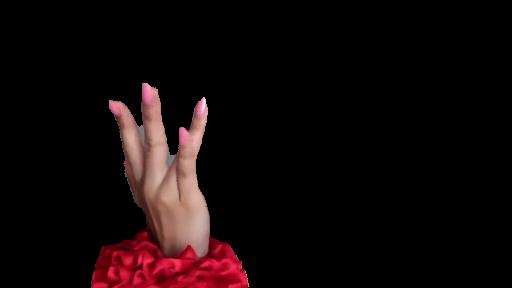} &
\includegraphics[width = 1.3in]{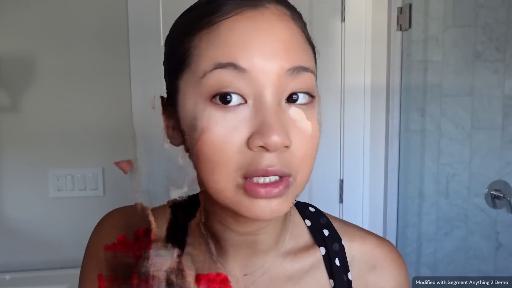} &
\includegraphics[width = 1.3in]{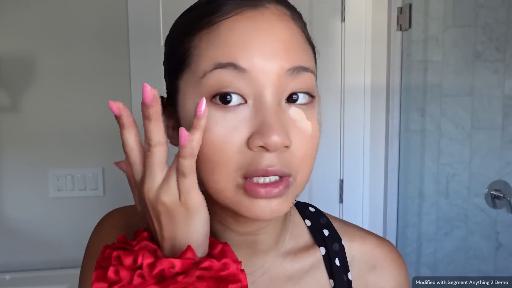} \\

\includegraphics[width = 1.3in]{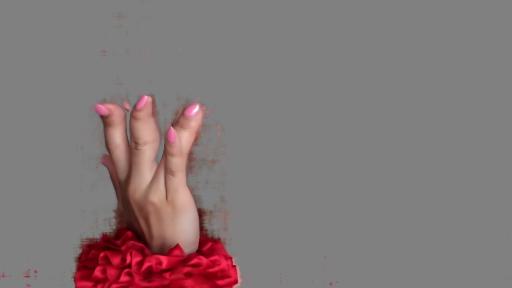} &
\includegraphics[width = 1.3in]{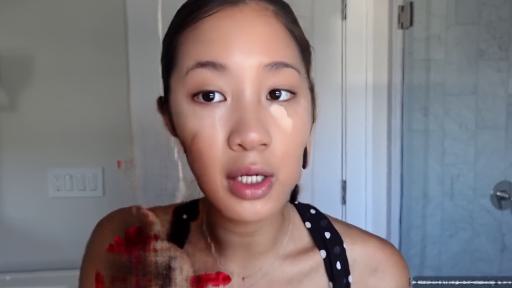} &
\includegraphics[width = 1.3in]{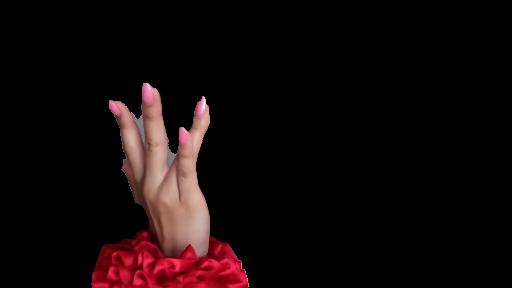} &
\includegraphics[width = 1.3in]{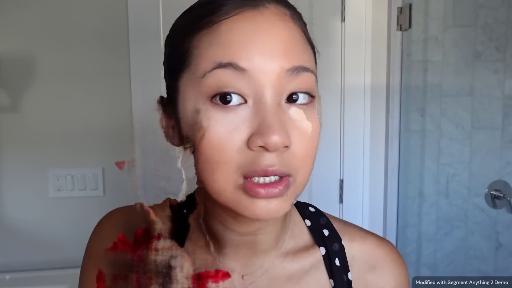} &
\includegraphics[width = 1.3in]{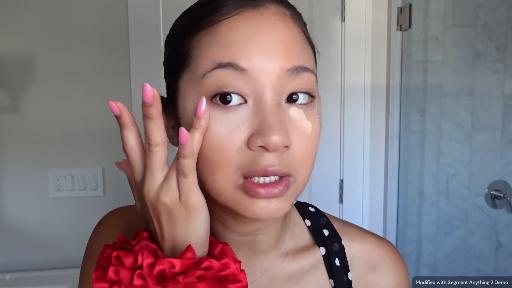} \\

\includegraphics[width = 1.3in]{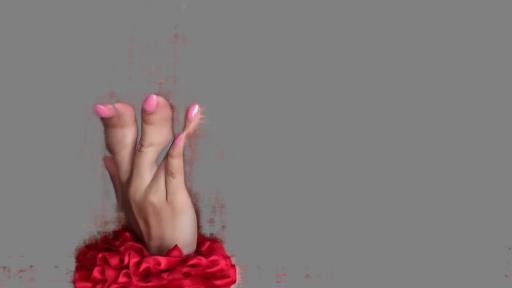} &
\includegraphics[width = 1.3in]{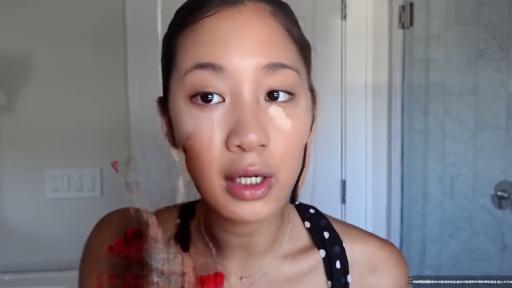} &
\includegraphics[width = 1.3in]{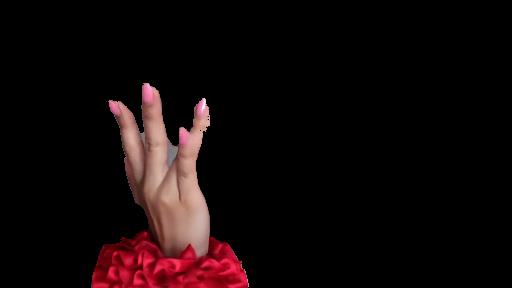} &
\includegraphics[width = 1.3in]{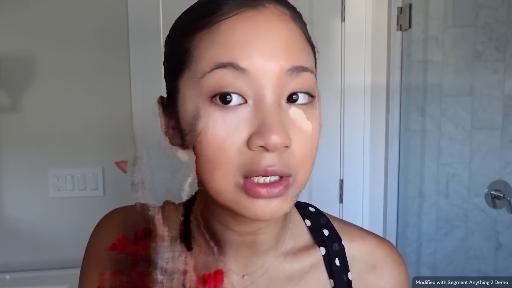} &
\includegraphics[width = 1.3in]{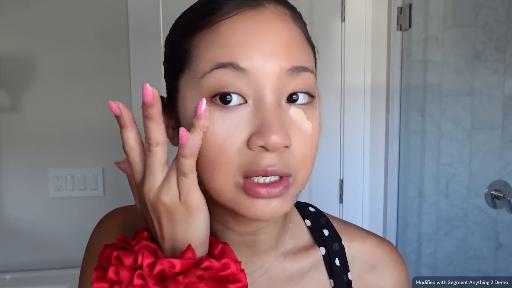} \\

\end{tabular}
  \caption{Additional ablation studies on number of control points. \textit{From top to bottom:} 2 (24.195dB), 4 (26.132dB), 8 (28.433dB), 16 (32.209dB), 20 (32.721dB), 30 (34.280dB), 41 (37.010dB), and 82 (36.606dB) control points for a video of 41 frames.}
  \label{fig:add_ablation_ncp}
\end{figure*}

Fig.~\ref{fig:add_ablation_ncp} provides additional ablation studies on the number of control points. While the best fit can be obtained with the number of control points equal to the number of frames, our method can also reasonably reconstruct the scene with fewer control points.

\subsection{Number of iterations}
\camready{While performance optimization was not the research focus of this work, we acknowledge the processing time can be accelerated using faster neural representations (e.g. hash encodings \cite{muller2022instant}), optimized learning libraries (e.g. PyTorch Lightning \cite{Falcon_PyTorch_Lightning_2019}), and quantization (half-precision). We provide additional ablation studies on the effects of training iterations in Fig. \ref{fig:r_iterations}. As observed, reasonable results can be obtained with 30K iterations ($<$30min), with only a 1dB drop w.r.t. the fully trained model ($\sim$90min).}

\section{Failure cases}
\label{sec:sup_failure}
\begin{figure*}[h]
  \centering
  \setlength{\tabcolsep}{0.5pt}
\begin{tabular}{cccccc}
Can. Foreground & Can. Background & Foreground & Background & Composited & GT\\

\includegraphics[width = 1.1in]{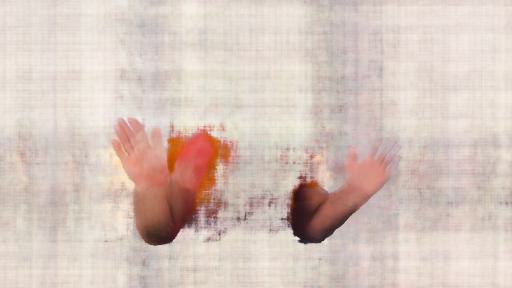} &
\includegraphics[width = 1.1in]{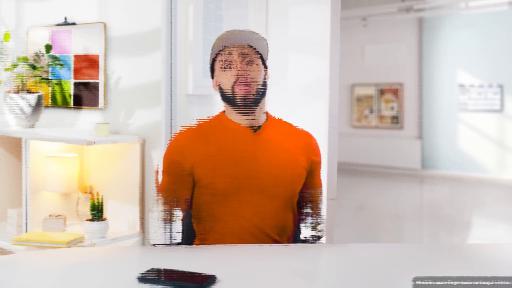} &
\includegraphics[width = 1.1in]{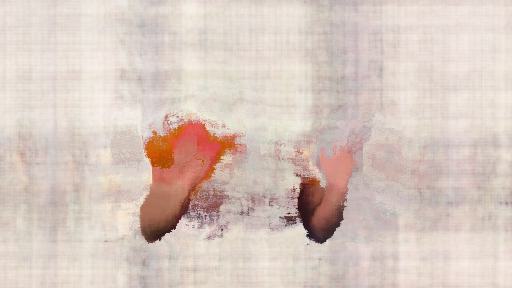} &
\includegraphics[width = 1.1in]{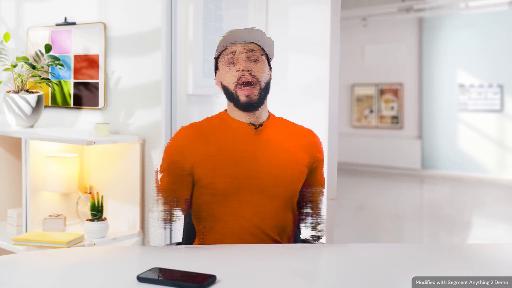} &
\includegraphics[width = 1.1in]{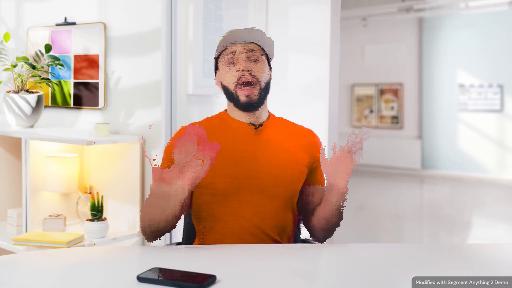} & 
\includegraphics[width = 1.1in]{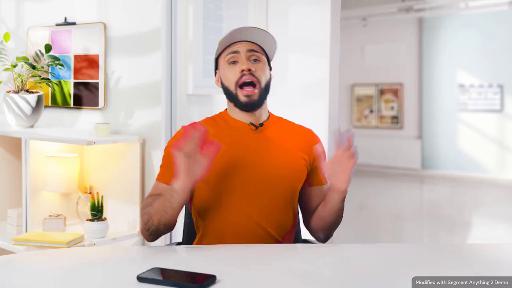} \\

\includegraphics[width = 1.1in]{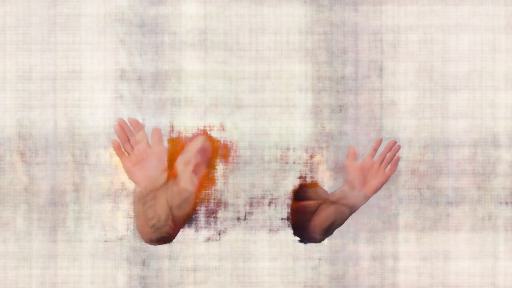} &
\includegraphics[width = 1.1in]{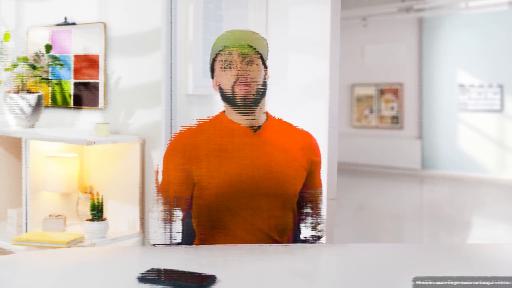} &
\includegraphics[width = 1.1in]{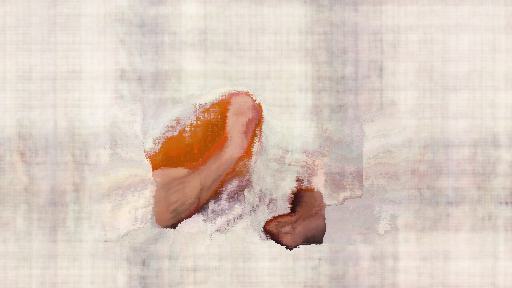} &
\includegraphics[width = 1.1in]{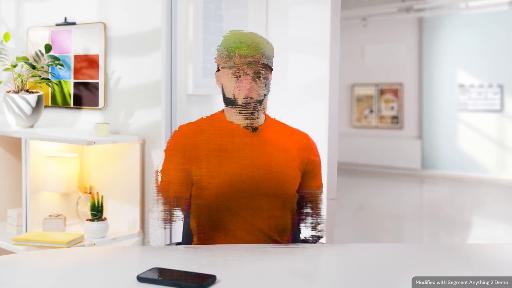} &
\includegraphics[width = 1.1in]{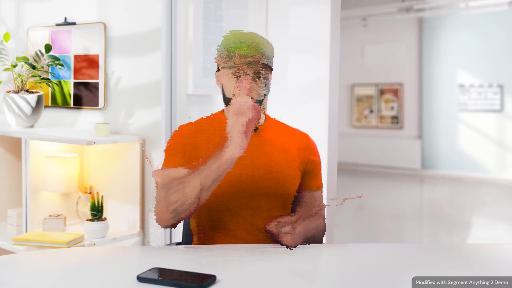} & 
\includegraphics[width = 1.1in]{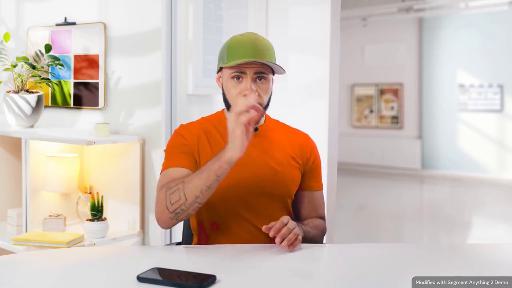} \\

\includegraphics[width = 1.1in]{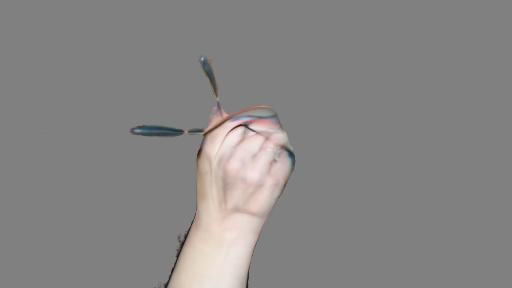} &
\includegraphics[width = 1.1in]{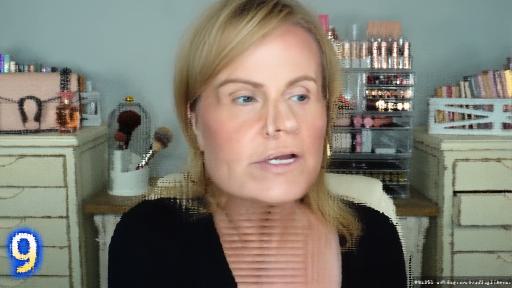} &
\includegraphics[width = 1.1in]{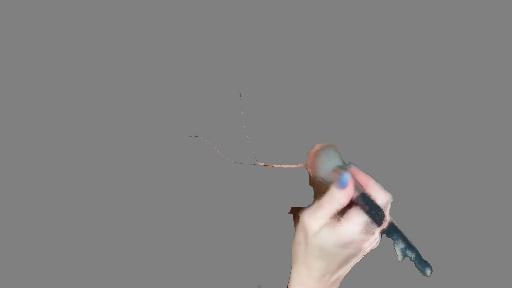} &
\includegraphics[width = 1.1in]{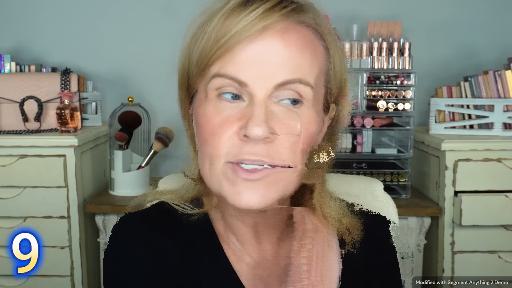} &
\includegraphics[width = 1.1in]{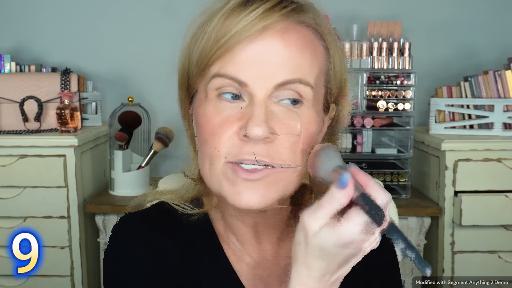} & 
\includegraphics[width = 1.1in]{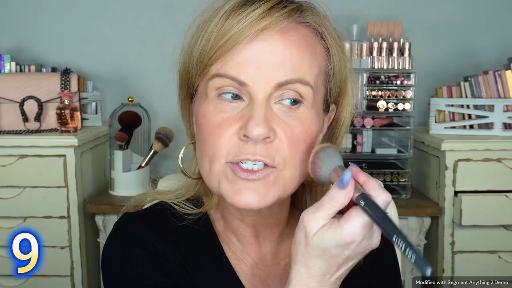} \\

\includegraphics[width = 1.1in]{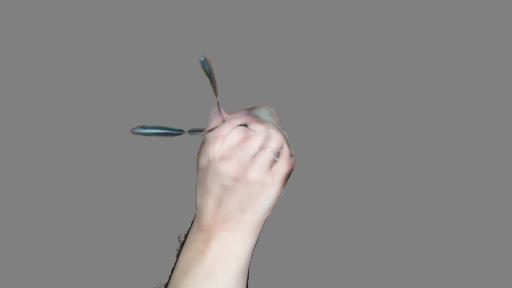} &
\includegraphics[width = 1.1in]{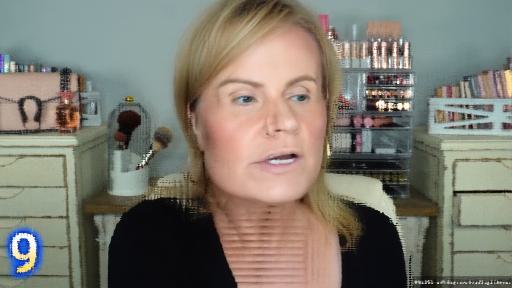} &
\includegraphics[width = 1.1in]{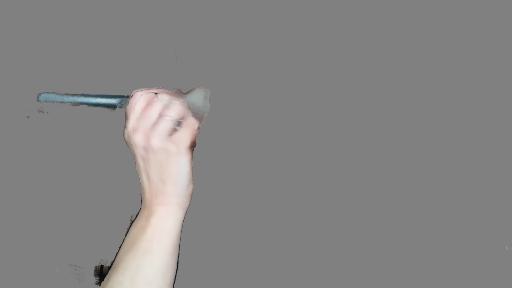} &
\includegraphics[width = 1.1in]{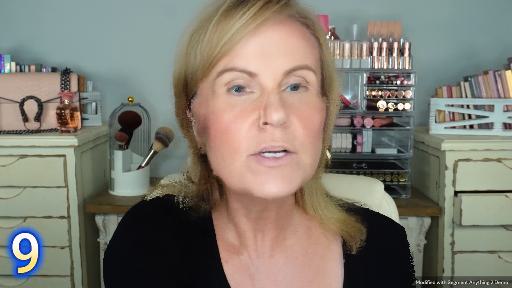} &
\includegraphics[width = 1.1in]{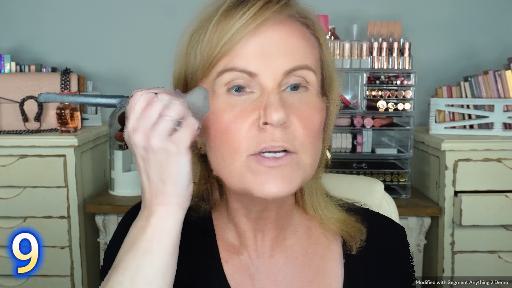} & 
\includegraphics[width = 1.1in]{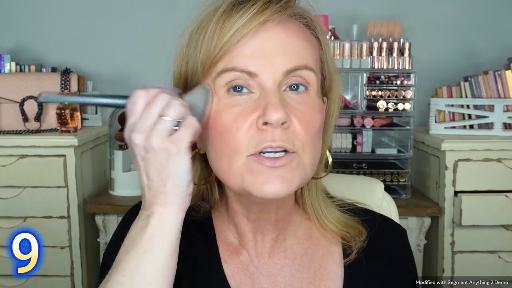} \\

\end{tabular}
  \caption{Failure cases. \textit{Top two rows:} The dynamic region in background image (face region) is too small. \textit{Bottom two rows:} Too large foreground motion and self-occlusion (opposite sides of brush) cause a double brush effect in foreground canonical space.}
  \label{fig:add_failure}
\end{figure*}

Fig. \ref{fig:add_failure} illustrates examples of failure cases. In the top two rows, our method fails to reconstruct a feasible canonical space for the background face. This is because the relative size of the facial region with respect the amount and complexity of the motion is very small. A work-around for this issue would be to crop the images around the face region and run our method again. In the bottom two rows, the amount of motion is too large for our model to capture. In these cases, the brush goes from one side to the other and also rotates showing different faces of it, inducing two brushes on our estimated canonical space. A potential solution would consist on modeling the brush with different layers when it is on one side or the other. 

\section{Ethical implications}
\label{sec:sup_ethical}

The use of ControlNet in conjunction with our proposed method to modify the appearance of video content may raise ethical concerns around authenticity and potential misuse, such as creating misleading information. To address these concerns, we advocate for the responsible and transparent use of this technology, ensuring that any modifications are clearly indicated and used ethically.

Our collected dataset from publicly available YouTube videos contains exclusively \textit{Creative Commons} licensed videos, with the corresponding URLs provided in the \texttt{\small{urls.json}} file. Authors of these videos are free to contact us upon publication (due to the anonymous nature of submission) to have their videos removed from this dataset or paper results.


\clearpage
\clearpage
{
\small
\bibliographystyle{ieeenat_fullname}
\bibliography{main}
}

\end{document}